\documentclass[11pt,english,aps,pra,onecolumn,tightenlines,superscriptaddress,notitlepage,floatfix,fleqn]{revtex4-1}
\pdfoutput=1
\usepackage[utf8]{inputenc}
\usepackage[T1]{fontenc}
\usepackage{amsmath}
\usepackage{amssymb}
\usepackage{amsfonts}
\usepackage{bm}
\usepackage{bbm}
\usepackage{epsfig}
\usepackage{times}
\usepackage{amsthm}

\usepackage{numprint}
\usepackage{enumitem}

\usepackage[usenames,dvipsnames]{color}
\definecolor{grey}{rgb}{.35,.35,.35}
\definecolor{dblue}{rgb}{0,0.1,.6}
\definecolor{dgreen}{rgb}{0,.6,0.1}

\usepackage[colorlinks=true,citecolor=dblue,linkcolor=dblue,urlcolor=dblue]{hyperref}
\usepackage[all]{hypcap}
\newcommand{\Footnote}[1]{\footnote{\unexpanded{#1}}}

\newcommand{\id}{\mathbbm{1}}
\newcommand{\Id}{\operatorname{Id}}
\newcommand{\bra}{\langle}
\newcommand{\ket}{\rangle}
\newcommand{\bbra}{\langle\!\langle}
\newcommand{\kket}{\rangle\!\rangle}
\newcommand{\Tr}{\operatorname{Tr}}
\renewcommand{\vec}[1]{{\boldsymbol{#1}}}
\newcommand{\ud}{\mathrm{d}}

\newcommand{\B}{\mc{B}}
\newcommand{\E}{\mc{E}}
\newcommand{\F}{\mc{M}}
\newcommand{\G}{\mc{G}}
\newcommand{\M}{\mc{M}}
\newcommand{\Q}{\mc{Q}}

\newcommand{\hH}{\hat{H}}
\newcommand{\hh}{\hat{h}}
\newcommand{\hA}{\hat{A}}
\newcommand{\hB}{\hat{B}}
\newcommand{\hG}{\hat{G}}
\newcommand{\hL}{\hat{L}}
\newcommand{\hl}{\hat{\ell}}
\newcommand{\hP}{\hat{P}}
\newcommand{\hR}{\hat{R}}
\newcommand{\hr}{\hat{r}}
\newcommand{\hU}{\hat{U}}
\newcommand{\tU}{\tilde{U}}
\newcommand{\hV}{\hat{V}}
\newcommand{\hW}{\hat{W}}
\newcommand{\hX}{\hat{X}}
\newcommand{\hx}{\hat{x}}
\newcommand{\hY}{\hat{Y}}
\newcommand{\hy}{\hat{y}}
\newcommand{\hZ}{\hat{Z}}
\newcommand{\hd}{\hat{d}}
\newcommand{\hg}{\hat{g}}
\newcommand{\hsigma}{\hat{\sigma}}
\newcommand{\dm}{{\hat{\rho}}}

\newcommand{\va} {\vec{a}}
\newcommand{\vs} {\vec{s}}
\newcommand{\vsT}{\vs^\intercal}
\newcommand{\vt} {\vec{t}}
\newcommand{\vtT}{\vt^\intercal}

\newcommand{\CC}{\mathbb{C}}
\renewcommand{\Re}{\operatorname{Re}}
\newcommand{\Span}{\operatorname{span}}
\newcommand{\End} {\operatorname{End}}
\newcommand{\groupU} {\operatorname{U}}
\newcommand{\Avg} {\operatorname{Avg}}
\newcommand{\Var} {\operatorname{Var}}
\newcommand{\Swap}{\operatorname{Swap}}
\newcommand{\doub}{{(2)}}
\newcommand{\Out}{{\operatorname{out}}}
\newcommand{\tb}{\text{b}}
\newcommand{\tR}{\text{R}}
\newcommand{\tL}{\text{L}}
\newcommand{\tC}{\text{C}}

\newcommand{\bin}{\text{bi}}
\newcommand{\ter}{\text{ter}}
\newcommand{\non}{\text{non}}

\newcommand{\mc}[1]{\mathcal{#1}}
\newcommand{\pdag}{{\phantom{\dag}}}

\renewcommand{\L}{\mc{L}}
\renewcommand{\S}{\mc{S}}
\newcommand{\veps}{\varepsilon}
\newcommand{\mri}{\mathrm{i}}

\newcommand  {\Pmatrix}[1]{\begin{pmatrix}#1\end{pmatrix}}
\newcommand  {\Psmatrix}[1]{\left(\begin{smallmatrix}#1\end{smallmatrix}\right)}

\newcommand{\Emph}[1]{\textbf{\emph{#1}}}

\newtheorem{theorem}{Theorem}
\newcommand{\propHead}[1]{\bfseries \emph{#1}}

\renewcommand{\thesection}{\Roman{section}}
\renewcommand{\thesubsection}{\thesection.\arabic{subsection}}
\makeatletter
\renewcommand{\p@subsection}{}
\renewcommand{\p@subsubsection}{}
\makeatother

\newcommand{\duke} {Department of Physics, Duke University, Durham, North Carolina 27708, USA}
\newcommand{\dqc}  {Duke Quantum Center, Duke University, Durham, North Carolina 27701, USA}
\newcommand{\tz}   {Tensor Center, Auf dem Dresch 15, 52152 Simmerath, Germany}

\begin{document}

\title{\texorpdfstring{Absence of barren plateaus and scaling of gradients in the energy optimization of\\ isometric tensor network states}{Absence of barren plateaus and scaling of gradients in the energy optimization of isometric tensor network states}}
\author{Thomas Barthel}
\affiliation{\duke}
\affiliation{\dqc}
\affiliation{\tz}
\author{Qiang Miao}
\affiliation{\duke}
\affiliation{\dqc}

\begin{abstract}
Vanishing gradients can pose substantial obstacles for high-dimensional optimization problems. Here we consider energy minimization problems for quantum many-body systems with extensive Hamiltonians and finite-range interactions, which can be studied on classical computers or in the form of variational quantum eigensolvers on quantum computers. Barren plateaus correspond to scenarios where the average amplitude of the energy gradient decreases exponentially with increasing system size. This occurs, for example, for quantum neural networks and for brickwall quantum circuits when the depth increases polynomially in the system size. Here we prove that the variational optimization problems for matrix product states, tree tensor networks, and the multiscale entanglement renormalization ansatz are free of barren plateaus. The derived scaling properties for the gradient variance provide an analytical guarantee for the trainability of randomly initialized tensor network states (TNS) and motivate certain initialization schemes. In a suitable representation, unitary tensors that parametrize the TNS are sampled according to the uniform Haar measure. We employ a Riemannian formulation of the gradient based optimizations which simplifies the analytical evaluation.
\end{abstract}

\date{June 8, 2023}

\maketitle

\renewcommand{\baselinestretch}{0.9}\normalsize
\tableofcontents
\renewcommand{\baselinestretch}{1}\normalsize

\section{Introduction and summary of results}
\vspace{-0.5em}
Plateau phenomena are a common feature of high-dimensional optimization problems, where optimization progress may be substantially hampered by vanishing gradients of the cost function \cite{Hochreiter1998-06,Fukumizu2000-13,Dauphin2014-2,Shalev2017-70}. Here, we are concerned with the energy minimization for quantum many-body states $|\Psi\ket$. In particular, given an ansatz $|\Psi\ket$ and a Hamiltonian $\hH$, the goal is to minimize $\bra\Psi|\hH|\Psi\ket$ under the constraint $\|\Psi\|=1$. In Ref.~\cite{McClean2018-9}, McClean \emph{et al.}\ considered states that are generated by a random unitary circuit. They found vanishing average expectation values for energy gradients and that the variance of the energy gradient decreases exponentially in the system size if relevant parts of the unitary circuit form 2-designs. This phenomenon is referred to as barren plateaus. It is prevalent in various variational quantum eigensolvers like quantum neural networks \cite{McClean2018-9,Cerezo2021-12,Ortiz2021-2,Uvarov2021-54,Sharma2022-128,Napp2022_03} and goes hand in hand with an exponential decay of the cost-function variance \cite{Arrasmith2022-7,Miao2024-9}. For specific cases, it has been established that the severity of the barren-plateau phenomenon is related to the expressiveness of the ansatz for $|\Psi\ket$ \cite{Cerezo2021-12,Patti2021-3,Holmes2022-3}.
Profiting from the excellent machine precision, small gradients are tolerable to some extent for optimizations on classical computers. The issue is more pressing when evaluating gradients on quantum computers. With $N_s$ measurement samples per term, the statistical error of the gradient and, hence,
the achievable energy accuracy improve slowly as $1/\sqrt{N_s}$. It is hence difficult to accurately determine small gradients, and one may end up doing a random walk in a flat region of the energy landscape.

In this work, we study how the amplitude of energy gradients in tensor network states (TNS) scale with the system size, distances, and bond dimensions. TNS \cite{Baxter1968-9,White1992-11,Niggemann1997-104,Verstraete2004-7,Vidal-2005-12,Schollwoeck2011-326,Orus2014-349} encode the state $|\Psi\ket$ as a network of partially contracted tensors, where non-contracted indices label local basis states, e.g., referring to the $z$ magnetization of a spin on a given lattice site. The other indices are referred to as virtual or bond indices and the dimension of the associated vector space is called the bond dimension \cite{Orus2014-349}. Tensor network methods are a powerful approach for the investigation of strongly-correlated quantum matter. While they are so far mostly used in classical simulations, they can also be employed in variational quantum eigensolvers \cite{McClean2016-18} for the study of quantum many-body systems on quantum computers \cite{Liu2019-1,Smith2019-5,Miao2021_08,Slattery2021_08,Barratt2021-7,FossFeig2021-3,Niu2022-3,Chertkov2022-18}.
Tensor networks are, in a sense, decidedly unexpressive in order to allow for an efficient optimization while capturing the relevant physics. For fixed bond dimension, they typically feature entanglement area or log-area laws, consistent with the scaling of entanglement entropies in ground states of (typical) quantum many-body systems \cite{Srednicki1993,Callan1994-333,Holzhey1994-424,Vidal2003-7,Jin2004-116,Latorre2004,Calabrese2004,Zhou2005-12,Plenio2005,Wolf2005,Gioev2005,Barthel2006-74,Li2006,Cramer2006-73,Hastings2007-76,Brandao2013-9,Cho2018-8,Kuwahara2020-11}. See Refs.~\cite{Eisert2008,Latorre2009,Laflorencie2016-646} for reviews on this topic.

We consider lattice systems and address three prominent classes of isometric TNS which, without loss of generality, can be entirely parametrized by unitary tensors -- matrix product states (MPS) \cite{Baxter1968-9,Accardi1981,Fannes1992-144,White1992-11,Rommer1997,PerezGarcia2007-7,Schollwoeck2011-326} with open boundary conditions, tree tensor network states (TTNS) \cite{Fannes1992-66,Otsuka1996-53,Shi2006-74,Murg2010-82,Tagliacozzo2009-80}, and the multiscale entanglement renormalization ansatz (MERA) \cite{Vidal-2005-12,Vidal2006}.
For extensive Hamiltonians $\hH=\sum_i\hh_i$, where the finite-range interaction term $\hh_i$ acts non-trivially in the vicinity of lattice site $i$, our analytical results on the scaling of Haar-averaged TNS energy-gradient amplitudes prove that the corresponding optimization problems are free of barren plateaus. The dependence on bond dimensions and the layer depth in TTNS and MERA bears implications for efficient initialization procedures. Note that this class of groundstate problems is Quantum-Merlin-Arthur (QMA) complete \cite{Kempe2006-35,Oliveira2008-8,Aharonov2009-287,Gottesman2009_05,Bausch2017-18}.

The absence of barren plateaus for the considered isometric TNS is consistent with years of successful classical simulations in condensed matter theory and other fields. One might also expect it based on results for the logarithmic-depth alternating layered ansatz \cite{Cerezo2021-12}, closely related circuits addressed in Ref.~\cite{Uvarov2021-54}, and prior results for local optimization problems on specific subclasses of TNS \cite{Liu2019-1,Liu2022-129,Garcia2023-2023,Zhao2021-5,Martin2023-7}, which are discussed in the following.
\vspace{-0.5em}

\subsection{Prior work on barren plateaus for tensor network states}\label{sec:priorWork}
\vspace{-0.5em}
Refs.~\cite{Liu2022-129,Garcia2023-2023} study a subclass of MPS $|\Psi\ket$ with periodic boundary conditions, isometric MPS tensors, and with variable norms $\|\Psi\|$ which are exponentially concentrated around one \cite{Haferkamp2021-2}. For the maximization of the overlap to a second state $|\Phi\ket$, Liu \emph{et al.}\ \cite{Liu2022-129} find that the Haar-averaged gradient amplitudes decrease exponentially with increasing system size. This is consistent with the orthogonality catastrophe, i.e., the exponential decrease of random-state overlaps with increasing system size. For the minimization of expectation values $\bra\Psi|\hh_i|\Psi\ket$ for a single-site Hamiltonian $\hh_i$ that acts non-trivially only on site $i$, Refs.~\cite{Liu2022-129,Garcia2023-2023} provide an upper bound on the Haar-averaged variance of the gradient with respect to the MPS tensor of site $j$. The obtained bound decays exponentially as $d^{-|i-j|}$ where $d$ is the single-site Hilbert-space dimension.

Zhao and Gao \cite{Zhao2021-5} employ ZX-calculus \cite{Coecke2011-13} to evaluate gradient variances for a subclass of right-orthonormal MPS with open boundary conditions, bond dimension $m=2$, and single-site dimension $d=2$ \footnote{The MPS, TTNS, and MERA discussed in Refs.~\cite{Zhao2021-5,Martin2023-7} with ZX-calculus have bond dimension $m=2$ and each tensor involves one CNOT gate. Hence, they form subclasses of the corresponding unconstrained $m=2$ tensor networks: The representation of a general two-qubit unitary in terms of CNOT and single-qubit gates requires three CNOT gates \cite{Vatan2004-69,Shende2004-69}.}.
For a single-site Hamiltonian $\hh_L$ acting on the last site, it is shown that the average variance of the gradient with respect to the MPS tensor $\hU_1$ of site $j=1$ decreases exponentially in the system size $L$. Cervero Martín \emph{et al.}\ \cite{Martin2023-7} extend this result, showing that the average gradient variance with respect to $\hU_1$ for a product operator $\hh_i\hh_{i+1}$ acting non-trivially on sites $i$ and $i+1$ also decays exponentially in $i$. Some further results turn out to be non-generic and due to the particular tensors chosen for these $m=2$ MPS. See Appx.~\ref{appx:ZX-MPS} for a detailed discussion.
Similarly, Cervero Martín \emph{et al.}\ \cite{Martin2023-7} apply ZX-calculus to subclasses of binary TTNS and MERA with bond dimension $\chi=2$ and single-site dimension $d=2$ \cite{Note1}. For a single-site Hamiltonian $\hh_i$, the average variance of the gradient with respect to the top tensor in layer $T$ is found to decrease algebraically with respect to the system size $L=2^T$ \footnote{Zhao and Gao \cite{Zhao2021-5} also discuss a ``tree tensor network ansatz'' and a (MERA-like) quantum convolutional neural network with ZX-calculus and $\chi=d=2$. However, the considered optimization problem has a single-site ``observable'' in the final layer of the network (renormalized lattice $\L_T$ in our notation of Sec.~\ref{sec:MERA}) and the reference state in the first layer ($\L_0$). This inverted problem has no direct relation to the energy minimization problem for TTNS and MERA that we address here.}.

Note that the TNS optimization problems for single-site Hamiltonians $\hh_i$ and two-site products $\hh_i\hh_{i+1}$ considered in Refs.~\cite{Liu2022-129,Garcia2023-2023,Zhao2021-5,Martin2023-7} and for general product Hamiltonians $\hh_1\otimes\dotsb\otimes \hh_L$ are solved by product states $|\Psi\ket=|\phi_1\ket\otimes\dotsb\otimes|\phi_L\ket$ composed of single-site eigenstates $|\phi_i\ket$ of the operators $\hh_i$. The same holds for sums $\sum_i\hh_i$ of single-site Hamiltonians.

These seminal works on single-site and two-site product Hamiltonians \cite{Liu2022-129,Garcia2023-2023,Zhao2021-5,Martin2023-7} motivate our investigation for extensive Hamiltonians $\hH=\sum_i\hh_i$ with finite-range interactions $\hh_i$.
For a specific model, the latter case has been analyzed numerically in Ref.~\cite{Liu2019-1}: For open spin-1/2 Heisenberg chains and a subclass of MPS generated from single-qubit and CNOT gates, Fig.~6 of Ref.~\cite{Liu2019-1} suggests a power-law decay of the energy-gradient variance. This turns out to be a finite-size effect, and the gradient variance ultimately converges to a system-size independent value at larger $L$ as suggested by Theorem~\ref{thrm:MPSnn} below. See Appx.~\ref{appx:QMPS} and Ref.~\cite{Miao2024-109} for numerical checks and a discussion of the finite-size effects.

\subsection{Main results and methods}
We extend the prior work, considering groundstate problems for extensive Hamiltonians $\hH=\sum_i\hh_i$ with finite-range interactions $\hh_i$, which are generally QMA complete. We show that the corresponding energy optimization with respect to MPS, TTNS, and MERA and any bond dimensions has no barren plateaus and elucidate the scaling of gradient variances with respect to the bond dimension and layer number.

Instead of employing particular parametrizations for the TNS tensors, we formulate the optimization problems in terms of Riemannian gradients.
All results are based on first- and second-moment Haar-measure integrals over the relevant unitary groups. The Haar-averaged energy gradients are zero and the scaling of gradient variances is deduced from the spectra of quantum channels that propagate in the spatial direction for MPS and in the preparation direction for TTNS and MERA.

In Sec.~\ref{sec:MPS}, we consider general heterogeneous MPS $|\Psi\ket$ with bond dimension $m$, single-site Hilbert space dimension $d$, norm one, and open boundary conditions. Exploiting their gauge freedom, we can bring such MPS into left-orthonormal form, where all MPS tensors are isometries \cite{Schollwoeck2011-326,Barthel2022-112}. To assess the question of barren plateaus, the isometries \Footnote{An operator $\hW:\CC^m\to\CC^n$ is a partial isometry if $\hW^\dag\hW=\id_m$. For brevity we refer to such operators as isometries and, for $m=n$, as unitaries.} are drawn uniformly from the corresponding Stiefel manifolds. Equivalently, the isometries can be realized as partially projected unitaries and the unitaries $\hU_j$ be drawn according to the Haar measure.
We begin by revisiting the optimization problem for single-site Hamiltonians, finding that $\Var \partial_{\hU_j} \bra\Psi|\hh_i|\Psi\ket \sim
2\Tr(\hh^2)\eta^{j-i}/(md)^2$ for $j\geq i$ and zero otherwise, where the decay factor is $\eta=d(m^2-1)/(m^2d^2-1)$ [Theorem~\ref{thrm:MPS-dist}]. Note that $\eta\sim 1/d $ is consistent with the bound from Ref.~\cite{Liu2022-129}.
A similar result holds for nearest-neighbor interaction terms $\hh_i$ [Theorem~\ref{thrm:MPSnn}]. For extensive Hamiltonians $\hH=\sum_i\hh_i$ with single-site terms $\hh_i$, we find $\Var \partial_{\hU_j} \bra\Psi|\hH|\Psi\ket\sim 2\Tr(\hh^2)/(md)^2$ [Theorem~\ref{thrm:MPSext-bond}]. Of course, this optimization problem is still trivially solved by product states. For the practically relevant case of nearest-neighbor interactions $\hh_i$, the energy gradient-variance is found to assume the system-size independent value $\Var \partial_{\hU_j} \bra\Psi|\hH|\Psi\ket \sim 4\big[\Tr(\hh^2)+2\Tr(\Tr^2_1\hh)\big]/(m^2 d^4)$ [Theorem~\ref{thrm:MPSnn}].

In Sec.~\ref{sec:MERA}, we consider heterogeneous TTNS and MERA $|\Psi\ket$ with bond dimension $\chi$ and extensive Hamiltonians with finite-range interactions. For simplicity, single-site Hilbert space dimensions are chosen as $d=\chi$, but results carry over to $d<\chi$ by interpreting our Hamiltonians as those arising from the physical Hamiltonians after a few coarse-graining or renormalization steps. 
TTNS and MERA \cite{Fannes1992-66,Otsuka1996-53,Shi2006-74,Murg2010-82,Tagliacozzo2009-80,Vidal-2005-12,Vidal2006} are hierarchical tensor networks consisting of unitary disentanglers and isometries that map $b$ renormalized sites of layer $\tau$ into one renormalized site in layer $\tau+1$.
For binary one-dimensional (1D) MERA, we establish that there is a finite fraction of unitaries $\hU_{\tau,k}$ in layer $\tau$ for which $\Var \partial_{\hU_{\tau,k}} \bra\Psi|\hH|\Psi\ket \sim \Theta\big((2\eta_\bin)^\tau\big)$ [Theorem~\ref{thrm:1dMERAbin}]. Here, $\eta_\bin =[\chi^2(1+\chi)^4]/[2(1+\chi^2)^4]$, the Landau symbol $\Theta(f)$ indicates that there exist upper and lower bounds scaling like $f$, and we have omitted $\tau$-independent prefactors ($\tau$-independent algebraic functions in $\chi$). Similarly, for ternary 1D MERA, $\Var \partial_{\hU_{\tau,k}} \bra\Psi|\hH|\Psi\ket \sim \Theta\big((3\eta_\ter)^\tau\big)$ with $\eta_\ter\sim 1/(3\chi^2)$ [Theorem~\ref{thrm:1dMERAter}], and $\Var \partial_{\hU_{\tau,k}} \bra\Psi|\hH|\Psi\ket \sim \Theta\big((9\eta_\non)^\tau\big)$ with $\eta_\non\sim 1/(9\chi^8)$ for nonary 2D MERA [Sec.~\ref{sec:MERA-nonary}]. In Sec.~\ref{sec:furtherMERA}, we explain why, generically, optimization problems for TTNS and MERA with extensive Hamiltonians have system-size independent gradient variances, i.e., do not feature barren plateaus.

In these analyses, we assume nearest-neighbor interactions $\hh_i$ for MPS and ternary 1D MERA and TTNS, up to next-nearest neighbor interactions for binary 1D MERA and TTNS, and interaction terms $\hh_i$ on $2\times 2$ site blocks for nonary 2D MERA and TTNS. The results can be generalized to systems with longer-ranged interactions: One can either reduce the interaction range by coarse-graining the lattice (blocking sites) or by explicit adaptation of the proofs.

\newpage
\section{Preliminaries: Haar measure averages and Riemannian gradients}\label{sec:Preliminaries}
\vspace{-0.5em}
In the following, let us discuss some preliminaries on Haar-measure integrals, corresponding quantum channels, and Riemannian gradients, and introduce the corresponding notations.

\subsection{Haar measure integrals}\label{sec:Haar}
\vspace{-0.5em}
We will frequently evaluate Haar-measure averages for certain expressions involving unitaries $\hU\in\groupU(N)$. The first and second moments are covered by the Weingarten formulas \cite{Weingarten1978-19,Collins2006-264}
\begin{subequations}\label{eq:Haar}
\begin{equation}
	\int_{\groupU(N)}\ud U\, U_{i_1,j_1} U^*_{m_1,n_1} = \frac{1}{N} \delta_{i_1,m_1}\delta_{j_1,n_1}
\end{equation}\vspace{-1em}
\begin{align}\nonumber
	\int_{\groupU(N)}\ud U\, U_{i_1,j_1}U_{i_2,j_2}& U^*_{m_1,n_1} U^*_{m_2,n_2}
	=\frac{1}{N^2-1}\big(\delta_{i_1,m_1}\delta_{i_2,m_2} \delta_{j_1,n_1}\delta_{j_2,n_2}
	                      +\delta_{i_1,m_2}\delta_{i_2,m_1} \delta_{j_1,n_2}\delta_{j_2,n_1}\big)\\
	 &-\frac{1}{N(N^2-1)}\big(\delta_{i_1,m_1}\delta_{i_2,m_2} \delta_{j_1,n_2}\delta_{j_2,n_1}
	                            +\delta_{i_1,m_2}\delta_{i_2,m_1} \delta_{j_1,n_1}\delta_{j_2,n_2}\big),
\end{align}
\end{subequations}
where 
\begin{equation}
	\int_{\groupU(N)}\ud U\,f(U)=:\Avg_U\,f(U)
\end{equation}
denotes the Haar-measure integral over the unitary group of degree $N$.
Using the Dirac bra-ket notation and tensor products, they can be written in the convenient form
\begin{subequations}
\begin{equation}\label{eq:Haar1}
	\int_{\groupU(N)}\ud U\, \hU\otimes \hU^\dag = \frac{1}{N}\sum_{i,j}|i,j\ket\bra j,i|=:\frac{1}{N}\Swap
\end{equation}\vspace{-1em}
\begin{align}\nonumber
	\int_{\groupU(N)}&\ud U\, \hU\otimes \hU\otimes \hU^\dag\otimes \hU^\dag\\\nonumber
	&\ =\quad\ \frac{1}{N^2-1}\quad\sum_{i_1,i_2,j_1,j_2}\Big(|i_1,i_2,j_1,j_2\ket\bra j_1,j_2,i_1,i_2| + |i_1,i_2,j_2,j_1\ket\bra j_1,j_2,i_2,i_1| \Big)\\\nonumber
	&\quad -\frac{1}{N(N^2-1)}\sum_{i_1,i_2,j_1,j_2}\Big(|i_1,i_2,j_2,j_1\ket\bra j_1,j_2,i_1,i_2| + |i_1,i_2,j_1,j_2\ket\bra j_1,j_2,i_2,i_1|\Big)\\
	\label{eq:Haar2}
	&\ = \frac{1}{N^2-1}\left( 1 - \frac{1}{N}\Swap_{3,4}\right)\left(\Swap_{1,3}\Swap_{2,4}+\Swap_{1,4}\Swap_{2,3}\right),
\end{align}
\end{subequations}
where $\Swap_{i,j}$ swaps the $i^\text{th}$ and $j^\text{th}$ components of $\CC^N\otimes\CC^N\otimes\CC^N\otimes\CC^N$.

\subsection{Simple quantum channels based on Haar integrals}\label{sec:channels}
In the evaluation of average expectation values, gradients, and gradient variances, we will encounter a number of channels that are based on Haar-measure integrals.

\Emph{The fully depolarizing channel.} --
The simplest one is
\begin{equation}\label{eq:G}
	\G(\hR):=\int_{\groupU(N)}\ud U\, \hU \hR \hU^\dag \stackrel{\eqref{eq:Haar1}}{=}
	 \frac{1}{N}\sum_{i,j}|i\ket\bra j|\hR|j\ket\bra i|=\frac{\id_N}{N}\Tr \hR.
\end{equation}
Based on the Hilbert-Schmidt inner product
\begin{equation}\label{eq:HS-innerProd}
	\bbra \hX|\hY\kket:=\Tr(\hX^\dag\hY)
\end{equation}
we can introduce a Dirac notation for operators with super-kets $|\hY\kket:=\hY$ and  super-bras $\bbra\hX|\dotsc:=\Tr(\hX^\dag\dotsc)$, where $\hX,\hY\in\End(\CC^N)$ are operators on the $N$-dimensional Hilbert space $\CC^N$. The channel $\G$ can then be written as
\begin{equation}\textstyle
	\G\stackrel{\eqref{eq:G}}{=}|\id'_N\kket\bbra\id^\pdag_N|\quad \text{with}\quad
	\id'_N:=\frac{1}{N}\id_N\quad\text{such that}\quad
	\bbra\id_N|\hY\kket=\Tr\hY.
\end{equation}

\Emph{The MPS channel.} --
Averaged TNS expectation values involve channels where we start from an operator $\hR$ on an $N$-dimensional Hilbert space, add an auxiliary system of dimension $d$ initialized in a reference state $|0_d\ket$, apply a Haar-random unitary on the composite system and then trace out the auxiliary system. This gives
\begin{equation}\label{eq:E}
	\E(\hR):=\int_{\groupU(Nd)}\ud U\,\Tr_d\big( \hU \big(\hR\otimes|0_d\ket\bra 0_d|\big) \hU^\dag\big) \stackrel{\eqref{eq:G}}{=}
	 \frac{\id_N}{N}\Tr \hR,
\end{equation}
which we will refer to as the \emph{MPS channel} and which coincides with the fully depolarizing channel \eqref{eq:G}.
\begin{figure*}[t]
	\label{fig:channelG}
	\includegraphics[width=\textwidth]{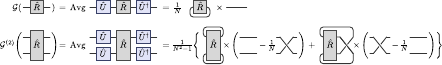}
	\caption{Diagrammatic representations for the fully depolarizing channel $\G$ and the doubled fully depolarizing channel $\G^\doub$ as defined in Eqs.~\eqref{eq:G} and \eqref{eq:G2}.}
\end{figure*}

\Emph{The doubled fully depolarizing channel.} --
When evaluating gradient variances for TNS expectation values, we will employ two copies $\CC^N\otimes\CC^N$ of an $N$-dimensional system, and will act on both with the same Haar-random unitary. The simplest resulting channel is
\begin{align}\nonumber
	\G^\doub(\hR):&=\int_{\groupU(N)}\ud U\, (\hU\otimes\hU) \hR (\hU^\dag\otimes\hU^\dag)\\\nonumber
	&\stackrel{\eqref{eq:Haar2}}{=}
	   \frac{1}{N^2-1}\left[ \left(\id_{N^2}-\frac{1}{N}\Swap    \right) \Tr \hR
	                        +\left(\Swap    -\frac{1}{N}\id_{N^2}\right) \Tr(\Swap\hR)\right]\\\label{eq:G2}
	&=\frac{2}{N(N+1)}\hP_+\Tr(\hP_+\hR) + \frac{2}{N(N-1)}\hP_-\Tr(\hP_-\hR),
\end{align}
where we have introduced projectors $\hP_\pm=\hP^2_\pm$ onto the permutation-symmetric and antisymmetric subspaces,
\begin{equation}\label{eq:hP}
	\hP_\pm:=\frac{1}{2}\left(\id_{N^2}\pm\Swap\right)\quad\Rightarrow\quad
	\Tr\hP_\pm=\frac{N(N\pm 1)}{2}.
\end{equation}
See Fig.~\ref{fig:channelG} for a diagrammatic representation.
In the dyadic notation, the channel simply reads
\begin{equation}\label{eq:G2dyadic}\textstyle
	\G^\doub \stackrel{\eqref{eq:G2}}{=} |\hP'_+\kket\bbra\hP_+|+|\hP'_-\kket\bbra\hP_-|=\G^{\doub\dag}
	\quad \text{with}\quad
	\hP'_\pm:=\frac{1}{\Tr\hP_\pm}\hP_\pm=\frac{2}{N(N\pm 1)}\hP_\pm
\end{equation}
such that we have biorthogonality in the sense that $\bbra\hP_\pm|\hP'_\pm\kket=1$ and $\bbra\hP_\pm|\hP'_\mp\kket=0$.
According Eq.~\eqref{eq:G2dyadic}, the \emph{doubled fully depolarizing channel} $\G^\doub$ has the doubly-degenerate steady-state eigenvalue 1 and all other eigenvalues are zero.

\Emph{The doubled MPS channel.} --
Similarly, we will need the \emph{doubled MPS channel} for two copies $\CC^N\otimes\CC^N$ of an $N$-dimensional system, where we add to each component one $d$-dimensional auxiliary system  $\CC^d$, initialized in reference state $|0_d\ket$, apply the same Haar-random unitary $\hU\in\groupU(Nd)$ on both composite spaces $\CC^N\otimes\CC^d$ and, finally, trace out the auxiliary systems.
\begin{align}\nonumber
	\E^\doub(\hR):&=\Tr_{d\times d}\Big(\int_{\groupU(Nd)}\ud U\, (\hU\otimes\hU) \big(\hR\otimes|0_d,0_d\ket\bra 0_d,0_d|\big) (\hU^\dag\otimes\hU^\dag)\Big)\\\nonumber
	&\textstyle\stackrel{\eqref{eq:G2}}{=}\Tr_{d\times d}\Big(\G^\doub\big(\hR\otimes|0_d,0_d\ket\bra 0_d,0_d|\big)\Big) \\\nonumber
	&= \ \ \ \frac{1}{N(Nd+1)}\left[(d+1)\hP_+ +(d-1)\hP_-\right]\Tr(\hP_+\hR)\\\label{eq:E2}
	&\quad + \frac{1}{N(Nd-1)}\left[(d-1)\hP_+ +(d+1)\hP_-\right]\Tr(\hP_-\hR),
\end{align}
So, in the biorthogonal left and right bases $\B_L=\big(\bbra\hP_+|,\bbra\hP_-|\big)$ and $\B_R=\big(|\hP'_+\kket,|\hP'_-\kket\big)$, the channel $\E^\doub$ has the matrix representation
\begin{equation}\label{eq:E2matrix}
	\big[\E^\doub\big]_\B= \Pmatrix{e_{+,+}&e_{+,-}\\e_{-,+}&e_{-,-}}
	= \frac{1}{2} \Pmatrix{N+1&0\\0& N-1}\Pmatrix{d+1&d-1\\d-1&d+1} \Pmatrix{\frac{1}{Nd+1}&0\\0&\frac{1}{Nd-1}}.
\end{equation}
Here the matrix elements are defined as $e_{\sigma,\sigma'}=\bbra \hP_{\sigma}|\E^\doub|\hP'_{\sigma'}\kket\equiv \Tr\big[\hP_{\sigma}\E^\doub(\hP'_{\sigma'})\big]$. The diagonalization of the matrix \eqref{eq:E2matrix}, yields the eigenvalues
\begin{align}\label{eq:E2eigen}
	&\textstyle 1\ \text{and} \ \eta:=\frac{1-1/N^2}{d-1/(N^2d)}
	  \quad\text{with corresponding left and right eigenvectors}\\
	&\textstyle\nonumber
	 \bbra\hl_1|:=\bbra\id_{N^2}|,\quad
	 |\hr_1\kket:=\frac{1}{2(N^2d+1)}\big[(Nd+1)(N+1)|\hP'_+\kket+(Nd-1)(N-1)|\hP'_-\kket\big],\\
	&\textstyle\nonumber
	 \bbra\hl_2|:=\frac{1}{{2(N^2 d+1)}}\big[(Nd-1)(N-1)\bbra\hP_+|-(Nd+1)(N+1)\bbra\hP_-|\big],\quad
	 |\hr_2\kket:=|\hP'_+\kket-|\hP'_-\kket
\end{align}
such that $\bbra\hl_i|\hr_j\kket= \delta_{i,j}$ and
\begin{equation}\label{eq:E2dyadic}
	\E^\doub= |\hr_1\kket\bbra\hl_1| + \eta\, |\hr_2\kket\bbra\hl_2|.
\end{equation}
For $d=1$, $\E^\doub$ coincides with $\G^\doub$.

\subsection{Riemannian gradients}\label{sec:Riemann}
In the variational optimization of quantum circuits or isometric tensor networks on quantum computers \cite{McClean2016-18}, one often employs an explicit parametrization for the unitaries that compose the circuit or TNS, e.g., by employing rotations generated by Pauli operators \cite{Vatan2004-69,Shende2004-69}. This has some disadvantages. For example, the sensitivity to small changes of rotation angles then strongly depends on the current values of the angles for purely geometric reasons; think, e.g., of the north pole on the Bloch sphere. Also, when studying barren plateaus, one then needs to make certain assumptions about factors in the representation forming unitary 2-designs \cite{McClean2018-9,Ortiz2021-2,Uvarov2021-54,Sharma2022-128} or go through a corresponding case analysis.

Instead of employing an explicit parametrizations for the involved unitaries, one can formulate the optimization problems directly over the manifold $\mc{M}$ formed by the direct product of the corresponding unitary groups in a representation-free form. In this Riemannian approach, gradients are elements of the tangent space of $\mc{M}$ at a given point, and one can implement line searches and Riemannian quasi-Newton methods through retractions and vector transport on $\mc{M}$. This is the program of Riemannian optimization as discussed generally in Refs.~\cite{Smith1994-3,Huang2015-25}. Recent applications for TNS and quantum circuits such as in Refs.~\cite{Hauru2021-10,Luchnikov2021-23,Miao2021_08,Wiersema2023-107,Miao2023_03} demonstrate favorable convergence properties.

\Emph{Riemannian gradients for isometric TNS.} --
As described in sections~\ref{sec:MPS} and \ref{sec:MERA}, the energy expectation values for MPS, TTNS, and MERA can be written in the form
\begin{equation}\label{eq:energy}
	E(\hU)=\bra\Psi(\hU)|\hH|\Psi(\hU)\ket=\Tr\left(\hX (\hU^\dag\otimes\id_M)\hY(\hU\otimes\id_M)\right),
\end{equation}
where, for now, we only consider the dependence on a single unitary $\hU\in\groupU(N)$ in the tensor network. The operator $\hX=\hX^\dag$ depends on further parts of the TNS, and $\hY=\hY^\dag$ depends on TNS tensors and the Hamiltonian. $\hX$ and $\hY$ act on $\CC^N\otimes\CC^M$. For most considerations on MPS, we have $M=1$ such that some expressions simplify. For the brevity of notation, we define $\tU:=\hU\otimes\id_M$.

The energy gradient in the Euclidean embedding space $\End(\CC^N)$ is
\begin{equation}\label{eq:embedd_grad}
	\hd=2\Tr_M(\hY\tU\hX)\equiv2\Tr_M\big(\hY(\hU\otimes\id_M)\hX\big)\in \End(\CC^N),
\end{equation}
where $\Tr_M$ denotes the partial trace over the second component of the tensor product space $\CC^N\otimes\CC^M$.
The gradient $\hd$ fulfills $\partial_\veps E(\hU+\veps\hW)|_{\veps=0}=(\hd,\hW)$ for all $\hW$, where
\begin{equation}
	(\hA,\hB):=\Re\Tr(\hA^\dag\hB)
\end{equation}
is the Euclidean metric on the embedding space (the real part of the Hilbert-Schmidt inner product).
An element $\hW$ of the tangent space $\mc{T}_{\hU}$ for $\groupU(N)$ at $\hU$ needs to obey $(\hU+\veps \hW)^\dag(\hU+\veps \hW)=\id+\mc{O}(\veps^2)$, i.e., $\hU^\dag\hW$ needs to be skew-Hermitian and, hence,
\begin{equation}\label{eq:Un_tangentSpace}
	\mc{T}_{\hU}=\{\hU\hG\,|\,\hG=-\hG^\dag\in\End(\CC^{N})\}.
\end{equation}
The Riemannian energy gradient $\hg$ for the manifold $\groupU(N)$ at $\hU$ is obtained by projecting $\hd$ onto the tangent space such that $(\hW,\hg)=(\hW,\hd)$ for all $\hW\in\mc{T}_{\hU}$. This gives
\begin{equation}\label{eq:Riem_grad}
	\hg = \frac{1}{2}(\hd - \hU\hd^\dag\hU)
	\stackrel{\eqref{eq:embedd_grad}}{=}
	 \Tr_M\big(\hY\tU\hX - \tU\hX\tU^\dag\hY\tU\big)
	 \ \in \  \mc{T}_{\hU}.
\end{equation}
That $\hg$ lies indeed in the tangent space \eqref{eq:Un_tangentSpace} can be seen by writing it as $\hg=\hU\hG$ with $\hG=(\hU^\dag\hd - \hd^\dag\hU)/2=-\hG^\dag$. Similarly, writing $\hW\in\mc{T}_{\hU}$ as $\hW=\hU\hG'$, we have $(\hW,\hU\hd^\dag\hU)=\Re\Tr(\hG^{'\dag}\hd^\dag\hU)=-\Re\Tr(\hd^\dag\hU\hG')=-(\hd,\hW)$ and, hence, $(\hW,\hg)=(\hW,\hd)$.
A resulting Riemannian version of the limited-memory Broyden–Fletcher–Goldfarb–Shanno (L-BFGS) algorithm \cite{Nocedal2006,Liu1989-45} is given in Ref.~\cite{Miao2021_08}.

The gradients $\hd$ and $\hg$ vanish when averaged over Haar-random $\hU\in\groupU(N)$, as
\begin{equation}\label{eq:grad_avg0}\textstyle
	\int\ud U\,\hd(\hU)=\frac{1}{2}\int\ud U\,\left[\hd(\hU)+\hd(-\hU)\right]=0 \quad\text{and}\quad
	\int\ud U\,\hg(\hU)=\frac{1}{2}\int\ud U\,\left[\hg(\hU)+\hg(-\hU)\right]=0.
\end{equation}

\Emph{Variance of Riemannian gradients.} --
As, according to Eq.~\eqref{eq:grad_avg0}, the average gradient is zero, we can quantify the variance of the Riemannian gradient by
\begin{equation}\label{eq:gradVar}
	\Var_{\hU} \hg(\hU):=\frac{1}{N}\int_{\groupU(N)}\ud\hU\,\Tr(\hg^\dag\hg).
\end{equation}
The factor $1/N$ in this definition is motivated as follows: We can expand $\hg$ in an orthonormal basis of Hermitian and unitary operators $\{\hsigma_n\,|\,\hsigma_n=\hsigma^\dag_n,\,\hsigma^2_n=\id_N \}$ for $\End(\CC^N)$ with $\Tr(\hsigma_n\hsigma_{n'})=N\delta_{n,n'}$. This gives the gradient in the form $\hg=\mri \hU\sum_{n=1}^{N^2}\alpha_n \hsigma_n/N$. On a quantum computer, the rotation-angle derivatives can be determined as energy differences $\alpha_n=E(\hU e^{{\mri\pi\hsigma_n}/{4}}) - E(\hU e^{{-\mri\pi\hsigma_n}/{4}})$ \cite{Miao2021_08,Wiersema2023-107}. The variance of the rotation-angle derivatives is then $\int\ud U\,\frac{1}{N^2}\sum_n\alpha_n^2=\frac{1}{N}\int\ud U\,\Tr(\hg^\dag\hg).$
\begin{figure*}[t]
	\label{fig:Riem_grad}
	\includegraphics[width=\textwidth]{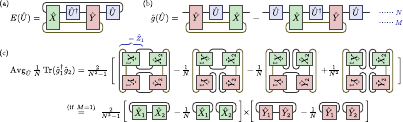}
	\caption{Diagrammatic representations for (a) the energy expectation value \eqref{eq:energy} depending on unitary $\hU\in\groupU(N)$, (b) the corresponding Riemannian gradient \eqref{eq:Riem_grad}, and (c) the variance \eqref{eq:gradVarProdM}. For the latter, we consider a slightly more general expression needed for extensive Hamiltonians, where we have $\Tr(\hg_1^\dag\hg_2)$ instead of $\Tr(\hg^\dag\hg)$ with $\hg_1$ and $\hg_2$ depending on the same unitary $\hU$, but possibly featuring different operators $\hX_1$, $\hY_1$ and $\hX_2$, $\hY_2$, respectively. The first line in (c) [Eq.~\eqref{eq:gradVarProdM}] simplifies to the form in the second line [Eq.~\eqref{eq:gradVarProd}] if dimension $M$ is $1$.}
\end{figure*}

Let us first address the case with $M=1$ in Eq.~\eqref{eq:energy} which covers most cases concerning MPS expectation values. Using the Haar measure integrals \eqref{eq:Haar2} and \eqref{eq:Haar2} or, equivalently, the quantum channels \eqref{eq:G} and \eqref{eq:G2}, the average over $\hU$ in Eq.~\eqref{eq:gradVar} then evaluates to
\begin{align}\nonumber
	\Var_{\hU} \hg(\hU)
	&\textstyle
	\stackrel{\eqref{eq:Riem_grad}}{=} \frac{2}{N}\int\ud U\,\Tr\left(\hY^2\hU\hX^2\hU^\dag-(\hY\hU\hX\hU^\dag)^2\right)\\\nonumber
	&\textstyle=\frac{2}{N}\Tr\left(\hY^2\G(\hX^2)\right) - \frac{2}{N}\Tr\left((\hY\otimes\hY)\,\G^\doub(\hX\otimes\hX)\Swap\right)\\\nonumber
	&\textstyle=\frac{2}{N^2}\Tr\hX^2\,\Tr\hY^2 
	  - \frac{2}{N(N^2-1)}\left(\Tr^2\hX\,\Tr\hY^2+\Tr\hX^2\,\Tr^2\hY\right)\\\nonumber
	&\textstyle\quad  + \frac{2}{N^2(N^2-1)}\left(\Tr\hX^2\,\Tr\hY^2+\Tr^2\hX\,\Tr^2\hY\right)\\
	\label{eq:gradVarProd}
	&\textstyle=\frac{2}{N^2-1}\Tr\left(\big(\hX-\frac{\id_N}{N}\Tr\hX\big)^2\right)\Tr\left(\big(\hY-\frac{\id_N}{N}\Tr\hY\big)^2\right)
\end{align}

Especially for TTNS and MERA, we have $M>1$ in Eq.~\eqref{eq:energy}. With $\tU\equiv\hU\otimes\id_M$ and $\Tr_M$ denoting the partial trace over the second component of $\CC^N\otimes\CC^M$, the variance \eqref{eq:gradVar} then evaluates to
\begin{subequations}\label{eq:gradVarProdM}
\begin{align}\nonumber
	\Var_{\hU} \hg(\hU)
	&\textstyle
	\stackrel{\eqref{eq:Riem_grad}}{=} \frac{1}{N}\int\ud U\,
	    \Tr\left(2\Tr_M(\hX\tU^\dag\hY)\Tr_M(\hY\tU\hX)-\Tr^2_M(\hX\tU^\dag\hY\tU)-\Tr^2_M(\tU^\dag\hY\tU\hX)\right)\\\nonumber
	&\textstyle=\frac{2}{N^2-1}\left(\Tr\hZ^2-\frac{1}{N}\Tr\big((\Tr_1\hZ)^2+(\Tr_2\hZ)^2\big)+\frac{1}{N^2}(\Tr\hZ)^2\right)\\
	\label{eq:gradVarProdMa}
	&\textstyle=\frac{2}{N^2-1}\Tr\left(\big[\Swap_{1,3}-\frac{1}{N}\id_{N^4}\big]\,
	                                    \big[\Swap_{2,4}-\frac{1}{N}\id_{N^4}\big]\,[\hZ\otimes\hZ]\right)
\end{align}
with $\hZ\in\End(\CC^N\otimes\CC^N)$, where 
\begin{equation}\label{eq:gradVarProdM-Z}\textstyle
	\bra i_1,i_2|\hZ|j_1,j_2\ket=\sum_{m,n=1}^M \bra i_1,m|\hX|j_1,n\ket\bra i_2,n|\hY|j_2,m\ket.
\end{equation}
\end{subequations}
The operators $\Swap_{1,3}$ and $\Swap_{2,4}$ swap the first with the third and the second with the fourth components of $\CC^N\otimes\CC^N\otimes\CC^N\otimes\CC^N$, respectively.
$\Tr_1\hZ$ and $\Tr_2\hZ$ denote the partial traces of $\hZ$ over the first and second components of $\CC^N\otimes\CC^N$, respectively. Diagrammatic representations of Eqs.~\eqref{eq:energy}, \eqref{eq:Riem_grad}, and \eqref{eq:gradVarProdM} are shown in Fig.~\ref{fig:Riem_grad}.

As pointed out below Eq.~\eqref{eq:energy}, $\hX$ and $\hY$ depends on further (unitary) TNS tensors and the Hamiltonian. Considering MPS, TTNS, and MERA where all these tensors are sampled according to the Haar resume, we will average $\Var_{\hU} \hg(\hU)$ over these tensors to obtain the Haar-variance of energy gradients and discuss (the absence of) barren plateaus. In this respect, an important property of both Eq.~\eqref{eq:gradVarProd} and Eq.~\eqref{eq:gradVarProdM} is that they vanish when $\hY=\id_N\otimes\id_M$. This is obvious for the case $M=1$ treated in Eq.~\eqref{eq:gradVarProd}. For Eq.~\eqref{eq:gradVarProdM}, note that $\hZ=\Tr_M(\hX)\otimes \id_N$ when $\hY=\id_N\otimes\id_M$. In this case, the four terms in the second line of Eq.~\eqref{eq:gradVarProdMa} become $\Tr\hZ^2=N\Tr\hx^2$, $\frac{1}{N}\Tr(\Tr^2_1\hZ)=\Tr^2\hx$, $\frac{1}{N}\Tr(\Tr^2_2\hZ)=N\Tr\hx^2$, and $\frac{1}{N^2}\Tr^2\hZ=\Tr^2\hx$, where $\hx:=\Tr_M(\hX)$. Hence, identity components of $\hY$ do not contribute to the variance \eqref{eq:gradVarProdM}.

\section{Quantum circuits with barren plateaus}\label{sec:Circuits}
The simplest conceivable setting for minimizing an energy expectation value $\bra\Psi|\hH|\Psi\ket$ is to choose 
\begin{equation}\label{eq:globalU}
	|\Psi\ket=\hU|0\ket\quad \text{with} \quad \hU\in\groupU(N),
\end{equation}
i.e., generating $|\Psi\ket$ by applying a global unitary to a reference state $|0\ket$ from the $N$-dimensional Hilbert space. This optimization problem is hampered by the barren plateau phenomenon.
\begin{theorem}[\propHead{Barren plateau for global unitary}]
\label{thrm:globalU}
Consider a system of $L$ sites with single-site Hilbert space dimension $d$ and a $k$-local Hamiltonian $\hH$, where $\Tr\hH=0$. With the unitary $\hU$ in Eq.~\eqref{eq:globalU} sampled according to the uniform Haar measure, the average of the Riemannian gradient $\partial_{\hU} \bra\Psi|\hH|\Psi\ket$ of the energy expectation value is zero and its variance decays exponentially in the system size $L$.
\begin{subequations}\label{eq:globalUgradDecay}
\begin{gather}\label{eq:globalUgradDecayAvg}
	\Avg \left(\partial_{\hU} \bra\Psi|\hH|\Psi\ket\right) = 0\quad\text{and}\\\label{eq:globalUgradDecayVar}
	\Var \left(\partial_{\hU} \bra\Psi|\hH|\Psi\ket\right)
	 =2\Tr(\hH^2)\frac{1}{d^L(d^L+1)}=\mc{O}(d^{-L}).
\end{gather}
\end{subequations}
\end{theorem}\noindent
\Emph{Proof:}
The energy expectation value $\bra 0|\hU^\dag \hH \hU| 0\ket$  can be written in the form of Eq.~\eqref{eq:energy} with $\hY=\hH$ and $\hX=| 0\ket\bra 0|$.
Equation~\eqref{eq:grad_avg0} then implies that the averaged Riemannian gradient [Eq.~\eqref{eq:Riem_grad}] is zero.
With the Hilbert space dimension$N=d^L$ in Eq.~\eqref{eq:gradVarProd}, the gradient variance can be assessed by studying the factors
\begin{equation}\label{eq:globalU_X-Y}\textstyle
	\Tr\big([\hX-\frac{\id_{N}}{N}\Tr\hX]^2\big)=1-\frac{1}{N}\quad\text{and}\quad
	\Tr\big([\hY-\frac{\id_{N}}{N}\Tr\hY]^2\big)=\Tr\hH^2.
\end{equation}
Here we have used that $\Tr\hX=\Tr\hX^2=1$ and $\Tr\hH=0$. This yields Eq.~\eqref{eq:globalUgradDecayVar}.
\qed

Due to the exponential growth $N=d^L$ of the Hilbert space dimension, for large systems, it is not practical to work with general unitaries $\hU$ in Eq.~\eqref{eq:globalU}. For 1D systems, a simple approach for variational quantum eigensolvers is to choose $\hU$ as a quantum circuit where nearest-neighbor two-site unitaries act alternatingly on all odd and all even bonds of the lattice. This is known as the alternating layered ansatz or a brickwall circuit. The gradient amplitudes for this ansatz also decrease exponentially in the system size $L$ as long as relevant parts of the circuit form unitary 2-designs \cite{McClean2018-9}. The latter condition is fulfilled approximately when increasing the number of layers linearly with the system size \cite{Dankert2009-80,Brandao2016-346,Harrow2023-05}. Ref.~\cite{Cerezo2021-12} provides a direct analysis of the gradient amplitudes for this case.

\section{Matrix product states}\label{sec:MPS}
\subsection{Setup}
\Emph{Isometric form of MPS with open boundary conditions.} --
For a 1D lattice of $L$ sites, each associated with a $d$-dimensional site Hilbert space $\CC^{d}$ with orthonormal basis $\{|s\ket\,|\,s=1,\dotsc,d\}$, every MPS with \emph{bond dimension} $m$ and open boundary conditions can be written in the form
\begin{subequations}\label{eq:MPS}
\begin{equation}\label{eq:MPS-state}\textstyle
	|\Psi\ket:=\sum_{s_1,\dotsc,s_L} \bra 0| \hA_1^{s_1}\hA_2^{s_2}\dotsb \hA_L^{s_L} |0\ket\, |s_1,s_2,\dotsc,s_L\ket
	\quad\in \ (\CC^{d})^{\otimes L}.
\end{equation}
It is characterized by the $L$ MPS tensors which are (multi-)linear maps
\begin{equation}
	\hat{A}_j:\CC^{m_j}\to\CC^{m_{j-1}}\otimes \CC^d,
\end{equation}
\end{subequations}
where $m_j\leq m$ is the bond dimension for bond $(j,j+1)$. According to the open boundary conditions, we set $m_0=m_L=1$ and choose $|0\ket$ as a normalized state spanning the trivial bond vector space $\CC^1$ at the left and right ends of the chain. The $\hA_j^s:\CC^{m_j}\to\CC^{m_{j-1}}$ with $\bra \alpha|\hA_j^s|\beta\ket:=\bra \alpha,s|\hA_j|\beta\ket$ and $s=1,\dotsc,d$ are linear operators mapping from the vector space for bond $(j,j+1)$ to that of bond $(j-1,j)$ such that $\hA_1^{s_1}\hA_2^{s_2}\dotsb \hA_L^{s_L}$ in Eq.~\eqref{eq:MPS-state} denotes the product of $L$ such operators -- the \emph{matrix product}.

The MPS \eqref{eq:MPS} is invariant under \emph{gauge transformations}
\begin{equation}\label{eq:gaugeTrafo}
	\left(\hA_j^s, \ \hA_{j+1}^{s'}\right) \ \mapsto \
	\left(\hA_j^s \hat{Z}^{-1}_j, \ \hat{Z}_j \hA_{j+1}^{s'}\right) \ \forall\,{s,s'} \quad\text{with} \quad j\in[1,L-1]
\end{equation}
and invertible operators $\hat{Z}_j$. This can be used to bring the matrix product into a so-called \emph{left-orthonormal form} with a sequence of QR decompositions \cite{Schollwoeck2011-326,Barthel2022-112} such that all MPS tensors become isometries with
\begin{equation}\label{eq:MPSisoConstr}
	\hA_j^\dag\hA_j=\sum_{s=1}^d \hA_j^{s \dag} \hA_j^s=\id_{m_j},
\end{equation}
where $\id_{m_j}$ denotes the identity on the bond vector space $\CC^{m_j}$. We employ this left-orthonormal form throughout the paper.

The Schmidt rank of the MPS \eqref{eq:MPS} for a bipartition of the system into blocks of sites $\{1,\dotsc,j\}$ and $\{j+1,\dotsc,L\}$ is bounded from above by $d^j$ and $d^{L-j}$. Hence, we can choose the first few bond dimensions as $m_0=1$, $m_1=d$, $m_2=d^2$ etc.\ until reaching the first bond $(b-1,b)$ for which
\begin{equation}
	d^{b-1}< m\quad\text{and}\quad d^{b}\geq m,\quad\text{i.e.,}\quad b:=\lceil\log_d m\rceil.
\end{equation}
From this bond on, we then have the full desired MPS bond dimension $m_j=m$. Proceeding analogously on sites $L-b+1,\dotsc,L$ at the right end of the chain and choosing isometric MPS tensors \eqref{eq:MPSisoConstr}, the resulting MPS has norm one;
\begin{align}\nonumber
	\|\Psi\|^2&\textstyle
	=\bra\Psi|\Psi\ket
	 \stackrel{\eqref{eq:MPS}}{=}
	  \sum_{s_1,\dotsc,s_L} \bra 0| \hA_L^{s_L\dag}\dotsb \ \hA_2^{s_2\dag}\hA_1^{s_1\dag} |0\ket\,\bra 0| \hA_1^{s_1}\hA_2^{s_2}\dotsb \hA_L^{s_L} |0\ket\\
	&\stackrel{\eqref{eq:MPSisoConstr}}{=}\textstyle
	  \sum_{s_2,\dotsc,s_L} \bra 0| \hA_L^{s_L\dag}\dotsb \ \hA_2^{s_2\dag}\hA_2^{s_2}\dotsb \hA_L^{s_L} |0\ket
	  =\dots = 1,
\end{align}
where we have used $|0\ket\bra 0|=\id_1$.

\Emph{Unitary parametrization of the MPS tensors.} --
For the bulk of the system, consisting of sites $j=b+1,\dotsc,L-b$, the MPS tensors with $m_j=m$ can be written in the form
\begin{subequations}\label{eq:MPSiso}
\begin{equation}\label{eq:MPSisoForm}
	\hA_j=\hU_j\,\big(\id_{m}\otimes |0_d\ket\big)\quad\text{with unitaries}\quad
	\hU_j\in \groupU(md)
\end{equation}
that act on the tensor product $\CC^m\otimes\CC^d$ of a bond vector space and the $d$-dimensional site Hilbert space with an (arbitrary) reference state $|0_d\ket$. See Fig.~\ref{fig:MPSsingle}. Assuming $m=d^{b}$, the tensors at the left boundary of the chain can be chosen as
\begin{equation}\label{eq:MPSisoForm-bL}
	\hA_j=\hU_j\in\groupU(m_j)\quad\text{with}\quad m_j=d^{j}\quad\text{for}\quad
	j=1,\dotsc,b.
\end{equation}
Similarly, the tensors at the right boundary of the chain can be chosen as
\begin{equation}\label{eq:MPSisoForm-bR}
	\hA_j=\hU_j\,\big(\id_{m_j}\otimes |0_d\ket\otimes |0_d\ket\big)\quad\text{with}\quad \hU_j\in\groupU(m_jd^2)
	\quad\text{and}\quad m_j=d^{L-j}
\end{equation}
for $j=L-b+1,\dotsc,L$
\end{subequations}

\Emph{Alternative choices for MPS.} --
The advantage of the MPS \eqref{eq:MPS-state} is that it has strictly norm $\|\Psi\|=1$ for any choice of the isometric tensors $\hA_j$, and that any MPS with bond dimension $m$ can be written in this form. A slight technical complication arises from the variation of the bond dimensions $m_j$ at the boundaries of the chain. For brevity, we will largely avoid this complication in the following by only considering energy gradients with respect to tensors that are in the bulk $\{b+1,\dotsc,L-b\}$ of the system.

An alternative would be to work with MPS of the form
\begin{equation}\label{eq:MPSunnorm}\textstyle
	|\Psi'\ket:=\sqrt{m}\,\sum_{s_1,\dotsc,s_L} \bra 0'_m| \hA_1^{s_1}\hA_2^{s_2}\dotsb \hA_L^{s_L} |0_m\ket\, |s_1,s_2,\dotsc,s_L\ket,
\end{equation}
where the bond dimension is constant, i.e., $m_j=m$ and $\hat{A}_j:\CC^m\to\CC^{md}$ for all $j$. Here $|0_m\ket$ and $|0'_m\ket$ are arbitrary normalized reference states from the bond vector space. One can again impose the isometry condition \eqref{eq:MPSisoConstr}. A drawback is that these MPS are not normalized, but only normalized on average in the sense that $\Avg \bra\Psi'|\Psi'\ket=1$ and $\Var \bra\Psi'|\Psi'\ket=\frac{m-1}{m^2d+1}\sim\frac{1}{md}$.

Another alternative would be to work with periodic boundary conditions, i.e., use
\begin{equation}\label{eq:MPSpbc}\textstyle
	|\Psi''\ket:=\sum_{s_1,\dotsc,s_L} \Tr\left( \hA_1^{s_1}\hA_2^{s_2}\dotsb \hA_L^{s_L} \right)\, |s_1,s_2,\dotsc,s_L\ket.
\end{equation}
But, in that setting, one cannot bring all MPS tensors into isometric form \eqref{eq:MPSisoForm}. When, nevertheless, imposing the isometry constraint \eqref{eq:MPSisoConstr} as done in Refs.~\cite{Liu2022-129,Garcia2023-2023}, one misses large classes of MPS with PBC. For the translation-invariant case, canonical forms of MPS with PBC are discussed in Refs.~\cite{Fannes1992-144,PerezGarcia2007-7}. In general, the MPS tensors then assume a block-diagonal structure. Also, MPS \eqref{eq:MPSpbc} with isometric tensors are in general not normalized \cite{Haferkamp2021-2} such that gradients also comprise contributions from the variable norm.

\subsection{Scaling of gradients for single-site Hamiltonians}\label{sec:MPS-single}
Let us consider the minimization problem for the cost function
\begin{equation}\label{eq:costLocal}
	\bra\Psi|\hh_i|\Psi\ket \quad\text{with}\quad
	\hh_i=\hh_i^\dag \quad\text{and}\quad
	\Tr\hh_i=0,
\end{equation}
where $\hh_i$ is a local observable with its spatial support restricted to the vicinity of site $i$. We will first consider single-site operators
\begin{equation}\label{eq:hiSite}
	\hh_i=\id_d^{\otimes(i-1)}\otimes\hh\otimes\id_d^{\otimes(L-i)}
	\quad\text{with}\quad \hh=\hh^\dag\in\End(\CC^d)
\end{equation}
and generalize below in Sec.~\ref{sec:MPS-nn}. In contrast to prior work, we consider a Riemannian version of the optimization problem, i.e., we do not introduce any particular parametrization for the unitaries $\hU_j$ (see Sec.~\ref{sec:Riemann}).
\begin{theorem}[\propHead{Exponential decay of MPS gradient variance with distance}]
\label{thrm:MPS-dist}
With MPS unitaries $\hU_j$ in Eq.~\eqref{eq:MPSiso} sampled according to the uniform Haar measure, the average of the Riemannian gradient $\partial_{\hU_j} \bra\Psi|\hh_i|\Psi\ket$ for the cost function \eqref{eq:costLocal} is zero and its variance decays exponentially in $|i-j|$. Specifically, for $i,j\in\{b+1,\dotsc,L-b\}$,
\begin{subequations}\label{eq:MPSgradDecay}
\begin{gather}\label{eq:MPSgradDecayAvg}
	\Avg \left(\partial_{\hU_j} \bra\Psi|\hh_i|\Psi\ket\right) = 0\quad\text{and, \ with}\quad
	\eta=\frac{1-1/m^2}{d-1/(m^2d)},\\\label{eq:MPSgradDecayVar}
	\Var \left(\partial_{\hU_j} \bra\Psi|\hh_i|\Psi\ket\right) =
	\begin{cases}
	 0 &\text{for}\quad j<i,\\
	 2\Tr(\hh^2)\,\frac{1}{d(m^2d+1)}\,\eta^{j-i}+\mc{O}(\eta^{L-i})
	 \quad&\text{for}\quad j\geq i.
	\end{cases}
\end{gather}
\end{subequations}
\end{theorem}\noindent
\Emph{Proof:} \textbf{(a)} Let us first consider the case $j<i$. Due to the isometry condition \eqref{eq:MPSisoConstr}, the expectation value $\bra\Psi|\hh_i|\Psi\ket$ simplifies to
\begin{align}
	\bra\Psi|\hh_i|\Psi\ket
	&\stackrel{\eqref{eq:MPS}}{=}\textstyle\nonumber
	 \sum_{s_1,\dotsc,s_L,s_i'}
	   \bra 0| \hA_L^{s_L\dag}\dotsb \hA_i^{s'_i\dag}\dotsb \hA_1^{s_1\dag} |0\ket\,
	   \bra 0| \hA_1^{s_1}\dotsb \hA_i^{s_i}\dotsb \hA_L^{s_L} |0\ket\, \bra s'_i|\hh|s_i\ket\\
	&\label{eq:MPS_expctIso}
	 \stackrel{\eqref{eq:MPSisoConstr}}{=}\textstyle
	 \sum_{s_i,\dotsc,s_L,s_i'}
	   \bra 0| \hA_L^{s_L\dag}\dotsb \hA_i^{s'_i\dag}  \hA_i^{s_i}\dotsb \hA_L^{s_L} |0\ket\, \bra s'_i|\hh|s_i\ket.
\end{align}
So, the cost function \eqref{eq:costLocal} is independent of $\hU_j$ which establishes the theorem for $j<i$.\\[0.4em]
\begin{figure*}[t]
	\label{fig:MPSsingle}
	\includegraphics[width=\textwidth]{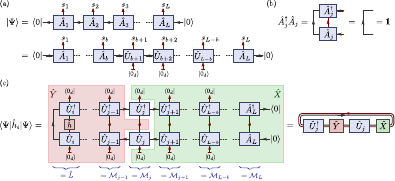}
	\caption{(a) The MPS \eqref{eq:MPS} for a system of $L$ sites with open boundary conditions. When imposing the left-orthonormal form \eqref{eq:MPSisoConstr}, the MPS tensors $\hA_j$ for sites $j=b+1,\dotsc,L-b$ can be parametrized in the form \eqref{eq:MPSisoForm} with unitaries $\hU_j\in\groupU(md)$. Arrows indicate the domains and co-domains of the tensors.
	(b) The MPS gauge freedom \eqref{eq:gaugeTrafo} can be used impose the left-orthonormality condition \eqref{eq:MPSisoConstr} on all sites.
	(c) Due to this orthonormality condition, the expectation value $\bra\Psi|\hh_i|\Psi\ket$ for a single-site operator $\hh_i$ is independent of $\{\hA_1,\dotsc,\hA_{i-1}\}$. It can be written in the form $\bra\Psi|\hh_i|\Psi\ket=\Tr_{m\times d}(\hX \hU_j^\dag\hY\hU_j)$ [Eq.~\eqref{eq:energy} and Fig.~\ref{fig:Riem_grad}a] with $M=1$ and $N=md$. The MPS transfer matrices correspond to quantum channels \eqref{eq:MPS_F}.}
\end{figure*}
\textbf{(b)} Next, consider the case $j>i$. We can bring the cost function into the form of Eq.~\eqref{eq:energy} with $M=1$,
\begin{align}\nonumber
	\bra\Psi|\hh_i|\Psi\ket
	&\stackrel{\eqref{eq:MPSisoConstr}}{=}
	 \sum_{s_i,\dotsc,s_L,s_i'} 
	   \Tr_m\Big( \hA_i^{s_i}\dotsb \hA_L^{s_L} |0\ket\,
	      \bra 0| \hA_L^{s_L\dag}\dotsb \hA_i^{s'_i\dag}\Big)\, \bra s'_i|\hh|s_i\ket\\
	&\stackrel{\eqref{eq:MPSiso}}{=}
	 \Tr_m\Big(\underbrace{\bra 0_d|\,\hU_i^\dag\big[\id_m\otimes\hh\big]
	                       \hU_i\,|0_d\ket}_{=:\hL}\,
	          \underbrace{\F_{i+1}\circ\dotsb\circ\F_L(|0\ket\bra 0|)}_{=:\F_{i+1}\circ\dotsb\circ\F_{j}(\hR)}\Big),
	\label{eq:MPS_L}
\end{align}
where $\hL$ and $\hR=\F_{j+1}\circ\F_{j+2}\circ\dotsb\circ\F_{L}(|0\ket\bra 0|)$ act on $\CC^m$, and we have defined the channels
\begin{subequations}
\begin{align}\label{eq:MPS_F}
	\F_n(\hW):&\textstyle
	=\sum_s \hA^s_n\hW \hA^{s\dag}_n
	\stackrel{\eqref{eq:MPSiso}}{=} \Tr_d\big(\hU_n\big[\hW\otimes |0_d\ket\bra 0_d|\big]\hU_n^\dag\big)\quad\text{with adjoints}\\
	\F_n^\dag(\hV)&\textstyle
	=\sum_s \hA^{s\dag}_n\hV \hA^s_n
	\;\stackrel{\eqref{eq:MPSiso}}{=}\bra 0_d|\hU_n^\dag\big(\hV\otimes \id_d\big)\hU_n|0_d\ket
\end{align}
\end{subequations}
such that $\Tr\big(\hV^\dag\F_n(\hW)\big)=\Tr\big([\F_n^\dag(\hV)]^\dag\hW\big)$. Using this relation for the cost function, we obtain
\begin{equation}
	\bra\Psi|\hh_i|\Psi\ket =
	\Tr_{m}\Big(\F^\dag_{j-1}\circ\dotsb\circ\F^\dag_{i+1}(\hL)\,
                       \Tr_d\big(\hU_j\big[\hR\otimes |0_d\ket\bra 0_d|\big]\hU_j^\dag\big)\Big).
\end{equation}
This is in fact of the form \eqref{eq:energy} with $M=1$: As indicated in Fig.~\ref{fig:MPSsingle}, $\bra\Psi|\hh_i|\Psi\ket=\Tr_{m\times d}(\hX \hU_j^\dag\hY\hU_j)$  with
\begin{subequations}\label{eq:MPS_XYdecay}
\begin{align}\label{eq:MPS_X}
	\hX&:=\hR\otimes |0_d\ket\bra 0_d|=\F_{j+1}\circ\F_{j+2}\circ\dotsb\circ\F_{L}(|0\ket\bra 0|)\otimes|0_d\ket\bra 0_d|
	\quad\text{and}\quad\\\label{eq:MPS_Y}
	\hY&:=\F^\dag_{j-1}\circ\dotsb\circ\F^\dag_{i+1}(\hL)\otimes\id_d,\quad
	\hL=\hA_i^\dag[\id_{m_{i-1}}\otimes\hh\big]\hA_i
	\stackrel{\eqref{eq:MPSisoForm}}{=} \bra 0_d|\,\hU_i^\dag\big[\id_m\otimes\hh\big]\hU_i\,|0_d\ket.
\end{align}
\end{subequations}
Equation~\eqref{eq:grad_avg0} then implies that the averaged Riemannian gradient [Eq.~\eqref{eq:Riem_grad}] is zero, $\Avg \partial_{\hU_j} \bra\Psi|\hh_i|\Psi\ket=0$.
In the second equality for $\hL$, we have used that $m_{i-1}=m$ when $i\in\{b+1,\dotsc,L-b\}$ and Eq.~\eqref{eq:MPSisoForm} applies.
\\[0.4em]
\textbf{(c)} 
With $N=md$ in Eq.~\eqref{eq:gradVarProd}, the gradient variance can be assessed by studying the factors
\begin{equation}\label{eq:MPS_X-Y}\textstyle
	\Tr\big([\hX-\frac{\id_{md}}{md}\Tr\hX]^2\big)=\Tr\hX^2-\frac{1}{md}\quad\text{and}\quad
	\Tr\big([\hY-\frac{\id_{md}}{md}\Tr\hY]^2\big)=\Tr\hY^2-\frac{1}{md}\Tr^2\hY,
\end{equation}
where we have used that $\Tr_{m\times d}\hX=\Tr_{1\times d}\big(|0\ket\bra 0|\otimes|0_d\ket\bra 0_d|\big)=1$.
These two terms depend on the MPS unitaries $\hU_{j+1},\dotsc,\hU_L$ and $\hU_i,\dotsc,\hU_{j-1}$, respectively. To obtain the variance, we need to execute the corresponding Haar-measure integrals for both terms.\\[0.4em]
\textbf{(d)} The Haar-measure average of $\Tr_{m\times d}\hX^2\stackrel{\eqref{eq:MPS_X}}{=}\Tr_m\hR^2=\Tr_{m\times m}\big(\Swap[\hR\otimes\hR]\big)$ is
\begin{align}\nonumber
	\Avg\Tr\hX^2
	&\stackrel{\eqref{eq:MPS_X}}{=}
	 \Avg\Tr_{m\times m}\left(\Swap \F^{\otimes 2}_{j+1}\circ\dotsb\circ\F^{\otimes 2}_{L}(|0,0\ket\bra 0,0|)\right)\\\label{eq:MPS_TrX2}
	&\stackrel{\eqref{eq:E2}}{=}\Tr_{m\times m}\left(\Swap\,(\E^\doub)^{L-j-b}(\dm_b)\right)
\end{align}
with $\dm_b:=\Avg \F^{\otimes 2}_{L-b+1}\circ\dotsb\circ\F^{\otimes 2}_{L}(|0,0\ket\bra 0,0|)$. We have used that, for $n\in\{b+1,\dotsc,L-b\}$, $\Avg\F^{\otimes 2}_n=\E^\doub$ according to Eq.~\eqref{eq:E2} with $N=m$. Also on the boundary sites, $\Avg \F^{\otimes 2}_n$ is a completely positive trace-preserving map  (quantum channel) such that $\dm_b$ is a trace-1 density operator. The doubled MPS channel $\E^\doub$ is strictly contractive. According to Eq.~\eqref{eq:E2dyadic}, its repeated application to $\dm_b$ converges exponentially fast to its steady state $\hr_1$, i.e.,
\begin{equation}\label{eq:E2_steadyState}\textstyle
	\Avg\hR\otimes\hR=\hr_1+\mc{O}(\eta^{L-j})\quad\text{with}\quad
	\hr_1\stackrel{\eqref{eq:E2eigen}}{=} \frac{1}{m^3d+m}\big[md\,\id_{m^2}+\Swap\big].
\end{equation}
With Eq.~\eqref{eq:MPS_X-Y}, Eq.~\eqref{eq:MPS_TrX2}, and $\Tr_{m\times m}(\Swap\,\hr_1)=m(d+1)/(m^2d+1)$, we arrive at
\begin{equation}\label{eq:MPS_TrX2part}\textstyle
	\Avg\Tr\big([\hX-\frac{1}{md}\Tr\hX]^2\big) = \frac{m(d+1)}{m^2 d+1}-\frac{1}{md} + \mc{O}( \eta^{L-j} )
	\quad\text{with}\quad
	\eta\stackrel{\eqref{eq:E2eigen}}{=} \frac{1-1/m^2}{d-1/(m^2d)}.
\end{equation}
\textbf{(e)} 
The operator $\Avg\hL\otimes\hL$ needed for the second term in Eq.~\eqref{eq:MPS_X-Y}, $\Tr\hY^2-\frac{1}{md}\Tr^2\hY$, evaluates to
\begin{align}\nonumber
	\Avg\hL\otimes\hL
	&\stackrel{\eqref{eq:MPS_Y}}{=}\textstyle
	 \Avg\,\bra 0_d,0_d|\hU_i^{\dag\otimes 2}\big([\id_m\otimes\hh]\otimes[\id_m\otimes\hh]\big) \hU^{\otimes 2}_i|0_d,0_d\ket\\
	&\,\,\stackrel{\eqref{eq:G2}}{=}\textstyle
	 \bra 0_d,0_d|\G^{\doub\dag}\big([\id_m\otimes\hh]\otimes[\id_m\otimes\hh]\big)  |0_d,0_d\ket
	 \stackrel{\eqref{eq:G2}}{=}\frac{\Tr\hh^2}{d\,(m^2d^2-1)}(md\Swap-\id_{m^2}). \label{eq:MPS_AvgLL}
\end{align}
Using $\Tr\hh=0$, it then follows that
\begin{align}\textstyle\nonumber
	\Avg\big(\Tr\hY^2-\frac{1}{md}\Tr^2\hY\big)
	&\textstyle =
	  \Tr\left((\Swap-\frac{1}{md}\id_{(md)^2})\,\Avg\hY\otimes \hY\right)\\ \nonumber
	&\textstyle =
	  d \Tr\left((\Swap-\frac{1}{m}\id_{m^2})\, (\E^{\doub\dag})^{j-i-1}\big(\Avg\hL\otimes\hL\big)\right)\\ \nonumber
	&\textstyle =
	  \frac{\Tr\hh^2}{m^2d^2-1}\eta^{j-i-1}\,\bbra\Swap-\frac{1}{m}\id_{m^2}|\hl_2\kket\,\bbra\hr_2|md\Swap-\id_{m^2}\kket\\
	&\textstyle =
	  \Tr\hh^2 \,\frac{md(m^2-1)}{m^2d^2-1}\eta^{j-i-1}=m\Tr\hh^2\, \eta^{j-i}.
	\label{eq:MPS_TrY2part}
\end{align}
For the second line, we have employed Eqs.~\eqref{eq:MPS_Y} and \eqref{eq:E2}. The third and fourth lines follow from
\begin{subequations}\label{eq:MPS_TrY2explain}
\begin{gather}\textstyle
	(\E^{\doub\dag})^n\stackrel{\eqref{eq:E2dyadic}}{=} |\hl_1\kket\bbra \hr_1| + \eta^n\,|\hl_2\kket\bbra \hr_2|,\quad
	\bbra\Swap-\frac{1}{m}\id_{m^2}|\hl_1\kket = \Tr(\Swap-\frac{1}{m}\id_{m^2}) = 0,\\\textstyle
	\bbra\Swap-\frac{1}{m}\id_{m^2}|\hl_2\kket = \frac{m^2-1}{2},\quad\text{and}\quad
	\bbra\hr_2|md\Swap-\id_{m^2}\kket =2md.
\end{gather}
\end{subequations}
Equations~\eqref{eq:MPS_TrX2part} and \eqref{eq:MPS_TrY2part} in conjunction with Eq.~\eqref{eq:gradVarProd} conclude the proof of Theorem~\ref{thrm:MPS-dist} for $j>i$.\\[0.4em]
\textbf{(f)} 
Lastly, we need to address the case $j=i$. The cost function \eqref{eq:costLocal} can again be written in the form $\bra\Psi|\hh_i|\Psi\ket=\Tr_{m\times d}(\hX \hU_i^\dag\hY'\hU_i)$ with $\hX$ as in Eq.~\eqref{eq:MPS_X} and
\begin{equation}\label{eq:MPS_Y2}
	\hY':=\id_m\otimes\hh.
\end{equation}
Equation~\eqref{eq:grad_avg0} then implies Eq.~\eqref{eq:MPSgradDecayAvg}.
For the variance, we again need to evaluate the two terms in Eq.~\eqref{eq:MPS_X-Y}. The one for $\hX$ results in Eq.~\eqref{eq:MPS_TrX2part} with $j=i$. Using $\Tr\hh=0$, the second term in Eq.~\eqref{eq:MPS_X-Y} evaluates to
\begin{equation}\label{eq:MPS_TrY2partLocal}
	\Avg\big(\textstyle\Tr\hY'^{2}-\frac{1}{md}\Tr^2\hY'\big)
	= \Avg\Tr\hY'^{2}=m\Tr\hh^2.
\end{equation}
Equations~\eqref{eq:MPS_TrX2part} and \eqref{eq:MPS_TrY2partLocal} in conjunction with Eq.~\eqref{eq:gradVarProd} prove Theorem~\ref{thrm:MPS-dist} for $j=i$. \qed

\subsection{Extension to extensive Hamiltonians}\label{sec:MPS-ext}
Let us now consider the practically more relevant extensive operators in the energy functional, i.e., cost functions of the form
\begin{equation}\label{eq:costGlobal}
	\sum_{i=1}^L\bra\Psi|\hh_i|\Psi\ket,
\end{equation}
But, for now, we will still restrict the $\hh_i$ to be single-site operators as defined in Eq.~\eqref{eq:hiSite} with $\Tr\hh=0$.
\begin{theorem}[\propHead{Bond-dimension dependence of the MPS gradient for extensive Hamiltonians}]
\label{thrm:MPSext-bond}
With MPS unitaries $\hU_j$ in Eq.~\eqref{eq:MPSiso} sampled according to the uniform Haar measure, the average of the Riemannian gradient $\partial_{\hU_j} \sum_i\bra\Psi|\hh_i|\Psi\ket$ for the cost function \eqref{eq:costGlobal} is zero and its variance
decays, for large $m$ as $2\Tr\hh^2/(md)^2$. Specifically, for $j\in\{b+1,\dotsc,L-b\}$,
\begin{subequations}\label{eq:varMPSext}
\begin{gather}\label{eq:varMPSextAvg}
	\Avg \Big(\partial_{\hU_j} \sum_i \bra\Psi|\hh_i|\Psi\ket\Big) = 0\quad\text{and}\\
	\label{eq:varMPSextVar}
	\Var \Big(\partial_{\hU_j} \sum_i \bra\Psi|\hh_i|\Psi\ket\Big)
	=2\Tr(\hh^2)\,\frac{m^2d^2-1}{d(d-1)(m^2d+1)^2}+\mc{O}(\eta^j)+\mc{O}(\eta^{L-j}).
\end{gather}
\end{subequations}
Finite-size effects decay exponentially in the distance of site $j$ from the boundaries and are controlled by the decay factor $\eta$ from Eq.~\eqref{eq:MPSgradDecay}.
\end{theorem}\noindent
\Emph{Proof:}
The global cost function \eqref{eq:costGlobal} can be written in the form
\begin{equation}\label{eq:costGlobal-XY}\textstyle
	\sum_{i=1}^L\bra\Psi|\hh_i|\Psi\ket
	 = \sum_{i=1}^j\Tr
	 (\hX_i \hU_j^\dag\hY_i \hU_j) + \sum_{i=j+1}^L\bra\Psi|\hh_i|\Psi\ket.
\end{equation}
Due to the left-orthonormality condition \eqref{eq:MPSisoConstr}, the second sum is independent of $\hU_j$ and does not contribute to the gradient. According to Eq.~\eqref{eq:grad_avg0}, the form \eqref{eq:costGlobal-XY} implies that the average Riemannian gradient \eqref{eq:varMPSextAvg} is zero.
For $i>b$, $\hX_i$ is given by Eq.~\eqref{eq:MPS_X}, and $\hY_i$ is given by the expression in Eq.~\eqref{eq:MPS_Y} for $j>i$ and  by Eq.~\eqref{eq:MPS_Y2} for $j=i$. Minor modifications occur when $i$ is close to the left end of the chain ($i\leq b$), but we consider $j$ in the bulk of the system and contributions to the gradient variance decay exponentially in $|i-j|$. For brevity, we will not discuss the boundary effects in detail and capture them with the terms $\mc{O}(\eta^j)+\mc{O}(\eta^{L-j})$ in Eq.~\eqref{eq:varMPSextVar}.
The Riemannian gradient [Eq.~\eqref{eq:Riem_grad}] now takes the form
\begin{equation}\label{eq:Riem_gradExt}\textstyle
	\partial_{\hU_j} \sum_{i=1}^L\bra\Psi|\hh_i|\Psi\ket
	\stackrel{\eqref{eq:costGlobal-XY}}{=} \sum_{i=1}^j\hg_i\quad\text{with}\quad
	\hg_i\stackrel{\eqref{eq:Riem_grad}}{=} \hY_i\hU_j\hX_i - \hU_j\hX_i\hU_j^\dag\hY_i\hU_j
\end{equation}
and, in generalization of Eq.~\eqref{eq:gradVarProd}, we find that its variance \eqref{eq:gradVar} is
\begin{align}\nonumber\textstyle
	\Var \partial_{\hU_j} \sum_i\bra\Psi|\hh_i|\Psi\ket
	&\textstyle =\frac{1}{N}\sum_{i_1,i_2=1}^j\Avg \Tr(\hg^\dag_{i_1}\hg_{i_2})\\\nonumber
	&\textstyle=\frac{2}{N^2-1}\sum_{i_1,i_2=1}^j
	   \Avg\Tr\left(\big(\hX_{i_1}-\frac{\id_N}{N}\Tr\hX_{i_1}\big)\big(\hX_{i_2}-\frac{\id_N}{N}\Tr\hX_{i_2}\big)\right)\\
	\label{eq:gradExtVarProd}
	&\textstyle\hspace{14ex}\times
	   \Avg\Tr\left(\,\big(\hY_{i_1}-\frac{\id_N}{N}\Tr\hY_{i_1}\big)\,\,\big(\hY_{i_2}-\frac{\id_N}{N}\Tr\hY_{i_2}\big)\,\right)
\end{align}
with $N=md$; see Fig.~\ref{fig:Riem_grad}c. The off-diagonal terms with $i_1\neq i_2$ in this expression vanish: In particular, consider the case $i_2<i_1\leq j$. Then, in analogy to Eq.~\eqref{eq:MPS_AvgLL}, we encounter the expression
\begin{gather}\nonumber
	 \Avg\,\bra 0_d,0_d|\hU_{i_2}^{\dag\otimes 2}\left( [\id_m\otimes\id_d]\otimes[\id_m\otimes\hh] \right) \hU^{\otimes 2}_{i_2}|0_d,0_d\ket\\
	\quad\,\,=
	 \bra 0_d,0_d|\G^{\doub}\left( [\id_m\otimes\id_d]\otimes[\id_m\otimes\hh] \right)  |0_d,0_d\ket
	 \stackrel{\eqref{eq:G2}}{=}0
\end{gather}
because $\Tr_{d\times d}(\id_d\otimes\hh)=d\Tr\hh=0$ and $\Tr_{d\times d}(\Swap\,[\id_d\otimes\hh])=\Tr\hh=0$. Hence, using Theorem~\ref{thrm:MPS-dist},
\begin{align}\nonumber\textstyle
	\Var \partial_{\hU_j} \sum_i\bra\Psi|\hh_i|\Psi\ket
	&\textstyle =\frac{1}{N}\sum_{i=1}^j\Avg \Tr(\hg^\dag_i\hg_i)\\
	&\textstyle =2\Tr(\hh^2)\,\frac{1}{d(m^2d+1)}\,\sum_{n=0}^\infty\eta^{n} + \mc{O}(\eta^j)+ \mc{O}(\eta^{L-j}).
\end{align}
With $\sum_{n=0}^\infty\eta^n=1/(1-\eta)\stackrel{\eqref{eq:MPSgradDecay}}{=}(m^2d^2-1)/[(d-1)(m^2d+1)]$, we arrive at Eq.~\eqref{eq:varMPSextVar}. \qed

\subsection{Extension to finite-range interactions}\label{sec:MPS-nn}
So far, we only considered single-site terms $\hh_i$ as in Refs.~\cite{Liu2022-129,Garcia2023-2023,Zhao2021-5,Martin2023-7}. The corresponding optimization problems are of course trivial in the sense that they are solved by product states $|\Psi\ket=|\phi_1\ket\otimes\dotsb\otimes|\phi_L\ket$ of the single-site ground states, corresponding to MPS with bond dimension $m=1$. The adaptation of the results to finite-range interactions is relatively straight forward. Specifically, consider nearest-neighbor interaction terms $\hh_i$ that act non-trivially on sites $i$ and $i+1$,
\begin{equation}\label{eq:hi-nn}
	\hh_i=\id_d^{\otimes(i-1)}\otimes\hh\otimes\id_d^{\otimes(L-i-1)}\ \ \text{with}\ \
	\hh=\hh^\dag\in\End(\CC^{d}\otimes\CC^{d}),\ \
	\Tr\hh=0,\ \text{and}\ \Tr_1\hh=\Tr_2\hh,
\end{equation}
where $\Tr_1 \hh$  and $\Tr_2 \hh$ denote the partial trace over the first component and second component of the two-site Hilbert space $\CC^{d}\otimes\CC^{d}$, respectively. So, the last constraint in Eq.~\eqref{eq:hi-nn} assumes that the two single-site components of $\hh$ agree.
\begin{theorem}[\propHead{Scaling of MPS gradient for Hamiltonians with nearest-neighbor interactions}]
\label{thrm:MPSnn}
With MPS unitaries $\hU_j$ in Eq.~\eqref{eq:MPSiso} sampled according to the uniform Haar measure, and nearest-neighbor interaction terms \eqref{eq:hi-nn} the averages of Riemannian gradients are
\begin{subequations}\label{eq:varMPSnn}
\begin{gather}\label{eq:varMPSnnAvg}\textstyle
	\Avg \Big(\partial_{\hU_j} \bra\Psi|\hh_i|\Psi\ket\Big) = 0,\quad\text{and}\quad
	\Avg \Big(\partial_{\hU_j} \sum_{i=1}^{L-1} \bra\Psi|\hh_i|\Psi\ket\Big) = 0.
\end{gather}
For a single Hamiltonian term \eqref{eq:hi-nn}, the gradient variance $\Var \big(\partial_{\hU_j} \bra\Psi|\hh_i|\Psi\ket\big)$ with $i,j\in\{b+1,\dotsc,L-b-1\}$ is
\begin{gather}\label{eq:varMPSnnVar-a}\textstyle
	0\quad\text{for}\ j<i,\\
	\label{eq:varMPSnnVar-c}\textstyle
	2\left[\Tr\hh^2+\left(\frac{d-1}{d}-\frac{m^2-1}{m^2d-1}\right)\Tr(\Tr^2_1\hh)\right]\,\frac{m^2d-1}{d^2(m^2d+1)(m^2d^2-1)}
	   +\mc{O}(\eta^{L-i})\quad\text{for}\ j=i,\ \text{and}\\
	\label{eq:varMPSnnVar-b}\textstyle
	2\left[\Tr\hh^2+\frac{d-1}{d}\Tr(\Tr^2_1\hh)\right]\,\frac{m^2d-1}{d^2(m^2d+1)(m^2d^2-1)}\,\eta^{j-i-1}
	   +\mc{O}(\eta^{L-i})
	\quad\text{for}\ j\geq i+1.
\end{gather}
For extensive Hamiltonians with nearest-neighbor interaction terms and large bond dimension $m$, the variance of the energy gradient scales as
\begin{equation}\label{eq:varMPSnnExtVar}\textstyle
	\Var \Big(\partial_{\hU_j} \sum_{i=1}^{L-1} \bra\Psi|\hh_i|\Psi\ket\Big)
	\sim \frac{4}{m^2 d^4}\left[\Tr(\hh^2)+2\Tr(\Tr^2_1\hh)\right] + \mc{O}(\eta^j) + \mc{O}(\eta^{L-j}).
\end{equation}
\end{subequations}
Finite-size effects decay exponentially in the distance of site $j$ from the boundaries and are controlled by the decay factor $\eta$ from Eq.~\eqref{eq:MPSgradDecay}.
\end{theorem}\noindent
The proof is given in Appendix~\ref{appx:MPSnn}. It also explicitly provides all terms contributing to Eq.~\eqref{eq:varMPSnnExtVar}; see Eqs.~\eqref{eq:varMPSnnExtVar-diag} and \eqref{eq:varMPSnnExtVar-offdiag}.

\section{\texorpdfstring{Multiscale entanglement renormalization ansatz\\ and tree tensor networks}{Multiscale entanglement renormalization ansatz and tree tensor networks}}\label{sec:MERA}
\subsection{Setup}
MERA \cite{Vidal-2005-12,Vidal2006} are hierarchical TNS motivated by real-space renormalization group schemes \cite{Kadanoff1966-2,Jullien1977-38,Drell1977-16}. In each renormalization step (layer), the system is partitioned into small cells. Those cells are to some extent disentangled from neighboring cells by local unitary transformations before the number of degrees of freedom per cell is reduced by application of isometries that map groups of $b$ sites into one renormalized site. The reduction factor is the so-called \emph{branching ratio} $b$. Every renormalized site is associated with a vector space $\CC^\chi$ of dimension $\chi$. The latter is also referred to as the \emph{bond dimension} of the MERA. The renormalization procedure $\tau\to\tau+1$ can be stopped after $T$ steps by applying a final layer ($\tau=T$) of isometries that map into one-dimensional spaces, i.e., by projecting onto some reference states. Seen in reverse, this renormalization procedure generates an entangled many-body state for the original lattice system.

TTNS \cite{Fannes1992-66,Otsuka1996-53,Shi2006-74,Murg2010-82,Tagliacozzo2009-80} are a subclass of MERA that contain no disentanglers. In this case, the network has no loops and assumes a tree structure.

The fact that all tensors in a MERA (or TTNS) $|\Psi\ket$ are isometries leads to strong simplifications in the evaluation of expectation values. A lot of tensors $\hW$ from $|\Psi\ket$ and their counterparts $\hW^\dag$ from $\bra\Psi|$ cancel to identities in expectation values like $\bra\Psi|\hh_i|\Psi\ket$ as $\hW^\dag\hW=\id$. Every block $\mc{A}$ of sites is associated with a \emph{causal cone}, containing only those tensors of the MERA that can influence observations on $\mc{A}$. The structure of the MERA implies that, in every layer $\tau$, there is only a system-size independent number of renormalized sites inside the causal cone. We will find that the gap of quantum channels that describe transitions from layer $\tau$ to layer $\tau-1$ inside the causal cone, result in an exponential decay of energy-gradient Haar variances with respect to $\tau$.

Let us first discuss the case of binary 1D MERA in detail. Then we will address the cases of ternary 1D MERA (Sec.~\ref{sec:MERA-ternary}) and nonary 2D MERA (Sec.~\ref{sec:MERA-nonary}). In Sec.~\ref{sec:furtherMERA}, we address the corresponding TTNS and explain why, generally, TTNS and MERA should not feature barren plateaus.

\subsection{Binary 1D MERA}
Consider a 1D lattice of $L=2^{T'}$ sites with periodic boundary conditions and a MERA $|\Psi\ket$ with branching ratio $b=2$ and $T$ layers. The physical lattice and the lattices of renormalized sites are
\begin{equation}
	\L_0=\{0,\dotsc,L-1=2^{T'}-1\},\ \ 
	\L_1=\{0,\dotsc,2^{T'-1}-1\},\ \dotsc\ ,
	\L_T=\{0,\dotsc,2^{T'-T}-1\}.
\end{equation}
For simplicity, let us assume that (a) $T\leq T'-2$ such that layer-transition maps for causal cones have all the same structure (are not impacted by the boundary conditions in the higher layers) and that (b) the dimension $d$ of each physical single-site Hilbert space agrees with the bond dimension $\chi$ of the MERA. In actual simulations, one would have a few initial layers where the bond dimension increases in steps from $d$ to the chosen bond dimension $\chi$.
\begin{figure*}[t]
	\label{fig:MERAbin}
	\includegraphics[width=1\textwidth]{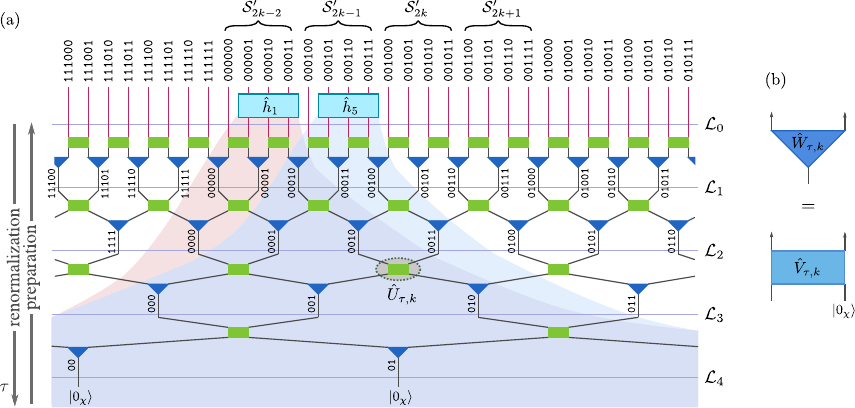}
	\caption{(a) Part of a binary 1D MERA with $T=4$ layers on $L=64$ sites. Shaded regions indicate causal cones for two three-site operators $\hh_1$ and $\hh_5$ acting on sites $\{1,2,3\}$ and $\{5,6,7\}$, respectively. The contained renormalized sites are determined by Eq.~\eqref{eq:1dMERAbin-cone}. The two cones merge after three renormalization steps. Site and renormalized-site numbers are shown in binary representation. The lowest bit of the first site $i_{\tau-1}\in\L_{\tau-1}$ in the causal cone determines whether the transition map $\M_{\tau,i_{\tau-1}}$ [Eq.~\eqref{eq:1dMERAbin-M}] of layer $\tau$ is left-moving (even $i_{\tau-1}$) or right-moving (odd $i_{\tau-1}$). The causal support $\S_k\subset\L_0$ of tensor $\hU_{\tau,k}$ is the set of physical sites $i$ with $\hU_{\tau,k}$ in the causal cone of $\hh_i$. In this example, the causal support for $\hU_{3,1}$ is $\S_k=\{0,1,\dotsc,15\}$. According to Eq.~\eqref{eq:1dMERAbin-Sk}, $\S_k$ can be written as the union of four disjoint neighboring four-site blocks $\S'_q$. All three-site terms $\hh_i$ with $i\in\S'_q$ have the same layer-$\tau$ transition map.
	(b) Isometries $\hW_{\tau,k}$ can be parametrized by unitaries $\hV_{\tau,k}$ that are, on one side, projected onto the reference state $|0_\chi\ket$; see Eq.~\eqref{eq:1dMERAbin-isometry}.}
\end{figure*}

For the first layer $\tau=1$, we apply unitary two-site nearest neighbor gates $\hU^\dag_{\tau,k}\in\groupU(\chi^2)$ (the so-called \emph{disentanglers}) on all even edges, i.e., pairs of sites $(2k,2k+1)$ from $\L_{\tau-1}$. Then, we apply isometries $\hW^\dag_{\tau,k}:\CC^{\chi\times\chi}\to\CC^{\chi}$ that map sites $(2k-1,2k)$ from $\L_{\tau-1}$ into the renormalized site $k\in\L_\tau$ with $\hW^\dag_{\tau,k}\hW_{\tau,k}=\id_\chi$. Repeating this for the remaining $T-1$ layers, we arrive at the lattice $\L_T$ containing $L/2^T=2^{T'-T}$ renormalized sites and end the procedure by projecting on every site onto an arbitrary reference state $|0_\chi\ket$ from $\CC^\chi$. Like the MPS tensors in Eq.~\eqref{eq:MPSiso}, the MERA isometries $\hW_{\tau,k}$ can be parametrized by unitaries $\hV_{\tau,k}\in \groupU(\chi^2)$ that are, on one side, projected onto the reference state $|0_\chi\ket$,
\begin{equation}\label{eq:1dMERAbin-isometry}
	\hW_{\tau,k}=\hV_{\tau,k}\,\big(\id_{\chi}\otimes |0_\chi\ket\big).
\end{equation}

Figure~\ref{fig:MERAbin} shows a binary 1D MERA and causal cones for two three-site blocks of $\L_0$ are indicated by the shaded regions.
They comprise three neighboring (renormalized) sites in each of the lattices $\L_{\tau\geq 1}$.
If we start with sites $\{i_0:=i,i_0+1,i_0+2\}\subset\L_0$ of the physical lattice, after $\tau$ renormalization steps, the causal cone contains only sites
\begin{equation}\label{eq:1dMERAbin-cone}
	\{i_\tau,i_\tau+1,i_\tau+2\}\subset\L_\tau\quad\text{where}\quad
	i_\tau=\lfloor i_{\tau-1}/2\rfloor.
\end{equation}

\begin{figure*}[t]
	\label{fig:MERAbin-expect}
	\includegraphics[width=1\textwidth]{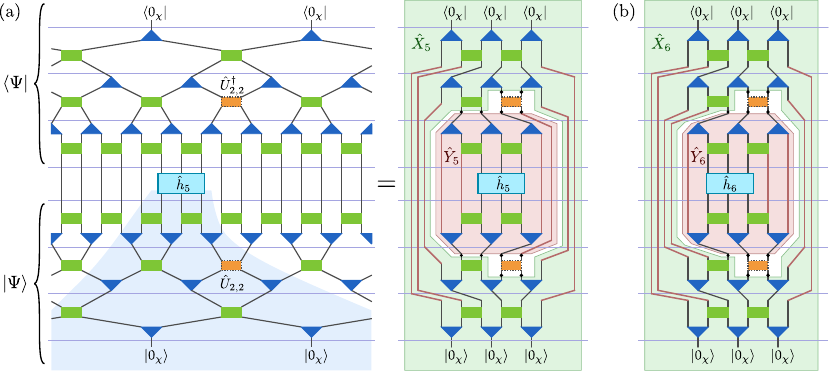}
	\caption{As all MERA tensors are isometric, expectation values of local operators $\bra\Psi|\hh_i|\Psi\ket$ only depend on the tensors in the causal cone of $\hh_i$. According to Eq.~\eqref{eq:1dMERAbin-expct}, they can be written in the form $\bra\Psi|\hh_i|\Psi\ket=\Tr\big[ \hX_{i} (\hU_{\tau,k}^\dag\otimes\id_M) \hY_{i} (\hU_{\tau,k}\otimes\id_M) \big]$. (a) Diagrammatic representation of $\bra\Psi|\hh_5|\Psi\ket=\Tr\big[ \hX_{5} (\hU_{2,2}^\dag\otimes\id_M) \hY_{5} (\hU_{2,2}\otimes\id_M) \big]$ for a binary 1D MERA with $T=3$ layers and bond dimension $\chi$. Here, $M=\chi^2$. (b) Diagrammatic representation of $\bra\Psi|\hh_6|\Psi\ket=\Tr\big[ \hX_6 (\hU_{2,2}^\dag\otimes\id_M) \hY_6 (\hU_{2,2}\otimes\id_M) \big]$ for the same MERA. Now, $M=\chi$ as discussed below Eq.~\eqref{eq:1dMERAbin-Edoub-b}.}
\end{figure*}

\begin{figure*}[b]
	\label{fig:MERAbin-layerTransit}
	\includegraphics[width=1\textwidth]{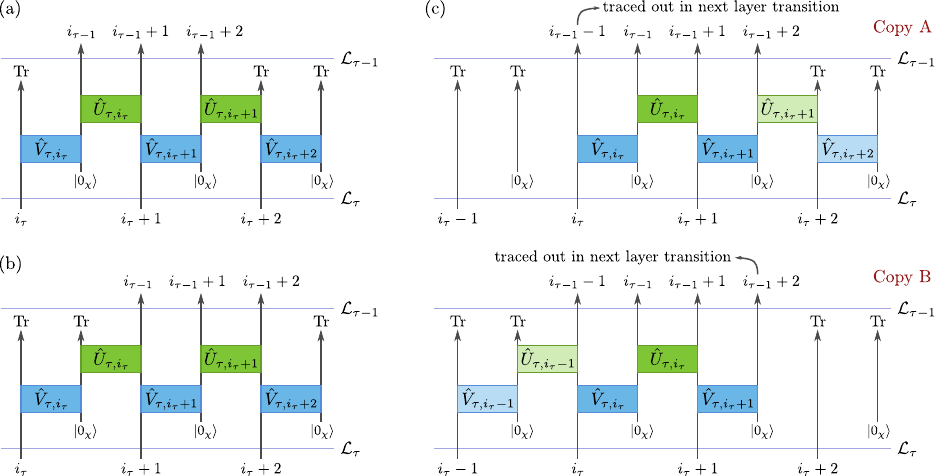}
	\caption{Graphical representations of layer transition maps for binary 1D MERA. (a) For the left-moving layer transition map $\M_{\tau,i_{\tau-1}}$ with even $i_{\tau-1}$ [Eq.~\eqref{eq:1dMERAbin-M}], we start with the density operator $\dm^{(\tau)}$ for the three-site intersection of the causal cone with lattice $\L_\tau$, apply three isometries \eqref{eq:1dMERAbin-isometry} followed by two disentanglers, and trace out one site on the left and two on the right to obtain the density operator $\dm^{(\tau-1)}$ for the three-site intersection of the causal cone with lattice $\L_{\tau-1}$.
	In the evaluation of the diagonal contributions to the gradient variance with $i=j$ in Eq.~\eqref{eq:1dMERAbin-gradVarProdM}, we apply these layer transition maps on two copies of the system. Taking then the Haar average over one of the involved tensors, it is basically replaced by the doubled fully depolarizing channel $\G^\doub$, in total yielding the doubled layer-transition channel $\E^\doub_{\bin,\tL}$; see Eqs.~\eqref{eq:1dMERAbin-Edoub-b} and \eqref{eq:1dMERAbin-EdoubLt-matrix}.
	(b) The right-moving layer transition map, occurring for odd $i_{\tau-1}$, only differs in the traced out sites.
	(c) In the evaluation of off-diagonal contributions to the gradient variance with $i\neq j$ in Eq.~\eqref{eq:1dMERAbin-gradVarProdM}, the causal cones in the two copies of the system have a relative shift. In the shown example, $i_\tau=j_\tau+1$ and $i_{\tau-1}=j_{\tau-1}+1$. The Haar average over tensors that are only applied in one of the two copies of the system yields fully depolarizing channels $\G$ in that copy only. The Haar average over the tensors that occur in both copies still yields $\G^{(2)}$. In total, one obtains the transition channel $\E^{(2)}_{\bin-1}$ as discussed in Appx.~\ref{appx:1dMERAbin-Edoub-matrix-offdiag}.}
\end{figure*}

Let us now consider cost functions
\begin{equation}\label{eq:1dMERAbin-cost}
	\sum_{i=0}^{L-1}\bra\Psi|\hh_i|\Psi\ket,
\end{equation}
of local extensive Hamiltonians, where interaction term $\hh_i$ acts non-trivially on sites $\{i,i+1,i+2\}$,
\begin{equation}\label{eq:hi-nnn}
	\hh_i=\id_\chi^{\otimes i}\otimes\hh\otimes\id_\chi^{\otimes(L-i-3)}\ \ \text{with}\ \
	\hh=\hh^\dag\in\End(\CC^{\chi^3}),\quad \Tr\hh=0,
\end{equation}
and, for simplicity, we assume $\hh$ to be symmetric under spatial reflection.
For $i>L-3$, the expression \eqref{eq:hi-nnn} for $\hh_i$ has to be adapted in accordance with the periodic boundary conditions ($i\equiv i\mod L$).
As discussed in the following, the optimization problem for the energy functional \eqref{eq:1dMERAbin-cost} is not hampered by barren plateaus.
\begin{theorem}[\propHead{Decay of energy gradients for binary 1D MERA}]
\label{thrm:1dMERAbin}
Consider 1D binary MERAs $|\Psi\ket$ with bond dimension $\chi$ and $T$ layers on $2^{T'}$ sites with $T'\geq T+2$.
With all disentanglers $\hU_{\tau,k}$ and unitaries $\hV_{\tau,k}$ for the isometries in Eq.~\eqref{eq:1dMERAbin-isometry} sampled according to the uniform Haar measure, the average of the Riemannian gradient $\partial_{\hU_{\tau,k}} \sum_i \bra\Psi|\hh_i|\Psi\ket$ for the cost function \eqref{eq:1dMERAbin-cost} is zero, and at least a finite fraction of the unitaries in layer $\tau$ has a gradient variance that scales as
\begin{equation}\label{eq:1dMERAbin-Var}
	\Var \Big(\partial_{\hU_{\tau,k}} \sum_i \bra\Psi|\hh_i|\Psi\ket\Big) = \Theta\big((2\eta_\bin)^\tau\big) +\mc{O}\big((2\lambda_3)^\tau\big)+\mc{O}(2^\tau\eta_\bin^T)
	\quad\text{with}\quad
	\eta_\bin =\frac{\chi^2(1+\chi)^4}{2(1+\chi^2)^4},
\end{equation}
where $\lambda_3=[\chi^2(1+\chi)^2]/[2(1+\chi^2)^3]$ controls small $\tau$ effects. The same applies for the $\hV_{\tau,k}$ gradients.
\end{theorem}\noindent
\Emph{Proof:}
\textbf{(a)}
As shown in Fig.~\ref{fig:MERAbin-expect}, the expectation value for a single Hamiltonian term $\hh_i$ can be evaluated by only contracting the tensors inside the causal cone \eqref{eq:1dMERAbin-cone} of $\hh_i$. Starting from the top layer $\tau=T$ with the reference state $|0_\chi\ket$ on sites $\{i_{T},i_{T}+1,i_{T}+2\}\subset\L_T$, i.e.,
\begin{equation}
	\dm^{(T)}=(|0_\chi\ket\bra 0_\chi|)^{\otimes 3},
\end{equation}
we can progress down layer by layer. In every step, $\dm^{(\tau)}\mapsto \dm^{(\tau-1)}$, we first apply one isometry on each of the three renormalized sites
\begin{subequations}\label{eq:1dMERAbin-transitionMap}
\begin{equation}
	\dm^{(\tau)}\mapsto
	\hat{\sigma}^{(\tau)}:=
	\left(\hW_{\tau,i_\tau}\otimes\hW_{\tau,i_\tau+1}\otimes\hW_{\tau,i_\tau+2}\right) \dm^{(\tau)}
	\left(\hW^\dag_{\tau,i_\tau}\otimes\hW^\dag_{\tau,i_\tau+1}\otimes\hW^\dag_{\tau,i_\tau+2}\right).
\end{equation}
For the resulting six-site state $\hat{\sigma}^{(\tau)}$, we then apply two unitary disentanglers on the central four sites and, finally, trace out three of the outer sites, either one on the left and two on the right or two on the left and one on the right, such that only sites $\{i_{\tau-1},i_{\tau-1}+1,i_{\tau-1}+2\}\subset\L_{\tau-1}$ as defined through Eq.~\eqref{eq:1dMERAbin-cone} remain. We will refer to the first case as a \emph{left-moving} and to the second as a \emph{right-moving} layer-transition map;
see Figs.~\ref{fig:MERAbin} and \ref{fig:MERAbin-layerTransit}.
Denoting the partial trace by $\Tr_\Out$,
\begin{equation}
	\hat{\sigma}^{(\tau)}\mapsto\dm^{(\tau-1)}=\Tr_\Out\left[
	\left(\id_\chi\otimes \hU_{\tau,i_\tau}\otimes\hU_{\tau,i_\tau+1}\otimes\id_\chi\right) \hat{\sigma}^{(\tau)}
	\left(\id_\chi\otimes \hU^\dag_{\tau,i_\tau}\otimes\hU^\dag_{\tau,i_\tau+1}\otimes\id_\chi\right)\right].
\end{equation}
We denote this layer-transition map as $\M_{\tau,i_{\tau-1}}$ such that
\begin{equation}\label{eq:1dMERAbin-M}
	\M_{\tau,i_{\tau-1}}(\dm^{(\tau)})=\dm^{(\tau-1)}
\end{equation}
\end{subequations}
Note that all $\M_{\tau,i_{\tau-1}}$ with even $i_{\tau-1}$ are left-moving and those with odd $i_{\tau-1}$ are right-moving transition maps.
With $i_0\equiv i$ as defined in the context of Eq.~\eqref{eq:1dMERAbin-cone}, the energy expectation value for the Hamiltonian term $\hh_i$ is
\begin{subequations}\label{eq:1dMERAbin-expct}
\begin{align}
	\bra\Psi|\hh_i|\Psi\ket
	&= \Tr\left[ \M_{1,i_{0}} \circ\dotsb\circ\M_{\tau-1,i_{\tau-2}}\circ\M_{\tau,i_{\tau-1}} \circ\dotsb\circ\M_{T,i_{T-1}} (\dm^{(T)})\, \cdot \,\hh\, \right]\\
	\label{eq:1dMERAbin-expct-b}
	&= \Tr\Big[ \M_{\tau,i_{\tau-1}}\circ
	             \underbrace{\M_{\tau+1,i_{\tau}}\circ\dotsb\circ\M_{T,i_{T-1}} (\dm^{(T)})}_{=:\hx_{i_\tau}}
	             \, \cdot \,
	             \underbrace{\M^\dag_{\tau-1,i_{\tau-2}}\circ\dotsb\circ\M^\dag_{1,i_{0}}(\hh)}_{=:\hy_i}
	             \Big]\\
	\label{eq:1dMERAbin-expct-c}
	&=: \Tr\left[ \hX_i (\hU^\dag\otimes\id_M) \hY_i (\hU\otimes\id_M) \right].
\end{align}
\end{subequations}
The third line has exactly the form of Eq.~\eqref{eq:energy}, where $\hU$ refers to one of the unitaries $\hU^\dag_{\tau,k}$ or $\hV^\dag_{\tau,k}$ from layer $\tau$ inside the causal cone, and we have decomposed $\M_{\tau,i_{\tau-1}}$ accordingly to get from line two to line three. To be specific, for the following, let us choose $\hU$ in these equations to be a disentangler $\hU_{\tau,k}$ in layer $\tau$; the argument works in exactly the same way for the isometries \eqref{eq:1dMERAbin-isometry}. Correspondingly, we will use labels $\hX_{k,i}$ and $\hY_{k,i}$ for the operators in Eq.~\eqref{eq:1dMERAbin-expct-c}.
Figure~\ref{fig:MERAbin-expect} shows for one particular disentangler $\hU_{\tau,k}$ and two sites $i$, the corresponding definitions of $\hX_{k,i}$ and $\hY_{k,i}$ in diagrammatic form.
For the binary 1D MERA, $\id_M$ is the identity on either one or two renormalized sites from $\L_{\tau-1}$ inside the causal cone, i.e., $M=\chi$ or $M=\chi^2$ as discussed in more detail below Eq.~\eqref{eq:1dMERAbin-Edoub-b}. Given the form \eqref{eq:1dMERAbin-expct}, the vanishing of the Haar-average Riemannian gradient follows from Eq.~\eqref{eq:grad_avg0}.
\\[0.4em]
\textbf{(b)} The scaling for the Haar-variance of the Riemannian gradient \eqref{eq:1dMERAbin-Var} can be analyzed on the basis of Eqs.~\eqref{eq:gradVarProdM} and \eqref{eq:1dMERAbin-expct}. For the extensive Hamiltonian $\sum_i\hh_i$ in the cost function \eqref{eq:1dMERAbin-cost}, the Riemannian gradient is
\begin{subequations}\label{eq:1dMERAbin-grad}
\begin{gather}
	\hg_k\equiv \partial_{\hU_{\tau,k}} \sum_i \bra\Psi|\hh_i|\Psi\ket=\sum_{i\in \S_k}\hg_{k,i}\quad\text{with}\\
	\hg_{k,i}\stackrel{\eqref{eq:1dMERAbin-expct},\eqref{eq:Riem_grad}}{:=}
	 \Tr_M\left[\hY_{k,i}(\hU_{\tau,k}\otimes\id_M)\hX_{k,i} - (\hU_{\tau,k}\otimes\id_M)\hX_{k,i}(\hU_{\tau,k}^\dag\otimes\id_M)\hY_{k,i}(\hU_{\tau,k}\otimes\id_M)\right],
\end{gather}
\end{subequations}
where $\S_k\subset\L_0$ denotes the set of physical sites $i$ with $\hU_{\tau,k}$ in the causal cone of $\hh_i$, i.e., $\S_k$ is the \emph{causal support} of tensor $\hU_{\tau,k}$. For extensive Hamiltonians, the expression \eqref{eq:gradVarProdM} for the gradient variance \eqref{eq:gradVar} needs to be generalized as discussed for MPS in Appendix~\ref{appx:MPSnn}. This leads to [cf.\ Eq.~\eqref{eq:gradExtVarProdM-b}]
\begin{subequations}\label{eq:1dMERAbin-gradVarProdM}
\begin{gather}\label{eq:1dMERAbin-gradVarProdM-a}
	\Var \partial_{\hU_{\tau,k}} \sum_i\bra\Psi|\hh_i|\Psi\ket
	 =\frac{1}{N}\Avg \Tr(\hg_k^\dag\hg_k^\pdag)
	 =\frac{1}{N}\sum_{i,j\in \S_k}\Avg \Tr(\hg^\dag_{k,i}\hg_{k,j}^\pdag)\quad\text{with}\\
	\label{eq:1dMERAbin-gradVarProdM-b}\textstyle
	\frac{1}{N}\Avg \Tr(\hg^\dag_{k,i}\hg_{k,j}^\pdag)=\frac{2}{N^2-1}
	  \Tr\left(\big[\Swap_{1,3}-\frac{1}{N}\id_{N^4}\big]
	           \big[\Swap_{2,4}-\frac{1}{N}\id_{N^4}\big]\,\Avg\,[\hZ_{k,i}\otimes\hZ_{k,j}]\right)
\end{gather}
\end{subequations}
with $N=\chi^2$ and $\hZ_{k,i}\in\End(\CC^N\otimes\CC^N)$ being linear in $\hX_{k,i}$ and $\hY_{k,i}$ as defined in Eq.~\eqref{eq:gradVarProdM-Z}. A diagrammatic representation for this expression is shown in Fig.~\ref{fig:Riem_grad}c.\\[0.4em]
\textbf{(c)} 
Let us discuss the diagonal contributions with $i=j$ in Eq.~\eqref{eq:1dMERAbin-gradVarProdM}. As illustrated in Fig.~\ref{fig:MERAbin}, the causal support
\begin{equation}\label{eq:1dMERAbin-Sk}
	\S_{k}=\S'_{2k-2}\cup\S'_{2k-1}\cup\S'_{2k}\cup\S'_{2k+1}\quad\text{with}\quad
	\S'_{q}:=\{i\in\L_0\,|\,i_\tau=q\}
\end{equation}
is the union of four disjoint and neighboring blocks of $|\S'_{q}|=2^{\tau-1}$ physical sites each. Specifically, $\hh_i$ with $i\in\S'_{q}$ are the Hamiltonian terms with $\{q,q+1,q+2\}\subset\L_{\tau-1}$ in their causal cone.
For $i\in\S'_{q}$, the only quantity in $\hg_{k,i}$ that varies with $i$ is $\hy_i$ as defined in Eq.~\eqref{eq:1dMERAbin-expct-b}. Hence, we can execute the corresponding part of the sum in Eq.~\eqref{eq:1dMERAbin-gradVarProdM-a} by evaluating the quantity
\begin{equation}\label{eq:1dMERAbin-yy}
	\sum_{i\in \S'_{q}}\!\!\!\Avg\hy_i\otimes\hy_i
	\stackrel{\eqref{eq:1dMERAbin-expct-b}}{=}
	\sum_{i\in \S'_{q}}\!\!\!\Avg\M^{\dag\otimes 2}_{\tau-1,i_{\tau-2}}\circ\dotsb\circ\M^{\dag\otimes 2}_{1,i_{0}}(\hh\otimes\hh)
	=2^{\tau-1}(\E^{\doub\dag}_\bin)^{\tau-1}(\hh\otimes\hh).
\end{equation}
Here,
\begin{gather}\label{eq:1dMERAbin-Edoub-a}
	\E^\doub_\bin:=\frac{1}{2}\Avg\left(\M^{\otimes 2}_{\tau,q}+\M^{\otimes 2}_{\tau,q+1}\right)
	= \frac{1}{2}\left(\E^\doub_{\bin,\tL}+\E^\doub_{\bin,\tR}\right)\quad
	\text{with}\\
	\label{eq:1dMERAbin-Edoub-b}
	\E^\doub_{\bin,q}:=
	\begin{cases}
	\Avg\M^{\otimes 2}_{\tau,q}=:\E^\doub_{\bin,\tL}&\text{for even $q$},\\
	\Avg\M^{\otimes 2}_{\tau,q}=:\E^\doub_{\bin,\tR}&\text{for odd $q$}.
	\end{cases}
\end{gather}
is the doubled layer-transition channel for the 1D binary MERA. It is the average over one left-moving and one right-moving transition map, which are further Haar-averaged over the comprising unitaries yielding $\E^\doub_{\bin,\tR}$ and $\E^\doub_{\bin,\tL}$.

We only need to consider the sum \eqref{eq:1dMERAbin-yy} for $q=2k-2$ and $q=2k-1$. The cases $q=2k$ and $q=2k+1$ follow by reflection symmetry.
For $q=2k-1$, the disentangler $\hU^\dag_{\tau,k}$ acts on the second and third sites ($2k,2k+1\in\L_{\tau-1}$) of the causal cone such that the $(M=\chi)$-dimensional space in Eqs.~\eqref{eq:1dMERAbin-expct} and \eqref{eq:1dMERAbin-grad} corresponds to the first site ($2k-1$) in the causal cone, and $\hY_{k,i}=\hy_i$.
For $q=2k-2$, the disentangler $\hU^\dag_{\tau,k}$ acts on the third site ($2k$) of the causal cone and the site $2k+1$ that leaves the causal cone such that the $(M=\chi^2)$-dimensional space then corresponds to the first and second sites ($2k-2,2k-1$) in the causal cone, and $\hY_{k,i}=\hy_i\otimes\id_\chi$.
See Fig.~\ref{fig:MERAbin-expect}.

In part (d) of this proof we will find that $\E^\doub_\bin$ has rank four, is diagonalizable, and is a strictly contractive channel, i.e., it has the non-degenerate eigenvalue $\lambda_1=1$ with left eigenvector $\bbra\id_{\chi^6}|$ and all others have amplitudes $|\lambda_{n>1}|<1$,
\begin{equation}\label{eq:1dMERAbin-Edoub-diag}
	\E^\doub_\bin=\sum_{n=1}^4 \lambda_n|\hr_n\kket\bbra\hl_n|\quad\text{with}\quad
	\bbra\hl_n|\hr_{n'}\kket=\delta_{n,n'},\quad
	\lambda_1=1,\quad\text{and}\quad \hl_1 = \id_{\chi^6}
\end{equation}
such that $\hr_1$ is its unique steady state. The second largest eigenvalue is
\begin{equation}\label{eq:1dMERAbin-eta}\textstyle
	\eta_\bin:=\lambda_2=\frac{\chi^2(1+\chi)^4}{2(1+\chi^2)^4}=\frac{1}{2\chi^2}+\frac{2}{\chi^3}+\mc{O}(\chi^{-4}).
\end{equation}

For large $\tau$, the leading term in Eq.~\eqref{eq:1dMERAbin-yy} is $2^{\tau-1}\bbra\hh\otimes\hh|\hr_1\kket\,\hl_1=2^{\tau-1}\bbra\hh\otimes\hh|\hr_1\kket\,\id_{\chi^6}$. But it does not contribute to the gradient variance \eqref{eq:1dMERAbin-gradVarProdM}, because, in the sum $\sum_{i\in \S'_q}\Avg\hY_{k,i}\otimes\hY_{k,i}$, it corresponds to a term $\propto\id_{\chi^8}$ for $q=2k-2,2k+1$ and a term  $\propto\id_{\chi^6}$ for $q=2k-1,2k$. These do not contribute to $\sum_{i\in \S'_q}\Avg\Tr(\hg_{k,i}^\dag\hg_{k,i})$ as explained below Eq.~\eqref{eq:gradVarProdM}. Hence, the leading contributing term in Eq.~\eqref{eq:1dMERAbin-yy} is
\begin{equation}
	(2\eta_\bin)^{\tau-1}\bbra\hh\otimes\hh|\hr_2\kket\,\hl_2.
\end{equation}

The second important component of $\Avg\hZ_{k,i}\otimes\hZ_{k,i}$ for the diagonal contributions in Eq.~\eqref{eq:1dMERAbin-gradVarProdM} is 
\begin{align}\nonumber
	\Avg\hx_{i_\tau}\otimes\hx_{i_\tau}
	&\stackrel{\eqref{eq:1dMERAbin-expct-b}}{=}
	  \Avg\M^{\otimes 2}_{\tau+1,i_{\tau}}\circ\dotsb\circ\M^{\otimes 2}_{T,i_{T-1}} (\dm^{(T)}\otimes\dm^{(T)})\\
	\label{eq:1dMERAbin-xx}
	&\stackrel{\eqref{eq:1dMERAbin-Edoub-b}}{=}
	  \E^\doub_{\bin,i_{\tau}}\circ\dotsb\circ \E^\doub_{\bin,i_{T-1}}(\dm^{(T)}\otimes\dm^{(T)}).
\end{align}
This is a density operator on $(\CC^\chi)^{\otimes 6}$ and identical for all $i$ within any of the four sets $\S'_q$.
Thus, we have established that the diagonal contributions to the gradient variance \eqref{eq:1dMERAbin-gradVarProdM} decay as
\begin{equation}\label{eq:1dMERAbin-gradVarDiag}
	\frac{1}{\chi^2}\sum_{i\in \S_k}\Avg \Tr(\hg^\dag_{k,i}\hg_{k,i}^\pdag)=\mc{O}\big((2\eta_\bin)^\tau\big).
\end{equation}
While this is an upper bound, we would also like to have a lower bound to exclude the occurrence of barren plateaus. Part (e) of the proof discusses a lower bound for the gradient variance averaged over all disentanglers $\hU_{\tau,k}$ in layer $\tau$, finding the same scaling as in Eq.~\eqref{eq:1dMERAbin-gradVarDiag}.\\[0.4em]
\textbf{(d)}
The spectrum of the doubled layer-transition channel \eqref{eq:1dMERAbin-Edoub-a} can be determined by obtaining its representations in a suitable operator basis and diagonalizing it.
For the right-moving layer-transition maps $\M_{\tau,i_{\tau-1}}$ with odd $i_{\tau-1}$ in Eq.~\eqref{eq:1dMERAbin-transitionMap},
as shown in Fig.~\ref{fig:MERAbin-layerTransit}b,
we start on the three neighboring sites $c_1,c_2,c_3\in\L_\tau$ of the causal cone, add three auxiliary sites $a_1,a_2,a_3$ initialized in state $|0_\chi\ket$, apply two-site unitaries on site groups $(c_1,a_1)$, $(c_2,a_2)$, and $(c_3,a_3)$ to implement the isometries \eqref{eq:1dMERAbin-isometry}, then apply unitaries (the disentanglers) on site groups $(a_1,c_2)$ and $(a_2,c_3)$, and finally trace out sites $c_1,a_1$, and $a_3$. We thus obtain a state on the three sites $c_2,a_2,c_3\in\L_{\tau-1}$.
The left-moving layer-transition maps $\M_{\tau,i_{\tau-1}}$ with even $i_{\tau-1}$ only differ in the final step, where we trace out sites $c_1,c_3$, and $a_3$.
For the right-moving and left-moving layer-transition channels $\E^\doub_{\bin,\tR}$ and $\E^\doub_{\bin,\tL}$, we apply $\M_{\tau,i_{\tau-1}}$ on two copies of the (three-site) system and take the Haar average over the five different unitaries. This averaging is equivalent to applying the doubled fully depolarizing channel $\G^\doub$ on each of the corresponding site groups. As seen in Eq.~\eqref{eq:G2dyadic}, the kernel and co-kernel of $\G^\doub$ are the orthogonal complement of the projection operators $\hP_\pm$. In the transition channels $\E^\doub_{\bin,\tR}$ and $\E^\doub_{\bin,\tL}$, we act on every (doubled) initial site and auxiliary site at least once with $\G^\doub$. We can hence express them and $\E^\doub_{\bin}$ using, for every doubled site, the biorthogonal left and right operator bases
\begin{equation}\label{eq:hP-basis}\textstyle
	\B_L=\{\bbra\hP_+|,\bbra\hP_-|\},\ \
	\B_R=\{|\hP'_+\kket,|\hP'_-\kket\}\quad \text{with}\quad
	\hP_\pm\stackrel{\eqref{eq:hP}}{=}\frac{1}{2}\left(\id_{\chi^2}\pm\Swap\right),\ 
	\hP'_\pm\stackrel{\eqref{eq:G2dyadic}}{=}\frac{1}{\Tr\hP_\pm}\hP_\pm
\end{equation}
such that $\bbra\hP_\pm|\hP'_\pm\kket=1$ and $\bbra\hP_\pm|\hP'_\mp\kket=0$. With this two-dimensional operator space for every doubled site, we obtain the $8\times8$ matrix representations for the doubled channels given in Appendix~\ref{appx:1dMERAbin-Edoub-matrix-diag} with matrix elements being functions of $\chi$. For $\E^\doub_{\bin,\tR}$ and $\E^\doub_{\bin,\tL}$, we find the spectrum
\begin{equation}\label{eq:1dMERAbin-EdoubR-spec}
	1,\quad
	\frac{\chi^2(1+\chi)^2}{(1+\chi^2)^3},\quad
	\frac{\chi^3(1+\chi)^2}{(1+\chi^2)^4},\quad
	\frac{\chi^3}{(1+\chi^2)^3},\quad
	0,\ \ \ 0,\ \ \ 0,\ \ \ 0.
\end{equation}
The channel \eqref{eq:1dMERAbin-Edoub-a} that describes the spatial average has the spectrum
\begin{equation}\label{eq:1dMERAbin-Edoub-spec}
	\lambda_1=1,\quad
	\lambda_2=\frac{\chi^2(1+\chi)^4}{2(1+\chi^2)^4},\quad
	\lambda_3=\frac{\chi^2(1+\chi)^2}{2(1+\chi^2)^3},\quad
	\lambda_4=\frac{\chi^3}{(1+\chi^2)^3},\quad
	0,\ \ \ 0,\ \ \ 0,\ \ \ 0.
\end{equation}
So, the second largest eigenvalue is $\eta_\bin$ as given in Eq.~\eqref{eq:1dMERAbin-eta}.\\[0.4em]
\textbf{(e)}
In part (c), we found the upper bound \eqref{eq:1dMERAbin-gradVarDiag} for the diagonal contributions to the gradient variance \eqref{eq:1dMERAbin-gradVarProdM}. To exclude the occurrence of barren plateaus, we now derive a lower bound for the gradient variance averaged over all disentanglers $\hU_{\tau,k}$ in layer $\tau$. Taking this spatial average corresponds to averaging over all trajectories $i_\tau$ in layers $\tau+1,\dotsc,T$ such that Eq.~\eqref{eq:1dMERAbin-xx} is replaced by
\begin{align}\nonumber
	\frac{1}{2^{T'-\tau}}\sum_{i_\tau\in\L_\tau}\Avg\hx_{i_\tau}\otimes\hx_{i_\tau}
	&\stackrel{\eqref{eq:1dMERAbin-xx}}{=}
	  \frac{1}{2^{T'-\tau}}\sum_{i_\tau\in\L_\tau}
	  \E^\doub_{\bin,i_{\tau}}\circ\dotsb\circ \E^\doub_{\bin,i_{T-1}}(\dm^{(T)}\otimes\dm^{(T)})\\
	\label{eq:1dMERAbin-xx2}
	&\stackrel{\eqref{eq:1dMERAbin-Edoub-a}}{=}
	  (\E^\doub_{\bin})^{T-\tau}(\dm^{(T)}\otimes\dm^{(T)}).
\end{align}
As the doubled layer-transition channel $\E^\doub_{\bin}$ is strictly contractive, for large $T-\tau$, the leading term in Eq.~\eqref{eq:1dMERAbin-xx2} is
\begin{equation}
	\bbra \hl_1|\dm^{(T)}\otimes\dm^{(T)}\kket\, \hr_1+\mc{O}(\eta_\bin^{T-\tau})
	\stackrel{\eqref{eq:1dMERAbin-Edoub-diag}}{=}
	\hr_1+\mc{O}(\eta_\bin^{T-\tau}).
\end{equation}
Explicit expressions for the steady state $\hr_1$ of $\E^\doub_{\bin}$, $\hl_2$, and $\hr_2$ can be found by the diagonalization of the $8\times 8$ matrix representation of $\E^\doub_{\bin}$ discussed in Appendix~\ref{appx:1dMERAbin-Edoub-matrix-diag}.

These allow us to evaluate the diagonal contributions to the gradient variance \eqref{eq:1dMERAbin-gradVarProdM}, averaged over all disentanglers $\hU_{\tau,k}$ in layer $\tau$
\begin{align}\nonumber
	&\frac{1}{2^{T'-\tau}}\sum_{k\in\L_\tau}\frac{1}{\chi^2}\sum_{i\in \S_k}\Avg \Tr(\hg^\dag_{k,i}\hg_{k,i}^\pdag)\\\nonumber
	&=\frac{1}{2^{T'-\tau}}\sum_{k\in\L_\tau}\frac{2}{\chi^2}\Avg\Big[\sum_{i\in \S'_{2k-2}}\Tr(\hg^\dag_{k,i}\hg_{k,i}^\pdag)+\sum_{i\in \S'_{2k-1}}\Tr(\hg^\dag_{k,i}\hg_{k,i}^\pdag)\Big]\\\nonumber
	&\textstyle\nonumber
	 =\frac{4\bbra\hh\otimes\hh|\hr_2\kket}{\chi^4-1}\,(2\eta_\bin)^{\tau-1}
	  \Big(\bbra\hl_2\otimes\id_{\chi^2}|(\Id_{a_1,c_2}\otimes\Q_{a_2,c_3})\tilde{\E}_{\bin,\tL}|\hr_1\kket
	      +\bbra\hl_2|(\Id_{c_2}\otimes\Q_{a_2,c_3})\tilde{\E}_{\bin,\tR}|\hr_1\kket\Big)\\
	\label{eq:1dMERAbin-gradVarDiagSpaceAvg}
	   &\quad +\mc{O}\big((2\lambda_3)^\tau\big)+\mc{O}(2^\tau\eta_\bin^T),
\end{align}
where the action of
\begin{equation}\label{eq:Q}\textstyle
	\Q:=|\Swap-\frac{1}{\chi^2}\id_{\chi^4}\kket\bbra\Swap-\frac{1}{\chi^2}\id_{\chi^4}|
\end{equation}
is equivalent to the two corresponding factors in Eq.~\eqref{eq:1dMERAbin-gradVarProdM-b}. The indices to $\Q$ and $\Id$ in Eq.~\eqref{eq:1dMERAbin-gradVarDiagSpaceAvg} indicate on which of the four doubled sites ($a_1,c_2,a_2,c_3$) they act in the first summand and on which of the three doubled sites ($c_2,a_2,c_3$) they act in the second summand.
The channel $\tilde{\E}_{\bin,\tR}$ is an adaptation of the right-moving channel $\E_{\bin,\tR}$, where we omit the second disentangler ($\hU_{\tau,k}$) that would have acted on sites $(a_2,c_3)$. The channel $\tilde{\E}_{\bin,\tL}$ is an adaptation of the left-moving channel $\E_{\bin,\tL}$, where we again omit the second disentangler ($\hU_{\tau,k}$) and also omit the trace over site $c_3$.

Plugging in the explicit expressions \eqref{eq:1dMERAbin-Edoub-eigenvec}, Eq.~\eqref{eq:1dMERAbin-gradVarDiagSpaceAvg} evaluates to
\begin{equation}\label{eq:1dMERAbin-gradVarDiagSpaceAvg-eval}
	\frac{4\bbra\hh\otimes\hh|\hr_2\kket}{\chi^4-1}\,(2\eta_\bin)^{\tau-1}
	\Big(\Big[\frac{1}{4}-\frac{1}{4\chi}\Big]+\Big[\frac{\chi}{4}+\frac{1}{8}-\frac{21}{8\chi}\Big]+\mc{O}(\frac{1}{\chi^2})\Big)
	+\mc{O}\big((2\lambda_3)^\tau\big)+\mc{O}(2^\tau\eta_\bin^T),
\end{equation}
where the Hamiltonian term in the prefactor is
\begin{align}\nonumber
	\bbra\hh\otimes\hh|\hr_2\kket\stackrel{\eqref{eq:1dMERAbin-Edoub-eigenvec-r2}}{=}
	\Tr\Big[(\hh\otimes\hh&)\big(\hP'_{+,1}\hP'_{+,2}\hP'_{+,3} - \hP'_{+,1}\hP'_{-,2}\hP'_{+,3}\\
	                          &- \hP'_{-,1}\hP'_{+,2}\hP'_{-,3} + \hP'_{-,1}\hP'_{-,2}\hP'_{-,3}\big)\Big]\textstyle
	+\mc{O}\Big(\frac{1}{\chi}\Big),
\end{align}
and the indices to $\hP'_\pm$ indicate on which of the three doubled sites the operator is acting. See Appendix~\ref{appx:1dMERAbin-diagSpaceAvg}. 
The result \eqref{eq:1dMERAbin-gradVarDiagSpaceAvg-eval} shows that the diagonal contributions to the spatially averaged gradient variance scale as $\Theta\big((2\eta_\bin)^\tau\big)$. The scaling in $\eta_\bin$ agrees with the upper bound \eqref{eq:1dMERAbin-gradVarDiag} for the individual variances.
\\[0.4em]
\begin{figure*}[b]
	\label{fig:MERAbin-shiftedCones}
	\includegraphics[width=0.45\textwidth]{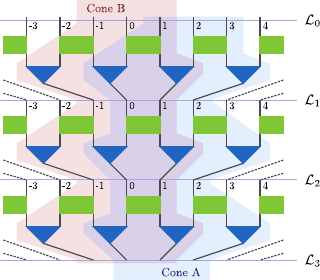}
	\caption{Two three-site causal cones in a binary 1D MERA, starting on sites $i$ (cone A) and $j$ (cone B) with $|i-j|\leq 3$ converge to having at most a shift by one site after two renormalization steps; see Eq.~\eqref{eq:1dMERAbin-coneConverge}. Equation~\eqref{eq:1dMERAbin-coneDist1} specifies in which layer two such cones will converge into one.}
\end{figure*}
\textbf{(f)}
We still need to discuss the off-diagonal contributions to the gradient variance with $i\neq j$ in Eq.~\eqref{eq:1dMERAbin-gradVarProdM}.
Similar to the situation for MPS addressed in Eq.~\eqref{eq:gradExtVarProd}, it can be shown that all off-diagonal terms with $|i-j|>3$ in Eq.~\eqref{eq:1dMERAbin-gradVarProdM} vanish:

$\hZ_i$ is linear in $\hy_i$ as defined in Eq.~\eqref{eq:1dMERAbin-expct-b}. The Haar average of $\hy_{i}\otimes\hy_{j}$ in Eq.~\eqref{eq:1dMERAbin-gradVarProdM} vanishes if the three-site supports of $\hh_{i}$ and $\hh_{j}$ are at least separated by one site, i.e., if $|i-j|>3$. In fact, we just need to look at the term
\begin{equation}
	\Avg \M^\dag_{1,i}(\hh)\otimes \M^\dag_{1,j}(\hh)
\end{equation}
which is the part of $\hy_i\otimes\hy_j$ that comprises the Hamiltonian terms and the relevant MERA tensors of layer $\tau=1$; cf.\ Eq.~\eqref{eq:1dMERAbin-expct}. 
As can be seen in Fig.~\ref{fig:MERAbin}, $\M^\dag_{1,i}$ and $\M^\dag_{1,j}$ have no disentanglers in common. Hence, the Haar average over the disentanglers in $\M^\dag_{1,i}$ follows the first-moment Weingarten formula and leads to application of the fully depolarizing channel $\G^\dag$ with $N=\chi^2$ in Eq.~\eqref{eq:G} on two of the sites in the support $\{i,i+1,i+2\}$ of $\hh_i$ and the application of the MPS channel $\E$ with $N=d=\chi$ in Eq.~\eqref{eq:E} on the third site. The same applies for $\M^\dag_{1,j}$ such that
\begin{equation}\textstyle
	\Avg \M^\dag_{1,i}(\hh)\otimes \M^\dag_{1,j}(\hh)
	=\left(\frac{\id_{\chi^3}}{\chi^3} \Tr\hh_i\right) \otimes \left(\frac{\id_{\chi^3}}{\chi^3} \Tr\hh_j\right) = 0
\end{equation}
as $\Tr\hh=0$. Thus, $\Avg\hy_i\otimes\hy_j=0$ for $|i-j|>3$.\\[0.4em]
\textbf{(g)}
Furthermore, the causal cones for nearby terms $\hh_i$ and $\hh_j$ converge. In particular, Eq.~\eqref{eq:1dMERAbin-cone} implies that
\begin{equation}\label{eq:1dMERAbin-coneConverge}
	|i_1-j_1|\leq 2\quad\text{and}\quad |i_2-j_2|\leq 1
	\quad\text{for all $i\equiv i_0$ and $j\equiv j_0$ with}\quad 1\leq |i-j|\leq 3.
\end{equation}
So, after two renormalization steps, the causal cones for off-diagonal contributions have converged ($|i_2-j_2|=0$) or have a distance of $|i_2-j_2|=1$. For the sum of the contributions with $|i_2-j_2|=0$, the same arguments as for the diagonal terms ($i=j$) apply such that
\begin{equation}\label{eq:1dMERAbin-gradVarOff-0}
	\frac{1}{N}\sum_{i,j\in \S_{\tau,k},|i_2-j_2|=0}\Avg \Tr(\hg^\dag_{k,i}\hg_{k,j}^\pdag)=\mc{O}\big((2\eta_\bin)^\tau\big).
\end{equation}
\textbf{(h)}
We finally, need to assess the contributions with $|i_2-j_2|=1$, choosing without loss of generality $i_2=j_2+1$:
As illustrated in Fig.~\ref{fig:MERAbin-shiftedCones},
we have
\begin{equation}\label{eq:1dMERAbin-coneDist1}
	i_t-j_t=\begin{cases}1&\text{for}\ \ 2\leq t <\tau',\\ 0&\text{for}\ \ t\geq\tau' \end{cases}
	\quad\text{for some $\tau'>2$ \ if and only if}\quad
	i_2 = q\cdot 2^{\tau'-2}
\end{equation}
with some integer $q$. For $2\leq t <\tau'$, we then apply left-moving transition maps $\M_{t,i_{t-1}}$ in the first component of the doubled system and right-moving transition maps $\M_{t,j_{t-1}}$ in the second component. Taking the Haar average over the unitaries in these transition maps, we obtain a corresponding layer-transition channel $\E^\doub_{\bin-1}$ for the doubled system which, as shown in Appendix~\ref{appx:1dMERAbin-Edoub-matrix-offdiag}, is diagonalizable with the two nonzero eigenvalues
\begin{equation}\label{eq:1dMERAbin-EdoubS1-spec}
	1,\quad
	\eta_{\bin-1}:=\frac{\chi^2(1+\chi)^2}{(1+\chi^2)^3}.
\end{equation}
For $t>\tau'$ the two causal cones have merged into one, and further layer transitions progress as for the diagonal contributions. In conclusion, off-diagonal terms  with $i\neq j$ contribute to the gradient variance \eqref{eq:1dMERAbin-gradVarProdM} with
\begin{equation}\label{eq:1dMERAbin-gradVarOff-1}
	\frac{1}{N}\sum_{i\neq j\in \S_{\tau,k}}\Avg \Tr(\hg_{k,i}^\dag\hg_{k,j}^\pdag)
	=\sum_{\tau'=2}^\tau\mc{O}\big(\eta_{\bin-1}^{\tau'-2}(2\eta_\bin)^{\tau-\tau'}\big)=\mc{O}\big((2\eta_\bin)^\tau\big)
\end{equation}
as $\eta_{\bin-1}<2\eta_\bin$ for all $\chi\geq 1$.\\[0.4em]
\textbf{(i)}
In part (e) of this proof, we found that the diagonal contributions to the spatially averaged gradient variance \eqref{eq:1dMERAbin-gradVarDiagSpaceAvg-eval} scale as $\Theta\big((2\eta_\bin)^\tau\big)$. As the total number of disentanglers in layer $\tau$ is proportional to $2^{T-\tau}$ and the individual gradient variances obey the upper bound \eqref{eq:1dMERAbin-gradVarDiag}, this implies that the diagonal contributions to the gradient variance of a finite fraction of disentanglers in layer $\tau$ scales as $\Theta\big((2\eta_\bin)^\tau\big)$. In parts (f-h) of the proof, we demonstrated the upper bound \eqref{eq:1dMERAbin-gradVarOff-1} on the off-diagonal contributions to the gradient variances and one can check explicitly that no cancellations occur. Thus, while gradient variances decay with $\mc{O}\big((2\eta_\bin)^\tau\big)$, they do not decay exponentially in the system size and the 1D binary MERA do not suffer from the barren plateau phenomenon. For the top layer with $\tau=T$, we have $(2\eta_\bin)^{T}= (2\eta_\bin)^{\log_2 L}$ if we have a MERA with $T'=T=\log_2 L$ layers.
\qed

\subsection{Ternary 1D MERA}\label{sec:MERA-ternary}
\begin{figure*}[t]
	\label{fig:MERAter}
	\includegraphics[width=1\textwidth]{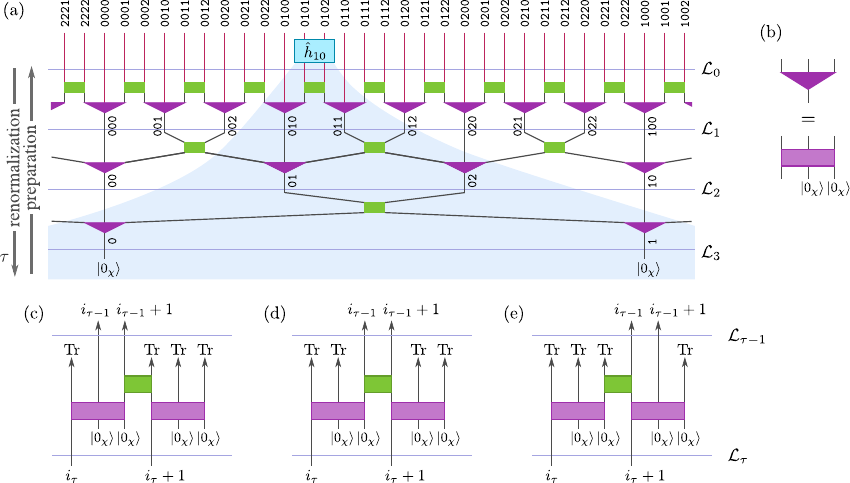}
	\caption{(a) Part of a ternary 1D MERA with $T=3$ layers on $L=81$ sites. The shaded region indicates the causal cones for the two-site operator $\hh_{10}$ acting on sites $\{10,11\}$. The renormalized sites contained in the causal cone are determined by Eq.~\eqref{eq:1dMERAter-cone}. Site and renormalized-sites numbers are shown in ternary representation. (b) The isometries can be parametrized by unitaries that are, on one side, projected onto the reference state $|0_\chi\ket$; see Eq.~\eqref{eq:1dMERAter-isometry}.
	(c)-(e) The lowest trit of the first site $i_{\tau-1}\in\L_{\tau-1}$ in the causal cone determines whether the transition map $\M_{\tau,i_{\tau-1}}$ of layer $\tau$ is left-moving ($i_{\tau-1}\!\mod\! 3=0$), central ($i_{\tau-1}\!\mod\! 3=1$), or right-moving ($i_{\tau-1}\!\mod\! 3=2$).}
\end{figure*}
The analysis for the binary 1D MERA can be extended to various different types of MERA and TTNS. As a concrete second example, consider a 1D lattice of $L=3^{T'}$ sites with periodic boundary conditions and a MERA $|\Psi\ket$ with branching ratio $b=3$ and $T$ layers. The physical lattice and the lattices of renormalized sites are now
\begin{equation}
	\L_0=\{0,\dotsc,L-1=3^{T'}-1\},\ \ 
	\L_1=\{0,\dotsc,3^{T'-1}-1\},\ \dotsc\ ,
	\L_T=\{0,\dotsc,3^{T'-T}-1\}.
\end{equation}
For simplicity, let us choose (a) $T\leq T'-1$ such that layer-transition maps for causal cones have all the same structure and that (b) the dimension $d$ of each physical single-site Hilbert space agrees with the bond dimension $\chi$ of the MERA.

For the first layer $\tau=1$, we apply unitary two-site nearest neighbor disentanglers $\hU^\dag_{\tau,k}\in\groupU(\chi^2)$ on all site groups $(3k+1,3k+2)$ from $\L_{\tau-1}$. Then, we apply isometries $\hW^\dag_{\tau,k}:\CC^{\chi^3}\to\CC^{\chi}$ that map sites $(3k-1,3k,3k+1)$ from $\L_{\tau-1}$ into the renormalized site $k\in\L_\tau$ with $\hW^\dag_{\tau,k}\hW_{\tau,k}=\id_\chi$. Repeating this for the remaining $T-1$ layers, we arrive at the lattice $\L_T$ containing $L/3^T=3^{T'-T}$ renormalized sites and end the procedure by projecting on every site onto an arbitrary reference state $|0_\chi\ket$ from $\CC^\chi$. The isometries $\hW_{\tau,k}$ can be parametrized by unitaries $\hV_{\tau,k}\in \groupU(\chi^3)$ that are, on one side, projected onto the reference state $|0_\chi\ket\otimes|0_\chi\ket$,
\begin{equation}\label{eq:1dMERAter-isometry}
	\hW_{\tau,k}=\hV_{\tau,k}\,\big(\id_{\chi}\otimes |0_\chi\ket\otimes |0_\chi\ket\big).
\end{equation}

Figure~\ref{fig:MERAter} shows a ternary 1D MERA, and the causal cone for two neighboring sites of $\L_0$ is indicated by the shaded region.
It comprises two neighboring (renormalized) sites in each of the lattices $\L_{\tau\geq 1}$.
If we start with sites $\{i_0:=i,i_0+1\}\subset\L_0$ of the physical lattice, after $\tau$ renormalization steps, the causal cone contains only sites
\begin{equation}\label{eq:1dMERAter-cone}
	\{i_\tau,i_\tau+1\}\subset\L_\tau\quad\text{where}\quad
	i_\tau=\lfloor i_{\tau-1}/3\rfloor.
\end{equation}

The cost functions to be optimized are expectation values
\begin{equation}\label{eq:1dMERAter-cost}
	\sum_{i=0}^{L-1}\bra\Psi|\hh_i|\Psi\ket,
\end{equation}
of local extensive Hamiltonians, where the interaction term $\hh_i$ acts non-trivially on sites $\{i,i+1\}$ as specified in Eq.~\eqref{eq:hi-nn}. In analogy to Theorem~\ref{thrm:1dMERAbin} for the binary MERA, we find the following.
\newpage
\begin{theorem}[\propHead{Decay of energy gradients for ternary 1D MERA}]
\label{thrm:1dMERAter}
Consider 1D ternary MERAs $|\Psi\ket$ with bond dimension $\chi$ and $T$ layers on $3^{T'}$ sites with $T'\geq T+1$.
With all disentanglers $\hU_{\tau,k}$ and unitaries $\hV_{\tau,k}$ for the isometries in Eq.~\eqref{eq:1dMERAter-isometry} sampled according to the uniform Haar measure, the average of the Riemannian gradient $\partial_{\hU_{\tau,k}} \sum_i \bra\Psi|\hh_i|\Psi\ket$ for the cost function \eqref{eq:1dMERAter-cost} is zero, and at least a finite fraction of the unitaries in layer $\tau$ has a gradient variance that scales as
\begin{gather}\label{eq:1dMERAter-Var}
	\Var \Big(\partial_{\hU_{\tau,k}} \sum_i \bra\Psi|\hh_i|\Psi\ket\Big) = \Theta\big((3\eta_\ter)^\tau\big) +\mc{O}\big((3\lambda_3)^\tau\big)+\mc{O}(3^\tau\eta_\ter^T)
	\quad\text{with}\\\nonumber
	\eta_\ter =\frac{\chi^2(1+8\chi^2+\chi^4+\sqrt{1+4\chi^2+54\chi^4+4\chi^6+\chi^8})}{6(1+\chi^2+\chi^4)^2},\quad\text{where}\quad
	\lambda_3 =\frac{\chi^2}{3(1+\chi^2+\chi^4)}
\end{gather}
controls small $\tau$ effects. The same applies for the $\hV_{\tau,k}$ gradients.
\end{theorem}\noindent
\Emph{Proof:}
The proof parallels the one for binary 1D MERA. The most significant difference is that we now have left-moving, central, and right-moving layer-transition maps. Correspondingly, one works with a ternary representation of the lattice sites as reflected in Eq.~\eqref{eq:1dMERAter-cone}. In analogy to Eq.~\eqref{eq:1dMERAbin-Edoub-a}, the doubled layer-transition channel for spatial averages now has the form
\begin{equation}\label{eq:1dMERAter-Edoub}
	\E^\doub_\ter:=\frac{1}{3}\Avg\left(\M^{\otimes 2}_{\tau,q-1}+\M^{\otimes 2}_{\tau,q}+\M^{\otimes 2}_{\tau,q+1}\right)
	= \frac{1}{3}\left(\E^\doub_{\bin,\tL}+\E^\doub_{\bin,\tC}+\E^\doub_{\bin,\tR}\right).
\end{equation}
Its $4\times 4$ matrix representation with respect to the operator basis \eqref{eq:hP-basis} is discussed in Appendix~\ref{appx:1dMERAter-Edoub-matrix}. It shows that $\E^\doub_\ter$ is diagonalizable with the spectrum
\begin{equation}\label{eq:1dMERAter-Edoub-spec}\textstyle
	1,\ \
	\frac{\chi^2(1+8\chi^2+\chi^4+\sqrt{1+4\chi^2+54\chi^4+4\chi^6+\chi^8})}{6(1+\chi^2+\chi^4)^2},\ \
	\frac{\chi^2}{3(1+\chi^2+\chi^4)},\ \
	\frac{\chi^2(1+8\chi^2+\chi^4-\sqrt{1+4\chi^2+54\chi^4+4\chi^6+\chi^8})}{6(1+\chi^2+\chi^4)^2}.
\end{equation}
The second and third largest of these eigenvalues are $\eta_\ter$ and $\lambda_3$ as already given in Eq.~\eqref{eq:1dMERAter-Var}. In the evaluation of gradient variances with respect to unitaries $\hU_{\tau,k}$ and $\hV_{\tau,k}$ from layer $\tau$, the eigenvalue 1 determines the leading effect of layers $\tau$ to $T$ with corrections controlled by $\eta_\ter$, and the eigenvalue $\eta_\ter$ determines the leading effect of layers 1 to $\tau$ with corrections controlled by $\lambda_3$. The number of local terms $\hh_i$ that contribute to the gradient variance of the considered tensor in layer $\tau$ scales as $3^\tau$ and hence leads to the factors 3 in the result \eqref{eq:1dMERAter-Var}.
\qed

\subsection{Nonary 2D MERA}\label{sec:MERA-nonary}
As a specific example for two spatial dimensions, we address the nonary 2D MERA introduced in Ref.~\cite{Evenbly2009-79}. In each renormalization step, it maps blocks of $3\times 3$ (renormalized) sites into one site such that the branching ratio is $b=9$. Consider a nonary MERA $|\Psi\ket$ with $T$ layers on a 2D lattice of $L\times L$ sites with $L=3^{T'}$ and periodic boundary conditions. The physical lattice and the lattices of renormalized sites are
\begin{equation}
	\L_0=\{0,\dotsc,L-1=3^{T'}-1\}^{\times 2},\ \ 
	\L_1=\{0,\dotsc,3^{T'-1}-1\}^{\times 2},\ \dotsc\ ,
	\L_T=\{0,\dotsc,3^{T'-T}-1\}^{\times 2}.
\end{equation}
As for the ternary 1D MERA, we choose $T\leq T'-1$ such that layer-transition maps for causal cones have all the same structure, and we choose the dimension $d$ of each physical single-site Hilbert space to agree with the bond dimension $\chi$ of the MERA.

\begin{figure*}[t]
	\label{fig:MERAnon}
	\includegraphics[width=1\textwidth]{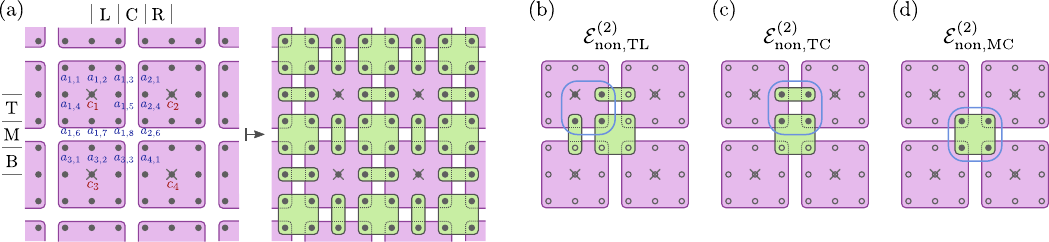}
	\caption{One layer transition in a nonary 2D MERA. (a) Seen in the preparation direction, we start from sites $c_1,c_2,c_3,c_4,\dotsc\in\L_\tau$ (crosses). First, we apply isometries mapping each site $c_i$ into nine sites $c_i,a_{i,1},\dotsc,a_{i,8}$. Then $2\times 2$ disentanglers are applied at the corners of four neighboring $3\times 3$ blocks, and $2\times 1$ disentanglers are applied on the edge centers of two neighboring blocks. The causal cones for operators acting on $2\times 2$ blocks contain $2\times 2$ site blocks in every renormalized lattice $\L_\tau$. There are nine possible $2\times 2$ blocks in $\L_{\tau-1}$ that get mapped to the block $\{c_1,c_2,c_3,c_4\}$ in $\L_\tau$: (b) block $\{c_1,a_{1,5},a_{1,7},a_{1,8}\}$, corresponding to the doubled layer-transition channel $\E^\doub_{\non,\text{TL}}$, (c) block $\{a_{1,5},a_{2,4},a_{1,8},a_{2,6}\}$, corresponding to the doubled layer-transition channel $\E^\doub_{\non,\text{TC}}$, (d) block $\{a_{1,8},a_{2,6},a_{3,3},a_{4,1}\}$, corresponding to the doubled layer-transition channel $\E^\doub_{\non,\text{MC}}$. All further layer-transition channels $\E^\doub_{\non,\text{TR}}$, $\E^\doub_{\non,\text{ML}}$ etc.\ are related to the former by symmetry transformations.}
\end{figure*}

Viewed in the renormalization direction, the layer transitions proceed as follows; see Fig.~\ref{fig:MERAnon}: The sites of lattice $\L_{\tau-1}$ are grouped into $3\times 3$ blocks. At their corners, $2\times 2$ unitary disentanglers are applied, and $2\times 1$ disentanglers are applied on the edge centers. Then, one $3\times 3\to 1$ isometry maps each block into one renormalized site of lattice $\L_\tau$. After $T$ renormalization steps, the procedure ends by projecting every site of lattice $\L_T$ onto an arbitrary reference state $|0_\chi\ket$ from $\CC^\chi$.

The causal cone for an operator $\hh_{i,j}$ acting on a $2\times 2$ block of sites from $\L_0$ comprises $2\times 2$ blocks of (renormalized) sites in each of the lattices $\L_{\tau\geq 1}$.
If we start with sites $\{i_0:=i,i_0+1\}\times \{j_0:=j,j_0+1\} \subset\L_0$ of the physical lattice, after $\tau$ renormalization steps, the causal cone contains only sites
\begin{equation}\label{eq:2dMERAnon-cone}
	\{i_\tau,i_\tau+1\}\times \{j_\tau,j_\tau+1\}\subset\L_\tau\quad\text{where}\quad
	i_\tau=\lfloor i_{\tau-1}/3\rfloor,\quad j_\tau=\lfloor j_{\tau-1}/3\rfloor.
\end{equation}
Hence, there are nine different types of layer-transition maps $\M_{\tau+1,i_\tau,j_\tau}$ and corresponding doubled layer-transition channels $\E^\doub_{\non,\text{TL}},\E^\doub_{\non,\text{TC}},\E^\doub_{\non,\text{TR}},\E^\doub_{\non,\text{ML}},\dotsc,\E^\doub_{\non,\text{BR}}$ where the causal cone continues to the top-left, top-center etc.

The cost functions to be optimized are expectation values
\begin{equation}\label{eq:2dMERAnon-cost}
	\sum_{i,j=0}^{L-1}\bra\Psi|\hh_{i,j}|\Psi\ket,
\end{equation}
of local extensive Hamiltonians.
In analogy to Eqs.~\eqref{eq:1dMERAbin-Edoub-a} and \eqref{eq:1dMERAter-Edoub}, the doubled layer-transition channel for spatial averages now has the form
\begin{equation}\label{eq:2dMERAnon-Edoub}
	\E^\doub_\non:=\frac{1}{9}\left(\E^\doub_{\non,\text{TL}}+\E^\doub_{\non,\text{TC}}+\E^\doub_{\non,\text{TR}}+\E^\doub_{\non,\text{ML}}+\E^\doub_{\non,\text{MC}}+\E^\doub_{\non,\text{MR}}+\E^\doub_{\non,\text{BL}}+\E^\doub_{\non,\text{BC}}+\E^\doub_{\non,\text{BR}}\right).
\end{equation}
Its matrix representation with respect to the operator basis \eqref{eq:hP-basis} is discussed in Appendix~\ref{appx:2dMERAnon-Edoub-matrix}. It shows that $\E^\doub_\non$ is diagonalizable with the four largest eigenvalues being
\begin{equation}\label{eq:2dMERAnon-Edoub-spec}\textstyle
	1,\quad
	\eta_\non=\frac{1}{9\chi^8}+\frac{7}{9\chi^{10}}-\frac{16}{9\chi^{12}}+\mc{O}(\chi^{-14}),\quad
	\lambda_3=\lambda_4=\frac{1}{9\chi^8}+\frac{1}{3\chi^{10}}-\frac{8}{9\chi^{12}}+\mc{O}(\chi^{-14}).
\end{equation}

With all disentanglers $\hU_{\tau,k}$ and unitaries $\hV_{\tau,k}$ for the isometries sampled according to the uniform Haar measure, the average of the Riemannian energy gradient $\partial_{\hU_{\tau,k}} \sum_{i,j} \bra\Psi|\hh_{i,j}|\Psi\ket$ is zero, and a finite fraction of the unitaries in layer $\tau$ has a gradient variance that scales as
\begin{gather}\label{eq:2dMERAnon-Var}
	\Var \Big(\partial_{\hU_{\tau,k}} \sum_{i,j} \bra\Psi|\hh_{i,j}|\Psi\ket\Big)
	= \Theta\big((9\eta_\non)^\tau\big) +\mc{O}\big((9\lambda_3)^\tau\big)+\mc{O}(9^\tau\eta_\non^T).
\end{gather}

\subsection{Further MERA and TTNS}\label{sec:furtherMERA}
The results for binary and ternary 1D MERA as captured by Theorems~\ref{thrm:1dMERAbin} and \ref{thrm:1dMERAter} as well as the nonary 2D MERA discussed in Sec.~\ref{sec:MERA-nonary} can be generalized to all MERA and TTNS. Their expectation values can always be written in the form \eqref{eq:energy} which, according to Eq.~\eqref{eq:grad_avg0}, implies that the Haar-averaged Riemannian gradients vanish. The central objects in the evaluation of the Haar-variance of the gradient are the doubled layer-transition channels like $\E^\doub_\bin$, $\E^\doub_\ter$, and $\E^\doub_\non$ in Eqs.~\eqref{eq:1dMERAbin-Edoub-a}, \eqref{eq:1dMERAter-Edoub}, and \eqref{eq:2dMERAnon-Edoub}. These are generally gapped channels with the unique amplitude-one eigenvalue 1. Let us call the eigenvalue with the second-largest amplitude $\eta$ and assume that it is nondegenerate and that there are no further eigenvalues of amplitude $\eta$. According to the derivation of Theorem~\ref{thrm:1dMERAbin}, the gradient variance for tensors in layer $\tau$ will then scale as $~(b\eta)^\tau$, where $b$ is the branching ratio of the MERA or TTNS. As the number of layers is bounded by $\log_b L$ with respect to the system size $L$, this implies that the optimization of such TNS is not hampered by barren plateaus. The eigenvalue $\eta$ decreases with increasing bond dimension $\chi$ such that $b\eta<1$ at least for sufficiently large $\chi$. For the three specific MERA analyzed above, we in fact have $b\eta<1$ for $\chi>1$.

As three concrete examples for TTNS, the spectra of the doubled layer-transition channels for the binary and ternary 1D TTNS as well as the nonary 2D TTNS are determined in Appendices~\ref{appx:1dTTNSbin-Edoub-matrix}, \ref{appx:1dTTNSter-Edoub-matrix}, and \ref{appx:2dMERAnon-Edoub-matrix}, finding the second largest eigenvalues
\begin{equation}
	\eta'_\bin=\frac{\chi}{1+\chi^2},\quad
	\eta'_\ter=\frac{\chi^2}{1+\chi^2+\chi^4},\quad\text{and}\quad
	\eta'_\non=\frac{\chi^8}{1+\chi^2(1+\chi^2)(1+\chi^4)(1+\chi^8)},
\end{equation}
respectively. Not surprisingly, these agree with the second largest eigenvalue $\eta={d(N^2-1)}/{(d^2N^2-1)}$ [Eq.~\eqref{eq:E2eigen}] of the doubled MPS channel $\E^\doub$ with $N=\chi$, when setting $d=\chi$ for the binary 1D TTNS, $d=\chi^2$ for the ternary 1D TTNS, and $d=\chi^8$ for the nonary 2D TTNS. This coincidence arises as, at least for specific sites $i$, local interaction terms $\hh_i$ get mapped to operators on a single renormalized site, and subsequent layer transitions then simply consist in applying $\E^{\doub\dag}$ multiple times.

\section{Discussion}\label{sec:Discuss}
The Hamiltonians for quantum many-body systems such as condensed matter systems are extensive and, while long-range interactions may exist, they are usually irrelevant for the long-range physics in the sense of the renormalization group \cite{Wilson1975,Wegner1972-5,Salmhofer1999}. High-dimensional optimization problems are often hampered by vanishing gradients of the cost functions and variational quantum algorithms can feature barren plateaus \cite{McClean2018-9}, where average gradient amplitudes decay exponentially in the system size.

We have found that the energy optimization problem for MPS, TTNS, and MERA with respect to extensive finite-range Hamiltonians $\hH=\sum_i\hh_i$ does not feature barren plateaus. We have formulated the results for general TNS tensors which, in a suitable gauge, are all (partial) isometries \cite{Note3} or unitaries. In averages over these tensors, we employed the uniform Haar measure. However, all results carry over to more constrained sets of TNS tensors as long as they are (approximate) 2-designs \cite{Dankert2009-80,Brandao2016-346,Harrow2023-05}.
The Clifford group forms a unitary 3-design \cite{Webb2016-16,Dankert2009-80} and, for $n=1,2,3$ qubits, it is of order $|\mc{C}_n|=24,\numprint{11520}$, and $\numprint{92897280}$, respectively. We have used it to check most of the presented analytical results for single-site Hilbert space dimension $d=2$ and bond dimensions $m=\chi=2$.

For heterogeneous MPS \cite{Baxter1968-9,Fannes1992-144,Schollwoeck2011-326}, the average energy-gradient amplitude scales for large bond dimensions $m$ as $\propto 1/(m^2 d^4)$, independent of the system size [Theorem~\ref{thrm:MPSnn}]. This allows us to initialize the optimization with random MPS. It is however advisable, to start with a small bond dimension and to then gradually increase it. Especially, for the computationally more demanding applications like strongly-correlated 2D systems such procedures are indeed considered best practice. See, for example, Ref.~\cite{Yan2011-332}. Note that translation invariance, assumed for the extensive Hamiltonians in Theorems~\ref{thrm:MPSext-bond} and \ref{thrm:MPSnn}, is not essential; the extension to heterogeneous systems is straightforward.

For heterogeneous TTNS \cite{Otsuka1996-53,Shi2006-74,Murg2010-82} and MERA \cite{Vidal-2005-12,Vidal2006} and extensive Hamiltonians, the average energy-gradient amplitude with respect to a tensor in layer $\tau$ scales as $\sim (b\eta)^\tau$ [Theorem~\ref{thrm:1dMERAbin}, Theorem~\ref{thrm:1dMERAter}, and Sec.~\ref{sec:furtherMERA}]. Here $b$ is the branching ratio of the TTNS or MERA and $\eta$ is the second largest eigenvalue-amplitude of a doubled layer transition channel. This eigenvalue is a decreasing algebraic function of the bond dimension $\chi$. For all considered MERA, we find $b\eta<1$ for all $\chi>1$ and, for all MERA and TTNS, $b\eta=\mc{O}\big(\chi^{-\gamma}\big)$ with some $\gamma>1$. It is an interesting question for future work, to establish such an upper bound on $b\eta$ for general TTNS and MERA. The scaling with respect to $\tau$ suggests to start the energy minimization by mostly optimizing tensors in the first few layers of the TNS (small $\tau$). When these start to converge, the energy gets sensitive to the longer-range correlations encoded by the tensors in higher-$\tau$ layers. For fast convergence, it may even be advisable to gradually increase the number of layers during the optimization. As for MPS, one can also start the optimization of TTNS and MERA with small bond dimensions $\chi$ and then gradually increase them. In any event, the number of layers $T$ is at most logarithmic in the system size, $T\leq \log_b L$, such that the optimization is not hampered by barren plateaus.

In the companion paper \cite{Miao2024-109}, we confirm the presented analytical results in numerical simulations for specific models, and extend them by also covering homogeneous TTNS and MERA, as well as Trotterized MERA \cite{Miao2021_08,Miao2023_03,Kim2017_11,Haghshenas2022-12,Haghshenas2023_05} for which the tensors are chosen as brickwall circuits to allow for an efficient optimization on quantum computers. The paper enlarges further on efficient initialization schemes. Finally, note that, in the optimization of quantum circuits, a decay of gradients goes hand in hand with a decay of cost-function variations \cite{Arrasmith2022-7,Miao2024-9}. In particular, the single-gate variance of the cost function is exactly half the variance of the Riemannian single-gate gradient, and the total variances of the cost function as well as its gradient are both bounded from above by the sum of single-gate variances and, conversely, bound single-gate variances from above \cite{Miao2024-9}. Hence, our results for the scaling of energy-gradient amplitudes also yield bounds on energy variances.

The term ``isometric TNS'' has also been used for an isometric subclass of PEPS \cite{Zaletel2020-124}. While the practical use of this class of states is still somewhat unclear, the techniques used here to determine energy-gradient variances also work for isometric PEPS, and the application would be a nice topic for further work.

\begin{acknowledgments}
We gratefully acknowledge discussions with Daniel Stilck Fran\c{c}a, Jin-Guo Liu, and Iman Marvian as well as support through US Department of Energy grant DE-SC0019449.
\end{acknowledgments}

\appendix
\renewcommand{\thesection}{\Alph{section}}
\renewcommand{\thesubsection}{\thesection.\arabic{subsection}}

\newpage
\section{Proof of Theorem~\ref{thrm:MPSnn}}\label{appx:MPSnn}\vspace{-0.5em}
For the proof of Eqs.~\eqref{eq:varMPSnnAvg}-\eqref{eq:varMPSnnVar-b}, let us consider a single nearest-neighbor interaction term \eqref{eq:hi-nn} and the Riemannian gradient $\partial_{\hU_j} \bra\Psi|\hh_i|\Psi\ket$.\\
\textbf{(a)} For $j<i$, the isometry condition \eqref{eq:MPSisoConstr} makes the expectation value $\bra\Psi|\hh_i|\Psi\ket$ independent of $\hU_j$. Hence, the average gradient and its variance are zero for $j<i$ [Eqs.~\eqref{eq:varMPSnnAvg} and \eqref{eq:varMPSnnVar-a}].\\[0.4em]
\textbf{(b)} Next, consider the case $j>i+1$. The MPS expectation value can be written in the form \eqref{eq:energy} with $M=1$, i.e., $\bra\Psi|\hh_i|\Psi\ket=\Tr_{m\times d}(\hX \hU_j^\dag\hY\hU_j)$ which implies that the average Riemannian gradient with respect to $\hU_j$ is zero [Eq.~\eqref{eq:varMPSnnAvg}]. The expressions for $\hX$ and $\hY$ are analogous to Eq.~\eqref{eq:MPS_XYdecay} and the proof proceeds very similarly to that of Theorem~\ref{thrm:MPS-dist}. Equation~\eqref{eq:MPS_TrX2part} applies unchanged. We now have
\begin{equation}\label{eq:MPSnn_Y}\textstyle
	\hY:=\F^\dag_{j-1}\circ\dotsb\circ\F^\dag_{i+2}(\hL)\otimes\id_d,\quad\text{with}\quad
	\hL=\sum_{s,t,s',t'}\hA_{i+1}^{t'\dag}\hA_i^{s'\dag}\hA^{s}_i\hA^{t}_{i+1}\bra s',t'|\hh|s,t\ket.
\end{equation}
In the evaluation of $\Tr\hY^2-\frac{1}{md}\Tr^2\hY$, Eq.~\eqref{eq:MPS_AvgLL} changes to
\begin{align}\textstyle\nonumber
	\Avg\hL\otimes\hL
	&\stackrel{\eqref{eq:MPSnn_Y}}{=}\textstyle
	  \bra 0_d|^{\otimes 4}\G_{\tb,2}^\doub\circ\G_{\tb,1}^\doub \big([\id_m\otimes\hh]\otimes[\id_m\otimes\hh]\big)|0_d\ket^{\otimes 4}\\\nonumber
	&\,\textstyle\stackrel{\eqref{eq:G2}}{=}
	  \frac{1}{d^2\,(m^2d^2-1)^2}
	   \Big(\big[(m^2d-1)\Tr\hh^2+m^2(d-1)\Tr(\Tr^2_1\hh)\big]\,(md\Swap-\id_{m^2})\\
	\label{eq:MPS_AvgLLb2}
	&\textstyle\hspace{15ex}
	        +m(d-1)\Tr(\Tr^2_2\hh)\,(md\id_{m^2}-\Swap) \Big).
\end{align}
The indices ``$\tb,1$'' and ``$\tb,2$'' to the doubled fully depolarizing channel $\G^{(2)}$ indicate that it acts on the bond vector space and the first or second site of the support of $\hh$, respectively. Note that we recover Eq.~\eqref{eq:MPS_AvgLL} if we set $\hh=\id_d\otimes\hh'$ with a single-site term $\hh'$ in Eq.~\eqref{eq:MPS_AvgLLb2}.
Using Eq.~\eqref{eq:MPS_TrY2explain} and $\bbra\hr_2|md\id_{m^2}-\Swap\kket=-2$, we find
\begin{align}\textstyle\nonumber
	\Avg\big(\Tr\hY^2-\frac{1}{md}\Tr^2\hY\big)
	&\textstyle =
	  \Tr\left((\Swap-\frac{1}{md}\id_{(md)^2})\,\Avg\hY\otimes \hY\right)\\
	&\textstyle =
	  \eta^{j-i-1}\,\frac{m(m^2d-1)}{d(m^2 d^2-1)}\left[\Tr\hh^2+\frac{d-1}{d}\Tr(\Tr^2_1\hh)\right].
	\label{eq:MPS_TrY2part2}
\end{align}
Equations~\eqref{eq:MPS_TrX2part} and \eqref{eq:MPS_TrY2part2} in conjunction with Eq.~\eqref{eq:gradVarProd} conclude the proof of Eq.~\eqref{eq:varMPSnnVar-b} for $j>i+1$.\\[0.4em]
\begin{figure*}[t]
	\label{fig:MPSnn-diag}
	\includegraphics[width=\textwidth]{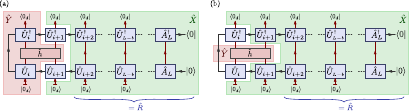}
	\caption{The MPS expectation value $\bra\Psi|\hh_i|\Psi\ket$ for a nearest-neighbor interaction term \eqref{eq:hi-nn} is independent of $\{\hA_1,\dotsc,\hA_{i-1}\}$.
	(a) For $j=i+1$, it can be written in the form $\bra\Psi|\hh_i|\Psi\ket=\Tr_{m\times d}(\hX \hU_{i+1}^\dag\hY\hU_{i+1})$ [Eq.~\eqref{eq:energy} and Fig.~\ref{fig:Riem_grad}a] with $N=md$ and $M=1$.
	(b) For $j=i$, we bring it into the form $\bra\Psi|\hh_i|\Psi\ket=\Tr_{m\times d\times d}\big(\hX [\hU_{i}^\dag\otimes\id_d]\hY[\hU_{i}\otimes\id_d]\big)$, corresponding to $N=md$ and $M=d$ in Eq.~\eqref{eq:energy}.}
\end{figure*}
\textbf{(c)}
For the case $j=i+1$, $\hX$ is as before and $\hY=(\hA^\dag_i\otimes\id_d)[\id_m\otimes\hh](\hA_i\otimes\id_d)$ as shown in Fig.~\ref{fig:MPSnn-diag}a. The evaluation of the gradient variance proceeds very similarly as for the case $j>i+1$, and shows that Eq.~\eqref{eq:varMPSnnVar-b} also holds for $j=i+1$.\\[0.4em]
\textbf{(d)}
For the case $j=i$, the MPS expectation value can be written in the form \eqref{eq:energy} with $M=d$, i.e.,
\begin{equation}
	\bra\Psi|\hh_i|\Psi\ket=\Tr_{m\times d\times d}\left(\hX (\hU_i^\dag\otimes\id_d)\hY(\hU_i\otimes\id_d)\right)
\end{equation}
which implies that the average Riemannian gradient with respect to $\hU_i$ is zero [Eq.~\eqref{eq:varMPSnnAvg}]. The operators $\hX,\hY\in\End(\CC^m\otimes\CC^d\otimes\CC^d)$, $\hx\in\End(\CC^m\otimes\CC^d)$ and $\hR\in\End(\CC^m)$ as indicated in Fig.~\ref{fig:MPSnn-diag}b are
\begin{subequations}\label{eq:MPSnni_YXR}
\begin{gather}\label{eq:MPSnni_YXRa}
	\hY=\id_m\otimes\hh\quad\text{and}\quad
	\hX=\hx_{\tb,2}\otimes|0_d\ket\bra 0_d|_1,\\
	\ \text{with}\quad\hx=\hU_{i+1}\big[\hR\otimes |0_d\ket\bra 0_d|\big] \hU_{i+1}^\dag\ \ \text{and}\ \
	\hR=\F_{i+2}\circ\dotsb\circ\F_{L}(|0\ket\bra 0|),
\end{gather}
\end{subequations}
where the indices ``$\tb,2$'' to $\hx$ in Eq.~\eqref{eq:MPSnni_YXRa} indicate that $\hx$ acts on the first and third components of the tensor product $\CC^m\otimes\CC^d\otimes\CC^d$, and the index ``1'' to $|0_d\ket\bra 0_d|$ indicates that it acts on the second component of the tensor product space, such that $\bra a,s,t|\hX|a',s',t'\ket=\bra a,t|\hx|a',t'\ket\bra s|0_d\ket\bra 0_d|s'\ket$.
As discussed in part (d) of the proof of Theorem~\ref{thrm:MPS-dist}, $\Avg\hR\otimes\hR = \hr_1 + \mc{O}(\eta^{L-i})$ [cf.\ Eq.~\eqref{eq:E2_steadyState}] and, hence,
\begin{equation}\label{eq:MPSnni_xx}\textstyle
	\Avg\hx\otimes\hx \stackrel{\eqref{eq:MPSnni_YXR},\eqref{eq:G2}}{=}
	 \frac{1}{md(m^2d+1)}\left(m\id_{(md)^2}+\Swap\right) + \mc{O}(\eta^{L-i}) .
\end{equation}
Plugging the resulting $\Avg\hX\otimes\hX$ and $\hY\otimes\hY$ from Eq.~\eqref{eq:MPSnni_YXRa} into Eq.~\eqref{eq:gradVarProdM}, we obtain the gradient variance \eqref{eq:varMPSnnVar-c}.\\[0.4em]
\textbf{(e)}
For the proof of Eq.~\eqref{eq:varMPSnnExtVar}, we now consider the Riemannian gradient for the MPS expectation value
\begin{equation}\label{eq:costGlobalnn-XY}\textstyle
	\sum_{i=1}^{L-1}\bra\Psi|\hh_i|\Psi\ket
	 = \sum_{i=1}^j\Tr\left(\hX_i(\hU_j^\dag\otimes\id_{M_i})\hY_i(\hU_j\otimes\id_{M_i})\right) + \sum_{i=j+1}^{L-1}\bra\Psi|\hh_i|\Psi\ket
\end{equation}
of an extensive Hamiltonian with two-site interactions \eqref{eq:hi-nn}. This expression is similar to Eq.~\eqref{eq:costGlobal-XY} from the discussion of single-site terms $\hh_i$.
Due to the left-orthonormality condition \eqref{eq:MPSisoConstr}, the second sum in Eq.~\eqref{eq:costGlobalnn-XY} is independent of $\hU_j$ and does not contribute to the gradient. According to Eq.~\eqref{eq:grad_avg0}, the form \eqref{eq:costGlobalnn-XY} implies that the average Riemannian gradient is zero [Eq.~\eqref{eq:varMPSnnAvg}].
For the gradient variance, we will only discuss contributions from terms with $i>b$. Generally, contributions decay exponentially in $|i-j|$, and we simply capture boundary effects with the terms $\mc{O}(\eta^j)+\mc{O}(\eta^{L-j})$ in Eq.~\eqref{eq:varMPSnnExtVar}. 
The gradient [Eq.~\eqref{eq:Riem_grad}] now takes the form
\begin{equation}\label{eq:Riem_gradExtM}\textstyle
	\partial_{\hU_j} \sum_{i=1}^{L-1}\bra\Psi|\hh_i|\Psi\ket
	\stackrel{\eqref{eq:costGlobal-XY}}{=} \sum_{i=1}^j\hg_i\quad\text{with}\quad
	\hg_i\stackrel{\eqref{eq:Riem_grad}}{=} \Tr_{M_i}\big(\hY_i\tU_j\hX_i - \tU_j\hX_i\tU_j^\dag\hY_i\tU_j\big),
\end{equation}
where $\tU_j:=\hU_j\otimes\id_{M_i}$.
In generalization of Eq.~\eqref{eq:gradVarProdM}, we find that its variance \eqref{eq:gradVar} is
\begin{subequations}\label{eq:gradExtVarProdM}
\begin{gather}\textstyle
	\Var \partial_{\hU_j} \sum_i\bra\Psi|\hh_i|\Psi\ket
	=\frac{1}{N}\sum_{i_1,i_2=1}^j\Avg \Tr(\hg^\dag_{i_1}\hg_{i_2})\quad\text{with}\\
	\label{eq:gradExtVarProdM-b}\textstyle
	\frac{1}{N}\Avg \Tr(\hg^\dag_{i_1}\hg_{i_2})=\frac{2}{N^2-1}
	  \Tr\left(\big[\Swap_{1,3}-\frac{1}{N}\id_{N^4}\big]\,
	           \big[\Swap_{2,4}-\frac{1}{N}\id_{N^4}\big]\,\Avg\,[\hZ_{i_1}\otimes\hZ_{i_2}]\right)
\end{gather}
with $N=md$ and $\hZ_i\in\End(\CC^N\otimes\CC^N)$ as defined in Eq.~\eqref{eq:gradVarProdM-Z}. A diagrammatic representation for these terms is shown in Fig.~\ref{fig:Riem_grad}c. Summands with $M_{i_1}=M_{i_2}=1$ simplify to the form given in Eq.~\eqref{eq:gradExtVarProd}, i.e.,
\begin{align}\nonumber\textstyle
	\frac{1}{N}\Avg \Tr(\hg^\dag_{i_1}\hg_{i_2})
	&\textstyle=\frac{2}{N^2-1}
	   \Avg\Tr\left(\big(\hX_{i_1}-\frac{\id_N}{N}\Tr\hX_{i_1}\big)\big(\hX_{i_2}-\frac{\id_N}{N}\Tr\hX_{i_2}\big)\right)\\
	&\label{eq:gradExtVarProdM-c} \textstyle\hspace{5.9ex}\times
	   \Avg\Tr\left(\,\big(\hY_{i_1}-\frac{\id_N}{N}\Tr\hY_{i_1}\big)\,\,\big(\hY_{i_2}-\frac{\id_N}{N}\Tr\hY_{i_2}\big)\,\right)
\end{align}
if $M_{i_1}=M_{i_2}=1$.
\end{subequations}
\newpage\noindent
\textbf{(f)}
The off-diagonal terms with $|i_1-i_2|\geq 2$ in Eq.~\eqref{eq:gradExtVarProdM} vanish. In particular, consider the case $i_1\geq i_2+2$. $\hY_{i_2}$ is then given by Eq.~\eqref{eq:MPSnn_Y} and, due to the left-orthonormality condition \eqref{eq:MPSisoConstr}, $\hY_{i_1}$ is independent of $\hU_{i_2}$ and $\hU_{i_2+1}$. Consequently, $\Avg\hZ_{i_1}\otimes \hZ_{i_2}=0$ as the term $\G_{\tb,2}\circ\G_{\tb,1}(\id_m\otimes\hh)\propto\Tr\hh=0$ from $\hY_{i_2}$ is zero.
In the following, we will first address the diagonal terms $i_1=i_2$ and then the terms with $|i_1-i_2|=1$\\[0.4em]
\textbf{(g)}
For the diagonal contributions ($i_1=i_2$) to the gradient variance \eqref{eq:gradExtVarProdM}, we simply need to sum the results \eqref{eq:varMPSnnVar-c} and \eqref{eq:varMPSnnVar-b} for $i\leq j$. With $\sum_{n=0}^\infty\eta^n=1/(1-\eta)\stackrel{\eqref{eq:MPSgradDecay}}{=}(m^2d^2-1)/[(d-1)(m^2d+1)]$, one obtains
\begin{align}\textstyle\nonumber
	\frac{1}{N}\sum_{i=1}^j\Avg \Tr(\hg^\dag_i\hg_i)
	=2\frac{m^2d-1}{d^2(m^2d+1)(m^2d^2-1)}\Big\{
	 &\textstyle\left(\frac{m^2d^2-1}{(d-1)(m^2d+1)}+1\right)\left[\Tr\hh^2+\frac{d-1}{d}\Tr(\Tr^2_1\hh)\right]\\
	 \label{eq:varMPSnnExtVar-diag}
	 &\textstyle-\frac{m^2-1}{m^2d-1}\Tr(\Tr^2_1\hh) \Big\} + \mc{O}(\eta^j)+ \mc{O}(\eta^{L-j}).
\end{align}
\begin{figure*}[b]
	\label{fig:MPSnn-offdiag}
	\includegraphics[width=\textwidth]{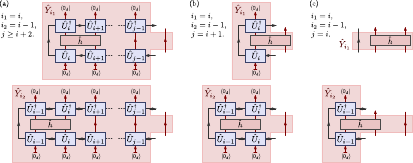}
	\caption{There are off-diagonal contributions $\frac{1}{N}\Avg \Tr(\hg^\dag_{i_1}\hg_{i_2\neq i_1})$ to the gradient variance \eqref{eq:gradExtVarProdM} for extensive nearest-neighbor Hamiltonians $\hH=\sum_i\hh_i$. These contributions have $|i_1-i_2|=1$ for the locations of the interaction terms. 
	For the gradient with respect to a unitary $\hU_j$ with $j> i_1,i_2$, we employ the operators $\hY_{i_1},\hY_{i_2}$ as shown in (a) and (b), as well as $\hX_{i_1}=\hX_{i_2}$ as shown in Fig.~\ref{fig:MPSnn-diag}a.
	For $j=\min(i_1,i_2)$, we employ $\hY_{i_1},\hY_{i_2}$ as shown in (c), as well as $\hX_{i_1}=\hX_{i_2}$ as shown in Fig.~\ref{fig:MPSnn-diag}b. In this latter case, $M=d$ in Eq.~\eqref{eq:energy}.}
\end{figure*}
\textbf{(h)}
Concerning the off-diagonal contributions with $|i_1-i_2|=1$ to the gradient variance \eqref{eq:gradExtVarProdM}, we choose $i_1=i_2+1$ without loss of generality and begin with the case $j\geq i_1+2$, where $M_{i_1}=M_{i_2}=1$. $\hX_{i_1}=\hX_{i_2}$ are then given by Eq.~\eqref{eq:MPS_X}. $\hY_{i_1}$, $\hY_{i_2}$, and the corresponding terms $\hL_{i_1}$ and $\hL_{i_2}$ are given by Eq.~\eqref{eq:MPSnn_Y} with diagrammatic representations shown in Fig.~\ref{fig:MPSnn-offdiag}a. With
\begin{align}\textstyle\nonumber
	\Avg\hL_{i_1}\otimes\hL_{i_2}
	&\stackrel{\eqref{eq:MPSnn_Y}}{=}\textstyle
	  \bra 0_d|^{\otimes 6}\G_{\tb,3}^\doub\circ\G_{\tb,2}^\doub\circ\G_{\tb,1}^\doub \big([\id_m\otimes(\id_d\otimes\hh)]\otimes[\id_m\otimes(\hh\otimes\id_d)]\big)|0_d\ket^{\otimes 6}\\
	\label{eq:MPS_AvgLLb3}
	&\,\textstyle\stackrel{\eqref{eq:G2}}{=}
	  \frac{\Tr(\Tr^2_1\hh)}{d^3\,(m^2d^2-1)^2}
	   \Big(\big[m^2d^2(d-1)-m^2d+1\big]\,\id_{m^2}+md^2(m^2-1)\Swap\Big).
\end{align}
and Eq.~\eqref{eq:MPS_TrY2explain} for the doubled MPS channel $\E^\doub$, we find
\begin{align}\textstyle\nonumber
	\Avg\Tr&\textstyle
	\left(\big(\hY_{i_1}-\frac{\id_{md}}{md}\Tr\hY_{i_1}\big)\big(\hY_{i_2}-\frac{\id_{md}}{md}\Tr\hY_{i_2}\big)\right)
	 = \Tr\left((\Swap-\frac{1}{md}\id_{(md)^2})\,\Avg\hY_{i_1}\otimes\hY_{i_2}\right)\\
	&\textstyle =
	  d\Tr\left((\Swap-\frac{1}{m}\id_{m^2})\,(\E^{\doub\dag})^{j-i_1-2}\big(\Avg\hL_{i_1}\otimes\hL_{i_2}\big)\right)
	  = \eta^{j-i_1}\,\frac{m}{d^2}\,\Tr(\Tr^2_1\hh).
	\label{eq:MPS_TrY2part2sh1}
\end{align}
Using this and Eq.~\eqref{eq:MPS_TrX2part} for the $\hX$ term in Eq.~\eqref{eq:gradExtVarProdM-c}, we arrive at
\begin{equation}\label{eq:gradExtVar_sh1a}\textstyle
	\frac{1}{N}\Avg \Tr(\hg^\dag_{i_1}\hg_{i_2})
	=\eta^{j-i_1}\,\frac{2}{d^3(m^2d+1)}\,\Tr(\Tr^2_1\hh)+\mc{O}(\eta^{L-i_1}).
\end{equation}
\textbf{(i)}
The off-diagonal contribution to the gradient variance \eqref{eq:gradExtVarProdM} with $i_1=i_2+1$ and $j=i_1+1$ can be treated very similarly to the previous case. $\hX_{i_1}=\hX_{i_2}$ are unchanged. $\hY_{i_1}$ and $\hY_{i_2}$ are shown diagrammatically in Fig.~\ref{fig:MPSnn-offdiag}b. One finds again Eqs.~\eqref{eq:MPS_TrY2part2sh1} and \eqref{eq:gradExtVar_sh1a}.\\[0.4em]
\textbf{(j)}
The evaluation of the off-diagonal contribution to the gradient variance \eqref{eq:gradExtVarProdM} with $i_1=i_2+1$ and $j=i_1$ is a bit more involved and similar to the derivation of Eq.~\eqref{eq:varMPSnnVar-c} in part (d) of this proof. We choose $\hX_{i_1}=\hX_{i_2}$ and the corresponding $\hx$ as defined in Eq.~\eqref{eq:MPSnni_YXR}, leading to $\Avg\hx\otimes\hx$ given in Eq.~\eqref{eq:MPSnni_xx}. We then have
\begin{equation}
	\hY_{i_1}=\id_m\otimes\hh\quad\text{and}\quad
	\hY_{i_2}=\big\{(\hA^\dag_{i_2}\otimes\id_d)[\id_m\otimes\hh](\hA_{i_2}\otimes\id_d)\big\}\otimes\id_d
\end{equation}
as shown in Fig.~\ref{fig:MPSnn-offdiag}c. This leads to the Haar average
\begin{equation}\textstyle
	\Avg\hY_{i_1}\otimes\hY_{i_2}\stackrel{\eqref{eq:MPSisoForm},\eqref{eq:G}}{=}
	\frac{1}{d}[\id_m\otimes\hh]\otimes[\id_m\otimes\Tr_1(\hh)\otimes\id_d].
\end{equation}
Plugging this and $\Avg\hX_{i_1}\otimes\hX_{i_2}$ as resulting from Eq.~\eqref{eq:MPSnni_xx} into Eq.~\eqref{eq:gradExtVarProdM-b}, we arrive at
\begin{equation}\label{eq:gradExtVar_sh1c}\textstyle
	\frac{1}{N}\Avg \Tr(\hg^\dag_{i_1}\hg_{i_2})
	=\frac{1}{N}\Avg \Tr(\hg^\dag_{j}\hg_{j-1})
	=\frac{2}{d^3(m^2d+1)}\,\Tr(\Tr^2_1\hh)+\mc{O}(\eta^{L-i_1})
\end{equation}
which is consistent with Eq.~\eqref{eq:gradExtVar_sh1a}.\\[0.4em]
\textbf{(k)}
We can now collect all off-diagonal contributions ($i_1\neq i_2$) to the gradient variance \eqref{eq:gradExtVarProdM}, summing the results \eqref{eq:gradExtVar_sh1a} for $i_1\leq j$ and multiply by two to cover $i_2-i_1=\pm 1$. With $\sum_{n=0}^\infty\eta^n=1/(1-\eta)\stackrel{\eqref{eq:MPSgradDecay}}{=}(m^2d^2-1)/[(d-1)(m^2d+1)]$, one obtains
\begin{equation}\label{eq:varMPSnnExtVar-offdiag}\textstyle
	\frac{1}{N}\sum_{i_1\neq i_2}\Avg \Tr(\hg^\dag_{i_1}\hg_{i_2})
	=4\frac{m^2d^2-1}{d^3(d-1)(m^2d+1)^2}\,\Tr(\Tr^2_1\hh)
	 + \mc{O}(\eta^j)+ \mc{O}(\eta^{L-j}).
\end{equation}
\textbf{(l)}
The large-$m$ (first), large-$d$ (second) scaling $\Var \big(\partial_{\hU_j} \sum_i \bra\Psi|\hh_i|\Psi\ket\big) \sim \frac{4}{m^2 d^4}\big[\Tr(\hh^2)+2\Tr(\Tr^2_1\hh)\big]$ stated in Eq.~\eqref{eq:varMPSnnExtVar} follows from the diagonal contributions \eqref{eq:varMPSnnExtVar-diag} scaling as $\sim \frac{4}{m^2 d^4}\big[\Tr(\hh^2)+\Tr(\Tr^2_1\hh)\big]$ and the off-diagonal contributions \eqref{eq:varMPSnnExtVar-offdiag} scaling as $\sim \frac{4}{m^2 d^4}\Tr(\Tr^2_1\hh)$.
\qed

\section{Primitives for doubled MERA and TTNS layer-transition channels}\label{appx:MERAdoubChanPrim}
According to the argument in the part (d) of the proof of Theorem~\ref{thrm:1dMERAbin}, the doubled layer-transition channels of all MERA and TTNS can be expressed, using the biorthogonal operator bases
\begin{subequations}\label{eq:hP-basis2}
\begin{gather}\textstyle
	\B_L=\{\bbra\hP_+|,\bbra\hP_-|\},\quad
	\B_R=\{|\hP'_+\kket,|\hP'_-\kket\}\quad \text{with}\\
	\label{eq:hP-trace}
	\hP_\pm\stackrel{\eqref{eq:hP}}{=}\frac{1}{2}\left(\id_{N^2}\pm\Swap\right),\ 
	\hP'_\pm\stackrel{\eqref{eq:G2dyadic}}{=}\frac{1}{\nu_\pm}\hP_\pm,\quad\text{and}\quad
	\nu_\pm:=\Tr\hP_\pm=\frac{1}{2}N(N\pm 1)
\end{gather}
\end{subequations}
for two copies of an $N$ dimensional Hilbert space.
We will discuss how to express all needed primitives using this basis for every doubled site.
The primitives are: appending auxiliary doubled sites initialized in the reference state $|0\chi,0\chi\ket$, executing (partial) traces over a doubled site, executing (partial) traces over a doubled site after swapping the two copies, applying Haar-averaged unitaries on one component of a doubled site, and applying Haar-averaged unitaries (acting on multiple sites) where the same unitary is applied to both components of the doubled system.
In following, we often refer to doubled sites simply as sites.

\subsection{Single-site operations}
Consider a single site with bond dimension $\chi$.
When applying isometries, we first append a (doubled) auxiliary site initialized in the reference state $|0_\chi,0_\chi\ket$. Due to subsequent applications of Haar-averaged unitaries or traces, we only need the projection of $|0,0\kket:=|0_\chi,0_\chi\ket\bra 0_\chi,0_\chi|$ onto the two-dimensional operator space spanned by $\hP_+$ and $\hP_-$. With
\begin{equation}\textstyle
	\bbra\hP_+|0,0\kket=\bra 0_\chi,0_\chi|\hP_+|0_\chi,0_\chi\ket
	\stackrel{\eqref{eq:hP-basis2}}{=}1\quad\text{and}\quad
	\bbra\hP_-|0,0\kket=0,	
\end{equation}
appending an auxiliary site corresponds to taking the tensor product with the vector
\begin{equation}\label{eq:a-matrix}
	\va:=\Pmatrix{1\\ 0}.
\end{equation}
The two components of this vector give the expansion coefficients of the projected reference state $|0,0\kket$ in the right single-site operator basis $\B_R$ with $N=\chi$ in Eq.~\eqref{eq:hP-basis2}.

Another operation primitive is to trace out a site. With
\begin{equation}
	\Tr\hR=\Tr\big[(\hP_++\hP_-)\hR\big]=\bbra\hP_+|\hR\kket+\bbra\hP_-|\hR\kket,
\end{equation}
this corresponds to multiplying in the relevant two-dimensional subspace with the transpose of the vector
\begin{equation}\label{eq:t-matrix}
	\vt:=\big[\bbra\id_{\chi^2}|\big]_{\B_L}=\big[\bbra\hP_+|+\bbra\hP_-|\big]_{\B_L}=\Pmatrix{1\\1},
\end{equation}
where $\B_L$ is the left single-site operator basis with $N=\chi$ in Eq.~\eqref{eq:hP-basis2}.
Similarly, we may want to trace out a site after applying a swap of the two copies. With
\begin{equation}
	\Tr(\Swap\hR)=\Tr\big[(\hP_+-\hP_-)\hR\big]=\bbra\hP_+|\hR\kket-\bbra\hP_-|\hR\kket,
\end{equation}
this corresponds to multiplying in the relevant two-dimensional subspace with the transpose of the vector
\begin{equation}\label{eq:s-matrix}
	\vs:=\big[\bbra\hP_+|-\bbra\hP_-|\big]_{\B_L}=\Pmatrix{1\\-1}.
\end{equation}

When considering off-diagonal contributions to the gradient variance, we need to apply a Haar-averaged unitary to only one component of a doubled system, i.e., we need to apply $\G\otimes\Id$ or $\Id\otimes\G$ with the fully depolarizing channel $\G$ from Eq.~\eqref{eq:G}. It turns out that, in the operator space $\Span \B_R$, both have the same effect, and we find the $2\times 2$ matrix representation
\begin{align}\nonumber
	g:&=\big[\G\otimes\Id\big]_\B=
	    \big[\Id\otimes\G\big]_\B
	   =\frac{1}{\chi^2}\left[\Big(|\hP_+\kket+|\hP_-\kket\Big)\Big(\bbra\hP_+|+\bbra\hP_-|\Big)\right]_\B\\
	\label{eq:G-matrix}
	  &=\frac{1}{\chi^2}\left[\Big(\nu_+|\hP'_+\kket+\nu_-|\hP'_-\kket\Big)\Big(\bbra\hP_+|+\bbra\hP_-|\Big)\right]_\B
	   =\frac{1}{\chi^2}\Pmatrix{\nu_+&\nu_+\\\nu_-&\nu_-}
	   =\frac{1}{\chi^2}\,\Omega\cdot\Pmatrix{1&1\\1&1},
\end{align}
where we have introduced the diagonal matrix
\begin{equation}\label{eq:Omega-matrix}
	\Omega=\Omega(\chi):=\Psmatrix{\nu_+&0\\0&\nu_-}
\end{equation}
which transforms $\B_R$ into $\B_L$.

\subsection{Two-site operations}
Now consider two sites with (bond) dimensions $\chi_1$ and $\chi_2$. The identity and swap operators on the joint system are tensor products of the corresponding single-site operators,
\begin{equation}
	\id=\id_1\otimes\id_2\quad\text{and}\quad \Swap=\Swap_1\otimes\Swap_2.
\end{equation}
With this, we can expand the projectors $\hP_\pm$ of the joint system in tensor products of the single-site projectors $\hP_{\pm,i}$,
\begin{align}\nonumber
	\hP_\pm&=\frac{1}{2}\left(\id\pm\Swap\right)
	=\frac{1}{2}\left[(\hP_{+,1}+\hP_{-,1})\otimes(\hP_{+,2}+\hP_{-,2})
	               \pm(\hP_{+,1}-\hP_{-,1})\otimes(\hP_{+,2}-\hP_{-,2})\right]\\
	\label{eq:hP-decomp}
	&=\hP_{+,1}\otimes\hP_{\pm,2}+\hP_{-,1}\otimes\hP_{\mp,2}.
\end{align}
This just reflects the fact that the tensor product of two symmetric or two antisymmetric states is symmetric and that the tensor product of a symmetric and an antisymmetric state is antisymmetric.
Similarly, we can express $\hP'_\pm\stackrel{\eqref{eq:hP-basis2}}{=}\hP_\pm/\nu_\pm$ in the form
\begin{equation}\label{eq:hPp-decomp}
	\hP'_\pm
	\stackrel{\eqref{eq:hP-decomp}}{=}
	\frac{1}{\nu_\pm}\left(\hP_{+,1}\otimes\hP_{\pm,2}+\hP_{-,1}\otimes\hP_{\mp,2}\right)
	=\frac{1}{\nu_\pm}\left(\nu_{+,1}\nu_{\pm,2}\hP'_{+,1}\otimes\hP'_{\pm,2}+\nu_{-,1}\nu_{\mp,2}\hP'_{-,1}\otimes\hP'_{\mp,2}\right)
\end{equation}
with
\begin{equation}\textstyle
	\nu_{\pm,i}=\Tr\hP_{\pm,i}\stackrel{\eqref{eq:hP-trace}}{=}\frac{1}{2}\chi_i(\chi_i\pm 1)
	\quad\text{and}\quad
	\nu_{\pm}=\Tr\hP_{\pm}\stackrel{\eqref{eq:hP-trace}}{=}\frac{1}{2}\chi_1\chi_2(\chi_1\chi_2\pm 1).
\end{equation}

With these properties of the of the projectors, we obtain, for example, the $4\times 4$ matrix representation of the doubled fully depolarizing channel \eqref{eq:G2} acting on two sites,
\begin{subequations}\label{eq:G2-matrix}
\begin{gather}
	G_{1,2}:=[\G^\doub]_{\B^{\otimes 2}} \stackrel{\eqref{eq:G2dyadic}}{=}
	\left[|\hP'_+\kket\bbra\hP^\pdag_+|+|\hP'_-\kket\bbra\hP^\pdag_-|\right]_{\B^{\otimes 2}} = J\cdot S\quad\text{with}\\
	\label{eq:S-J}
	J=\Pmatrix{\nu_{+,1}\nu_{+,2}&0\\0&\nu_{+,1}\nu_{-,2}\\ 0&\nu_{-,1}\nu_{+,2}\\ \nu_{-,1}\nu_{-,2}&0 }
	   \cdot\Pmatrix{1/\nu_+&0\\0&1/\nu_-}
	\quad\text{and}\quad
	S=\Pmatrix{1&0&0&1\\0&1&1&0}.
\end{gather}
\end{subequations}
Here, we have ordered the biorthogonal left and right operator bases for the two sites as
\begin{subequations}\label{eq:hP-basisProd2}
\begin{gather}
	\B_L^{\otimes 2}=\big(\bbra\hP_{+,1},\hP_{+,2}|,\bbra\hP_{+,1},\hP_{-,2}|,\bbra\hP_{-,1},\hP_{+,2}|,\bbra\hP_{-,1},\hP_{-,2}|\big)
	\quad \text{and}\\
	\B_R^{\otimes 2}=\big(|\hP'_{+,1},\hP'_{+,2}\kket,|\hP'_{+,1},\hP'_{-,2}\kket,|\hP'_{-,1},\hP'_{+,2}\kket,|\hP'_{-,1},\hP'_{-,2}\kket\big).
\end{gather}
\end{subequations}
In Eq.~\eqref{eq:S-J}, the matrix $S$ corresponds to the expansion \eqref{eq:hP-decomp} of $\hP_+$ and $\hP_-$ in the basis $\B_L^{\otimes 2}$, and matrix $J$ corresponds to the expansion \eqref{eq:hPp-decomp} of $\hP'_+$ and $\hP'_-$ in the basis $\B_R^{\otimes 2}$.

\subsection{The doubled MPS channel}
Using the operator basis \eqref{eq:hP-basis2} with  $N=m$ for the doubled bond vector space, as well as $\chi_1=m$ and $\chi_2=d$ for the fully depolarizing channel $G_{1,2}=:G_{1,2}(m,d)$ in Eq.~\eqref{eq:G2-matrix}, we obtain a matrix representation for the doubled MPS channel,
\begin{equation}
	[\E^\doub]_{\B}
	=\big[\Psmatrix{1&0\\0&1}\otimes\vtT\big]\cdot
	 G_{1,2}(m,d)\cdot
	 \big[\Psmatrix{1&0\\0&1}\otimes\va\big].
\end{equation}
This agrees with the expression in Eq.~\eqref{eq:E2matrix}.

\subsection{Three-site operations}
Similar to $G_{1,2}$, we determine the $8\times 8$ matrix representation $G_{1,2,3}$ of the doubled fully depolarizing channel \eqref{eq:G2} acting on three sites with bond dimension $\chi$ in the left and right operator bases $\B_L^{\otimes 3}$ and $\B_R^{\otimes 3}$. To this purpose, we can iterate the decomposition from Eq.~\eqref{eq:G2-matrix}, first decomposing from the $\chi^3$-dimensional space into two sites with dimensions $\chi_1=\chi$ and $\chi_2=\chi^2$, and then decomposing the second site into two with dimensions $\chi_{2'}=\chi_{3'}=\chi$. Using the label $J=J(\chi_1,\chi_2)$ for the matrix in Eq.~\eqref{eq:S-J}, we obtain
\begin{equation}\label{eq:G3-matrix}
	G_{1,2,3}:=[\G^\doub]_{\B^{\otimes 3}} \stackrel{\eqref{eq:G2dyadic},\eqref{eq:G2}}{=}
	\big[\Psmatrix{1&0\\0&1}\otimes J(\chi,\chi)\big] \cdot 
	\big[J(\chi,\chi^2)\cdot S \big]\cdot 
	\big[\Psmatrix{1&0\\0&1}\otimes S\big].
\end{equation}

\section{Matrix representations of doubled MERA and TTNS layer-transition channels}\label{appx:MERAdoubChan}
Using the operator basis \eqref{eq:hP-basis2}, we can deduce compact matrix representations for the doubled layer-transition channels of MERA and TTNS. The required primitive operations were discussed in Appendix~\ref{appx:MERAdoubChanPrim}. In the following, we again refer to each doubled site simply as a site.

\subsection{Binary 1D MERA -- diagonal contributions}\label{appx:1dMERAbin-Edoub-matrix-diag}
Let us first discuss the layer-transition channels for the diagonal contributions \eqref{eq:1dMERAbin-gradVarDiag} to the gradient variance for binary 1D MERA. For the doubled right-moving layer-transition channel $\E^\doub_{\bin,\tR}$ in Eq.~\eqref{eq:1dMERAbin-Edoub-b}, we
\begin{itemize}[itemsep=0em]
 \item
 start on the three neighboring sites $c_1,c_2,c_3\in\L_\tau$ of the causal cone and append three auxiliary sites $a_1,a_2,a_3$.
 \item
 Then, we apply the doubled fully depolarizing channel $\G^\doub$ [Eq.~\eqref{eq:G2}] on site groups $(c_1,a_1)$, $(c_2,a_2)$, and $(c_3,a_3)$ to implement the isometries \eqref{eq:1dMERAbin-isometry}.
 \item
 Next, we apply $\G^\doub$ (the disentanglers) on site groups $(a_1,c_2)$ and $(a_2,c_3)$.
 \item
 Finally, we trace out sites $c_1$, $a_1$, and $a_3$ such that $c_2,a_2,c_3\in\L_{\tau-1}$ represent the causal cone after the layer transition.
\end{itemize}
An illustration is given in Fig.~\ref{fig:MERAbin-layerTransit}b.
With the matrix and vector representations of the operation primitives from Eqs.~\eqref{eq:a-matrix}, \eqref{eq:t-matrix}, and \eqref{eq:G2-matrix}, we have
\begin{align}\nonumber
	\left[\E^\doub_{\bin,\tR}\right]_{\B^{\otimes 3}}
	=&\left(\vtT_{c_1}\otimes \vtT_{a_1}\otimes\id_{c_2,a_2,c_3}\otimes \vtT_{a_3}\right)
	  \cdot\left(\id_{c_1}\otimes G_{a_1,c_2}\otimes G_{a_2,c_3}\otimes\id_{a_3}\right)\\
	\label{eq:1dMERAbin-EdoubR-matrix}
	 &\times\left(G_{c_1,a_1}\otimes G_{c_2,a_2}\otimes G_{c_3,a_3}\right)
	  \cdot\left(\id_{c_1}\otimes\va_{a_1}\otimes \id_{c_2}\otimes\va_{a_2}\otimes \id_{c_3}\otimes\va_{a_3}\right),
\end{align}
where the employed biorthogonal eight-dimensional left and right operator bases are $\B_L^{\otimes 3}$ and $\B_R^{\otimes 3}$ with the single-site bases specified in Eq.~\eqref{eq:hP-basis2}. The eigenvalues of the $8\times 8$ matrix \eqref{eq:1dMERAbin-EdoubR-matrix} are given in Eq.~\eqref{eq:1dMERAbin-EdoubR-spec}.

Similarly, we obtain the matrix representation of the left-moving layer-transition channel $\E^\doub_{\bin,\tL}$ [Eq.~\eqref{eq:1dMERAbin-Edoub-b} and Fig.~\ref{fig:MERAbin-layerTransit}a] as
\begin{align}\nonumber
	\left[\E^\doub_{\bin,\tL}\right]_{\B^{\otimes 3}}
	=&\left(\vtT_{c_1}\otimes \id_{a_1,c_2,a_2}\otimes \vtT_{c_3}\otimes\vtT_{a_3}\right)
	  \cdot\left(\id_{c_1}\otimes G_{a_1,c_2}\otimes G_{a_2,c_3}\otimes\id_{a_3}\right)\\
	\label{eq:1dMERAbin-EdoubL-matrix}
	 &\times\left(G_{c_1,a_1}\otimes G_{c_2,a_2}\otimes G_{c_3,a_3}\right)
	  \cdot\left(\id_{c_1}\otimes\va_{a_1}\otimes \id_{c_2}\otimes\va_{a_2}\otimes \id_{c_3}\otimes\va_{a_3}\right).
\end{align}
As it is related to $\E^\doub_{\bin,\tR}$ by a site permutation, it has the same spectrum~\eqref{eq:1dMERAbin-EdoubR-spec}.

The diagonalization \eqref{eq:1dMERAbin-Edoub-diag} of the channel $\E_\bin^\doub=\big(\E^\doub_{\bin,\tL}+\E^\doub_{\bin,\tR}\big)/2$ leads to the spectrum given in Eq.~\eqref{eq:1dMERAbin-Edoub-spec} and the following vector representations for the first few left and right eigenvectors:
\begin{subequations}\label{eq:1dMERAbin-Edoub-eigenvec}
\begin{align}
	\vec{\ell}_1&:=[\hl_1]_{\B_L^{\otimes 3}}=[\id_{\chi^6}]_{\B_L^{\otimes 3}}
	  =(1,1,1,1,1,1,1,1)^\intercal,\\
	\vec{r}_1&:=[\hr_1]_{\B_R^{\otimes 3}}\textstyle
	  =\frac{1}{8}\left( 1+\frac{3}{\chi}, 1+\frac{1}{\chi}, 1+\frac{1}{\chi}, 1-\frac{1}{\chi},
	                     1+\frac{1}{\chi}, 1-\frac{1}{\chi}, 1-\frac{1}{\chi}, 1-\frac{3}{\chi} \right)^\intercal+\mc{O}\Big(\frac{1}{\chi^2}\Big),\\
	\vec{\ell}_2&:=[\hl_2]_{\B_L^{\otimes 3}}\textstyle
	  =\frac{1}{4}\left(3-\frac{8}{\chi},1-\frac{4}{\chi},-1,-1,1-\frac{4}{\chi},-1,-1,-1\right)^\intercal+\mc{O}\Big(\frac{1}{\chi^2}\Big),\\
	\label{eq:1dMERAbin-Edoub-eigenvec-r2}
	\vec{r}_2&:=[\hr_2]_{\B_R^{\otimes 3}}\textstyle
	  =\left(1+\frac{4}{\chi},-\frac{2}{\chi},-1,\frac{2}{\chi},-\frac{2}{\chi},-1+\frac{4}{\chi},\frac{2}{\chi},1-\frac{8}{\chi}\right)^\intercal+\mc{O}\Big(\frac{1}{\chi^2}\Big).
\end{align}
\end{subequations}

\subsection{Binary 1D MERA -- diagonal contributions to spatially averaged variance}\label{appx:1dMERAbin-diagSpaceAvg}
Equation \eqref{eq:1dMERAbin-gradVarDiagSpaceAvg} expresses the diagonal contributions to the Haar-variance of the Riemannian gradient for a disentangler $\hU_{\tau,k}$, spatially averaged over all $k\in\L_\tau$, in terms of the three-site Hamiltonian term $\hh$, eigenvalues and eigenvectors of $\E^\doub_\bin$, the map $\Q$, and adapted layer-transition channels $\tilde{\E}^\doub_{\bin,\tR}$ and $\tilde{\E}^\doub_{\bin,\tL}$ as described below Eq.~\eqref{eq:Q}. The matrix representations of the latter are
\begin{align}\nonumber
	\left[\tilde{\E}^\doub_{\bin,\tR}\right]_{\B^{\otimes 3}}
	=&\left(\vtT_{c_1}\otimes \vtT_{a_1}\otimes\id_{c_2,a_2,c_3}\otimes \vtT_{a_3}\right)
	  \cdot\left(\id_{c_1}\otimes G_{a_1,c_2}\otimes\id_{a_2,c_3,a_3}\right)\\
	\label{eq:1dMERAbin-EdoubRt-matrix}
	 &\times\left(G_{c_1,a_1}\otimes G_{c_2,a_2}\otimes G_{c_3,a_3}\right)
	  \cdot\left(\id_{c_1}\otimes\va_{a_1}\otimes \id_{c_2}\otimes\va_{a_2}\otimes \id_{c_3}\otimes\va_{a_3}\right)
\end{align}
and
\begin{align}\nonumber
	\left[\tilde{\E}^\doub_{\bin,\tL}\right]_{\B_R^{\otimes 4},\B_L^{\otimes 3}}
	=&\left(\vtT_{c_1}\otimes \id_{a_1,c_2,a_2,c_3}\otimes\vtT_{a_3}\right)
	  \cdot\left(\id_{c_1}\otimes G_{a_1,c_2}\otimes\id_{a_2,c_3,a_3}\right)\\
	\label{eq:1dMERAbin-EdoubLt-matrix}
	 &\times\left(G_{c_1,a_1}\otimes G_{c_2,a_2}\otimes G_{c_3,a_3}\right)
	  \cdot\left(\id_{c_1}\otimes\va_{a_1}\otimes \id_{c_2}\otimes\va_{a_2}\otimes \id_{c_3}\otimes\va_{a_3}\right).
\end{align}
Note that, while ${\E}^\doub_{\bin,\tR}$, ${\E}^\doub_{\bin,\tL}$, and $\tilde{\E}^\doub_{\bin,\tR}$ map from three sites to three sites, $\tilde{\E}^\doub_{\bin,\tL}$ maps to operators on four (doubled) sites as we omit the final trace over site $c_3$.
A matrix representation for the two-site map $\Q$ can be given using the primitives from Eqs.~\eqref{eq:t-matrix}, \eqref{eq:s-matrix}, and \eqref{eq:Omega-matrix},
\begin{align}\nonumber
	Q:&=[\Q]_{\B^{\otimes 2}}\stackrel{\eqref{eq:Q}}{=} \textstyle
	\Big[|\Swap-\frac{1}{\chi^2}\id_{\chi^4}\kket\bbra\Swap-\frac{1}{\chi^2}\id_{\chi^4}|\Big]_{\B^{\otimes 2}}\\
	&\textstyle
	 =(\Omega\otimes\Omega)
	  \cdot\Big(\vs\otimes\vs-\frac{1}{\chi^2}\vt\otimes\vt\Big)
	  \cdot\Big(\vsT\otimes\vsT-\frac{1}{\chi^2}\vtT\otimes\vtT\Big).
\end{align}

With this and the vector representations \eqref{eq:1dMERAbin-Edoub-eigenvec} for the relevant left and right $\E^\doub_\bin$ eigenvectors,
Eq.~\eqref{eq:1dMERAbin-gradVarDiagSpaceAvg} evaluates to the expression given in Eq.~\eqref{eq:1dMERAbin-gradVarDiagSpaceAvg-eval}.

\subsection{Binary 1D MERA -- off-diagonal contributions}\label{appx:1dMERAbin-Edoub-matrix-offdiag}
For the off-diagonal contributions to the gradient variance with $i\neq j$ in Eq.~\eqref{eq:1dMERAbin-gradVarProdM}, as discussed in part (e) of the proof for Theorem~\ref{thrm:1dMERAbin}, we only need to consider the layer-transition channel $\E^\doub_{\bin-1}$, where the causal cones in the two components of the doubled system are shifted by one site. For this channel,
\begin{itemize}[itemsep=0em]
 \item
 we start on four neighboring sites $c_{-2},c_{-1},c_0,c_1\in\L_\tau$ that comprise both of the three-site causal cones of the two components. In the first step, four auxiliary sites $a_{-2},a_{-1},a_0,a_1$ are appended.
 \item
 Then, we apply the depolarizing channel $\G$ [Eq.~\eqref{eq:G}] on the second components of sites $c_{-2}$ and $a_{-2}$, 
 the doubled fully depolarizing channel $\G^\doub$ [Eq.~\eqref{eq:G2}] on site groups $(c_{-1},a_{-1})$ and $(c_0,a_0)$, and
 $\G$ on the first components of sites $c_{1}$ and $a_{1}$ to implement the isometries \eqref{eq:1dMERAbin-isometry}.
 \item
 Next, we apply $\G$ on the second components of sites $a_{-2}$ and $c_{-1}$, 
 $\G^\doub$ on site group $(a_{-1},c_0)$, and $\G$ on the first components of sites $a_0$ and $c_1$ to implement the disentanglers.
 \item
 Finally, we trace out sites $c_{-2},a_{-2},c_1$, and $a_1$ such that $c_{-1},a_{-1},c_0,a_0\in\L_{\tau-1}$ compose the causal cone after  the layer transition.
\end{itemize}
An illustration is given in Fig.~\ref{fig:MERAbin-layerTransit}c.
With the matrix and vector representations of the operation primitives from Eqs.~\eqref{eq:a-matrix}, \eqref{eq:t-matrix}, \eqref{eq:G-matrix}, and \eqref{eq:G2-matrix}, we have
\begin{align}\nonumber
	\left[\E^\doub_{\bin-1}\right]_{\B^{\otimes 4}}
	=\ \ \ &\left(\vtT_{c_{-2}}\otimes \vtT_{a_{-2}}\otimes\id_{c_{-1},a_{-1},c_0,a_0}\otimes \vtT_{c_1}\otimes \vtT_{a_1}\right)\\\nonumber
	\times&\left(\id_{c_{-2}}\otimes g_{a_{-2}}\otimes g_{c_{-1}} \otimes G_{a_{-1},c_0}\otimes g_{a_0}\otimes g_{c_1}\otimes\id_{a_1}\right)\\\nonumber
	\times&\left(g_{c_{-2}} \otimes g_{a_{-2}} \otimes G_{c_{-1},a_{-1}}\otimes G_{c_0,a_0}\otimes g_{c_1} \otimes g_{a_1}\right)\\
	\label{eq:1dMERAbin-EdoubS1-matrix}
	\times&\left(\id_{c_{-2}}\otimes\va_{a_{-2}}\otimes\id_{c_{-1}}\otimes\va_{a_{-1}}\otimes\id_{c_0}\otimes\va_{a_0}\otimes\id_{c_1}\otimes\va_{a_1}\right),
\end{align}
This $16\times 16$ matrix has the two non-zero eigenvalues given in Eq.~\eqref{eq:1dMERAbin-EdoubS1-spec}.

\subsection{Ternary 1D MERA}\label{appx:1dMERAter-Edoub-matrix}
For the right-moving layer-transition channel $\E^\doub_{\ter,\tR}$ of a ternary 1D MERA, we 
\begin{itemize}[itemsep=0em]
 \item
 start on the two neighboring sites $c_1,c_2\in\L_\tau$ of the causal cone and append four auxiliary sites $a_1,b_1,a_2,b_2$.
 \item
 Then, we apply the doubled fully depolarizing channel $\G^\doub$ [Eq.~\eqref{eq:G2}] on site groups $(c_1,a_1,b_1)$ and $(c_2,a_2,b_2)$ to implement the isometries \eqref{eq:1dMERAter-isometry}.
 \item
 Next, we apply $\G^\doub$ (the disentangler) on site group $(b_1,c_2)$.
 \item
 Finally, we trace out sites $c_1,a_1,b_1$, and $b_2$ such that $c_2,a_2\in\L_{\tau-1}$ represent the causal cone after the layer transition.
\end{itemize}
An illustration is given in Fig.~\ref{fig:MERAter}e.
With the matrix and vector representations of the operation primitives from Eqs.~\eqref{eq:a-matrix}, \eqref{eq:t-matrix}, \eqref{eq:G2-matrix}, and \eqref{eq:G3-matrix}, we have
\begin{align}\nonumber
	\left[\E^\doub_{\ter,\tR}\right]_{\B^{\otimes 2}}
	=&\left(\vtT_{c_1}\otimes\vtT_{a_1}\otimes\vtT_{b_1}\otimes\id_{c_2,a_2}\otimes \vtT_{b_2}\right)
	  \cdot\left(\id_{c_1,a_1}\otimes G_{b_1,c_2}\otimes \id_{b_2,a_2}\right)\\
	\label{eq:1dMERAter-EdoubR-matrix}
	 &\times\left(G_{c_1,a_1,b_1}\otimes G_{c_2,a_2,b_2}\right)
	  \cdot\left(\id_{c_1}\otimes\va_{a_1}\otimes\va_{b_1}\otimes \id_{c_2}\otimes\va_{a_2}\otimes\va_{b_2}\right),
\end{align}
where the employed biorthogonal four-dimensional left and right operator bases are $\B_L^{\otimes 2}$ and $\B_R^{\otimes 2}$ as given in Eq.~\eqref{eq:hP-basisProd2}. The eigenvalues of the $4\times 4$ matrix \eqref{eq:1dMERAter-EdoubR-matrix} are
\begin{equation}\label{eq:1dMERAter-EdoubR-spec}\textstyle
	1,\quad
	\frac{\chi^2}{1+\chi^2+\chi^4},\quad
	\frac{\chi^4}{(1-\chi+\chi^2)^2(1+\chi^2)(1+\chi+\chi^2)},\quad\text{and}\quad
	\frac{\chi^4}{(1-\chi+\chi^2)(1+\chi^2)(1+\chi+\chi^2)^2}.
\end{equation}

The left-moving and central layer-transition channels $\E^\doub_{\ter,\tL}$ and $\E^\doub_{\ter,\tC}$ [Eq.~\eqref{eq:1dMERAter-Edoub} and Figs.~\ref{fig:MERAter}d, \ref{fig:MERAter}e] only differ from $\E^\doub_{\ter,\tR}$ in terms of the sites that are traced out.
For $\E^\doub_{\ter,\tC}$, we trace out sites $c_1,a_1,a_2$, and $b_2$ such that the first term in \eqref{eq:1dMERAter-EdoubR-matrix} is replaced by $\vtT_{c_1}\otimes\vtT_{a_1}\otimes\id_{b_1,c_1}\otimes\vtT_{a_2}\otimes \vtT_{b_2}$. The $\E^\doub_{\ter,\tC}$ spectrum is
\begin{equation}\label{eq:1dMERAter-EdoubC-spec}\textstyle
	1,\quad
	\frac{3\chi^4}{(1+\chi^2+\chi^4)^2},\quad
	0,\quad 0.
\end{equation}
For $\E^\doub_{\ter,\tL}$, we trace out sites $c_1,c_2,a_2$, and $b_2$ such that the first term in \eqref{eq:1dMERAter-EdoubR-matrix} is replaced by $\vtT_{c_1}\otimes\id_{a_1,b_1}\otimes\vtT_{c_2}\otimes\vtT_{a_2}\otimes \vtT_{b_2}$. The spectrum is given by Eq.~\eqref{eq:1dMERAter-EdoubR-spec} as $\E^\doub_{\ter,\tL}$ is related to $\E^\doub_{\ter,\tR}$ by a site permutation.
The spectrum of $\E_\ter^\doub=\big(\E^\doub_{\ter,\tL}+\E^\doub_{\ter,\tC}+\E^\doub_{\ter,\tR}\big)/3$ is given in Eq.~\eqref{eq:1dMERAter-Edoub-spec}.

\subsection{Binary 1D TTNS}\label{appx:1dTTNSbin-Edoub-matrix}
For the right-moving layer-transition channel $\E^{\prime\doub}_{\bin,\tR}$ of a binary 1D TTNS, we 
\begin{itemize}[itemsep=0em]
 \item
 start on the three neighboring sites $c_1,c_2,c_3\in\L_\tau$ of the causal cone, trace out site $c_1$, and append two auxiliary sites $a_2,a_3$.
 \item
 Then, we apply the doubled fully depolarizing channel $\G^\doub$ [Eq.~\eqref{eq:G2}] on site groups $(c_2,a_2)$ and $(c_3,a_3)$ to implement the isometries \eqref{eq:1dMERAbin-isometry}.
 \item
 Finally, we trace out site $a_3$ such that $c_2,a_2,c_3\in\L_{\tau-1}$ represent the causal cone after the layer transition.
\end{itemize}
See also the illustration in Fig.~\ref{fig:MERAbin-layerTransit}b.
With the matrix and vector representations of the operation primitives from Eqs.~\eqref{eq:a-matrix}, \eqref{eq:t-matrix}, and \eqref{eq:G2-matrix}, we have
\begin{equation}\label{eq:1dTTNbin-EdoubR-matrix}
	\left[\E^{\prime\doub}_{\bin,\tR}\right]_{\B^{\otimes 3}}
	=\left(\id_{c_2,a_2,c_3}\otimes \vtT_{a_3}\right)
	 \cdot\left(G_{c_2,a_2}\otimes G_{c_3,a_3}\right)
	 \cdot\left(\vtT_{c_1}\otimes \id_{c_2}\otimes\va_{a_2}\otimes \id_{c_3}\otimes\va_{a_3}\right),
\end{equation}
where the employed biorthogonal eight-dimensional left and right operator bases are $\B_L^{\otimes 3}$ and $\B_R^{\otimes 3}$ with the single-site bases specified in Eq.~\eqref{eq:hP-basis2}.
Similarly, we obtain the matrix representation of the left-moving layer-transition channel $\E^{\prime\doub}_{\bin,\tL}$ as
\begin{equation}\label{eq:1dTTNbin-EdoubL-matrix}
	\left[\E^\doub_{\bin,\tL}\right]_{\B^{\otimes 3}}
	=\left(\vtT_{c_1}\otimes \id_{a_1,c_2,a_2}\right)
	 \cdot\left(G_{c_1,a_1}\otimes G_{c_2,a_2}\right)
	 \cdot\left(\id_{c_1}\otimes\va_{a_1}\otimes \id_{c_2}\otimes\va_{a_2}\otimes \vtT_{c_3}\right).
\end{equation}

The channel $\E_\bin^{\prime\doub}=\big(\E^{\prime\doub}_{\bin,\tL}+\E^{\prime\doub}_{\bin,\tR}\big)/2$ governs the $\tau$ dependence of the Riemannian gradient variance. Its nonzero eigenvalues are
\begin{equation}
	1,\quad
	\eta'_\bin:=\frac{\chi}{1+\chi^2},\quad
	\frac{\chi}{2+2\chi^2},\quad
	\frac{\chi}{2+2\chi^2},\quad
	\frac{\chi^2}{2(1+\chi^2)^2},\quad\text{and}\quad
	\frac{\chi^2}{2(1+\chi^2)^2}.
\end{equation}

\subsection{Ternary 1D TTNS}\label{appx:1dTTNSter-Edoub-matrix}
For the right-moving layer-transition channel $\E^{\prime\doub}_{\ter,\tR}$ of a ternary 1D TTNS, we 
\begin{itemize}
 \item
 start on the two neighboring sites $c_1,c_2\in\L_\tau$ of the causal cone, trace out site $c_1$, and append two auxiliary sites $a_2,b_2$.
 \item
 Then, we apply the doubled fully depolarizing channel $\G^\doub$ [Eq.~\eqref{eq:G2}] on site group $(c_2,a_2,b_2)$ to implement the isometry \eqref{eq:1dMERAter-isometry}.
 \item
 Finally, we trace out site $b_2$ such that $c_2,a_2\in\L_{\tau-1}$ represent the causal cone after the layer transition.
\end{itemize}
See also the illustration in Fig.~\ref{fig:MERAter}e.
With the matrix and vector representations of the operation primitives from Eqs.~\eqref{eq:a-matrix}, \eqref{eq:t-matrix}, and \eqref{eq:G3-matrix}, we have
\begin{equation}\label{eq:1dTTNter-EdoubR-matrix}
	\left[\E^{\prime\doub}_{\ter,\tR}\right]_{\B^{\otimes 2}}
	=\left(\id_{c_2,a_2}\otimes \vtT_{b_2}\right) \cdot G_{c_2,a_2,b_2}
	  \cdot\left(\vtT_{c_1}\otimes \id_{c_2}\otimes\va_{a_2}\otimes\va_{b_2}\right),
\end{equation}
where the employed biorthogonal four-dimensional left and right operator bases are $\B_L^{\otimes 2}$ and $\B_R^{\otimes 2}$ as given in Eq.~\eqref{eq:hP-basisProd2}. 
Similarly, we obtain the matrix representation of the left-moving layer-transition channel $\E^{\prime\doub}_{\ter,\tL}$ as
\begin{equation}\label{eq:1dTTNter-EdoubL-matrix}
	\left[\E^{\prime\doub}_{\ter,\tL}\right]_{\B^{\otimes 2}}
	=\left(\vtT_{c_1}\otimes\id_{a_1,b_1}\right) \cdot G_{c_1,a_1,b_1}
	  \cdot\left(\id_{c_1}\otimes\va_{a_1}\otimes\va_{b_1}\otimes \vtT_{c_2}\right)
\end{equation}
and the matrix representation of the central layer-transition channel $\E^{\prime\doub}_{\ter,\tC}$ as
\begin{align}\nonumber
	\left[\E^{\prime\doub}_{\ter,\tC}\right]_{\B^{\otimes 2}}
	=&\left(\vtT_{c_1}\otimes\vtT_{a_1}\otimes\id_{b1,c_2}\otimes\vtT_{a_2}\otimes \vtT_{b_2}\right)
	  \cdot\left(G_{c_1,a_1,b_1}\otimes G_{c_2,a_2,b_2}\right)\\
	\label{eq:1dTTNter-EdoubC-matrix}
	 &\times\left(\id_{c_1}\otimes\va_{a_1}\otimes\va_{b_1}\otimes \id_{c_2}\otimes\va_{a_2}\otimes\va_{b_2}\right)
	=E(\chi,\chi^2)\otimes E(\chi,\chi^2),
\end{align}
where $E(\chi,\chi^2)$ refers to the matrix representation \eqref{eq:E2matrix} of the doubled MPS channel with $N=\chi$ and $d=\chi^2$.

The channel $\E_\ter^{\prime\doub}=\big(\E^{\prime\doub}_{\ter,\tL}+\E^{\prime\doub}_{\ter,\tC}+\E^{\prime\doub}_{\ter,\tR}\big)/3$ governs the $\tau$ dependence of the Riemannian gradient variance. Its nonzero eigenvalues are
\begin{equation}
	1,\quad
	\eta'_\ter:=\frac{\chi^2}{1+\chi^2+\chi^4},\quad
	\frac{1}{3}\,\frac{\chi^2}{1+\chi^2+\chi^4},\quad\text{and}\quad
	\frac{1}{3}\,\frac{\chi^4}{(1+\chi^2+\chi^4)^2}.
\end{equation}

\subsection{Nonary 2D MERA and TTNS}\label{appx:2dMERAnon-Edoub-matrix}
For the top-left layer-transition channel $\E^\doub_{\non,\text{TL}}$ of the nonary 2D MERA discussed in Sec.~\ref{sec:MERA-nonary}, we 
\begin{itemize}[itemsep=0em]
 \item
 start on the $2\times 2$ block of sites $c_1,c_2,c_3,c_4\in\L_\tau$ of the causal cone and append eight auxiliary sites $a_{i,1},\dotsc,a_{i,8}$ as indicated in Fig.~\ref{fig:MERAnon}.
 \item
 Then, we apply the doubled fully depolarizing channel $\G^\doub$ on the four $3\times 3$ blocks $(c_1,a_{1,1},\dotsc,a_{1,8})$, $(c_2,a_{2,1},\dotsc,a_{2,8})$, $(c_3,\dotsc)$, and $(c_4,\dotsc)$ to implement the isometries.
 \item
 Next, we apply doubled fully depolarizing channels $\G^\doub$ on site groups $(a_{1,8},a_{2,6},a_{3,3},a_{4,1})$, $(a_{1,5},a_{2,4})$, and $(a_{1,7},a_{3,2})$.
 \item
 Finally, we trace out all sites except $c_1,a_{1,5},a_{1,7},a_{1,8}\in\L_{\tau-1}$ which represent the causal cone the layer transition.
\end{itemize}

For the top-center layer-transition channel $\E^\doub_{\non,\text{TC}}$, we follow the same procedure but can omit the disentangler on $(a_{1,7},a_{3,2})$ and trace out all sites except $\{a_{1,5},a_{2,4},a_{1,8},a_{2,6}\}$. For the middle-center layer-transition channel $\E^\doub_{\non,\text{MC}}$, we follow the same procedure but can omit the disentanglers on $(a_{1,5},a_{2,4})$ as well as $(a_{1,7},a_{3,2})$, and trace out all sites except $\{a_{1,8},a_{2,6},a_{3,3},a_{4,1}\}$. All further layer-transition channels $\E^\doub_{\non,\text{TR}}$, $\E^\doub_{\non,\text{ML}}$ etc.\ are related to the former by symmetry transformations.
Using the operator basis \eqref{eq:hP-basis2}, we find $16\times 16$ matrix representations of these channels. The obtained spectrum of the doubled layer-transition channel for spatial averages \eqref{eq:2dMERAnon-Edoub} is given in Eq.~\eqref{eq:2dMERAnon-Edoub-spec}.

For the nonary 2D TTNS, we simply omit all disentanglers. The five largest eigenvalues of its layer-transition channel for spatial averages are
\begin{equation}
	1,\quad
	\eta'_\non:=\frac{\chi^8}{1+\chi^2(1+\chi^2)(1+\chi^4)(1+\chi^8)},\quad
	\frac{\eta'_\non}{3},\quad \frac{\eta'_\non}{3},\quad\text{and}\quad
	\frac{\eta'_\non}{9}.
\end{equation}

\section{Discussion of deviating results for ZX-MPS and QMPS}\label{appx:discrepancies}
\subsection{Subclass of ZX-MPS considered in prior work}\label{appx:ZX-MPS}
\begin{figure*}[t]
	\label{fig:ZX-MPS_circuit}
	\includegraphics[width=\columnwidth]{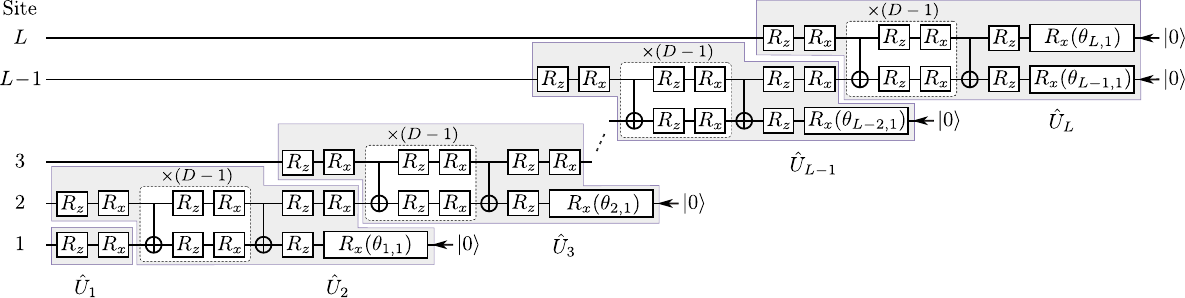}
	\caption{For depth-1 ZX-MPS as considered in Refs.~\cite{Zhao2021-5,Martin2023-7}, the MPS tensors \eqref{eq:ZX-MPS-1} are built from two-qubit circuits of $x$ and $z$ single-qubit rotations and one CNOT gate. In a generalization to depth-$D$ ZX-MPS, each tensor \eqref{eq:ZX-MPS-D} features $D$ CNOTs.}
\end{figure*}
As mentioned in Sec.~\ref{sec:priorWork}, Refs.~\cite{Zhao2021-5,Martin2023-7} discuss MPS gradient variances for a specific subclass of states that we call depth-1 ZX-MPS. They are MPS with open boundary conditions, bond dimension $m=2$, and single-site dimension $d=2$, where each tensor is defined by a circuit 
\begin{equation}\label{eq:ZX-MPS-1}
	\hU_j=\left[\id\otimes\hR_z\right]\cdot
	      \left[\id\otimes\hR_x\right]\cdot
	      \mathrm{CNOT}\cdot
	      \left[\hR_z\otimes\hR_z\right]\cdot
	      \left[\hR_x\otimes\hR_x\right]
	\quad\text{for}\quad j\geq 2,
\end{equation}
composed of single-qubit $x$ and $z$ rotations as well as one CNOT gate. Figure~\ref{fig:ZX-MPS_circuit} shows a graphical representation and a generalization to depth-$D$ ZX-MPS with $D$ CNOT gates per MPS tensor, i.e.,
\begin{equation}\label{eq:ZX-MPS-D}
	\hU_j=\left[\id\otimes\hR_z\right]\cdot
	      \left[\id\otimes\hR_x\right]\cdot
	      \left[\mathrm{CNOT}\cdot
	      (\hR_z\otimes\hR_z)\cdot
	      (\hR_x\otimes\hR_x)\right]^D
	\quad\text{for}\quad j\geq 2.
\end{equation}
\begin{figure*}[b]
	\label{fig:ZX-MPS}
	\includegraphics[width=0.485\columnwidth]{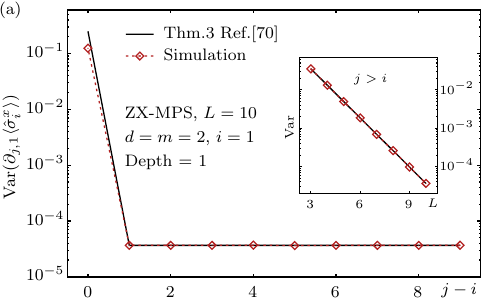}\hspace{0.02\textwidth}
	\includegraphics[width=0.485\columnwidth]{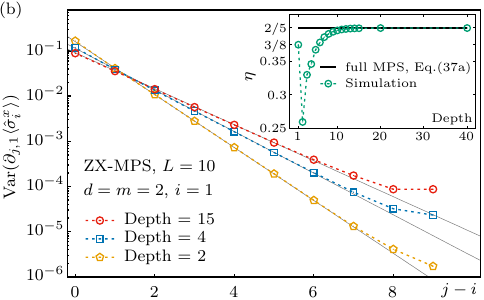}
	\caption{Gradient variance in ZX-MPS: (a) Simulation results from $10^7$ samples of depth-1 ZX-MPS as considered in Refs.~\cite{Zhao2021-5,Martin2023-7} with the tensor structure \eqref{eq:ZX-MPS-1}. We sample single-qubit gate angles uniformly from the interval $(-\pi,\pi]$ and evaluate the derivative variance for a rotation angle of $\hU_j$ and the Pauli matrix $\hsigma^x_i$ on site $i$. The data confirm the $j$-independent value of theorem~3 in Ref.~\cite{Martin2023-7} when $j > i$. The inset confirms the scaling $\sim\eta^{L-i}$.
	(b) This behavior changes drastically when we alleviate the restrictions of the depth-1 ZX-MPS \cite{Note1} by generalizing to depth-$D$ ZX-MPS \eqref{eq:ZX-MPS-D}. For $D\geq 2$, we recover the generic scaling $\sim\eta^{j-i}$ with $\eta$ approaching the value from Theorem~\ref{thrm:MPS-dist} with increasing depth. Each dotted line results from sampling $10^6\dotsc 10^7$ depth-$D$ ZX-MPS.}
\end{figure*}

At first sight, theorem~3 of Ref.~\cite{Martin2023-7} appears to contradict our Theorem~\ref{thrm:MPS-dist}: For unconstrained MPS, the variance of the energy-gradient for the MPS tensor $\hU_j$ of site $j$ and a single-site Hamiltonian $\hh_i$ on site $i$ decays exponentially in $|i-j|$ instead of just depending on $i$ as found for the depth-1 ZX-MPS \footnote{Note that we employ left-orthonormal MPS \cite{Schollwoeck2011-326,Barthel2022-112} such that gradients with respect to $\hU_j$ are exactly zero for $j<i$. In contrast, Refs.~\cite{Zhao2021-5,Martin2023-7} employ right-orthonormal MPS.}.
The reason for this interesting discrepancy is the specific ansatz \eqref{eq:ZX-MPS-1} of the depth-1 ZX-MPS.

The numerical data in Fig.~\ref{fig:ZX-MPS}a confirms theorem~3 of Ref.~\cite{Martin2023-7} for random depth-1 ZX-MPS. The doubled site-transition channel can be diagonalized
\begin{equation}\label{eq:ZX-MPS_E2}
	\E^\doub = \Avg\,\M_n^{\otimes 2}=|\hr_1\kket\bbra\id_4| + \eta\,|\hr_2\kket\bbra\hl_2| + \dots
\end{equation}
with $\eta$ denoting the second-largest-amplitude eigenvalue and $\M_n$ being the site-transition map \eqref{eq:MPS_F}.
For depth $D=1$, we find it to have rank two with $\eta= 3/8$. In the limit $D\to\infty$, one recovers the full-MPS channel \eqref{eq:E2}, which also has rank two but $\eta=2/5$ according to an evaluation of Eq.~\eqref{eq:MPSgradDecayAvg} with $d=m=2$.

Analogous to the discussion in Sec.~\ref{sec:MPS-single}, the essential step in the evaluation of the gradient variance is to determine the averages
\begin{subequations}
\begin{alignat}{3}
	\label{eq:ZX-MPS-XXavg}
	&\Avg\M^{\otimes 2}_{j+1}\circ\dotsb\circ\M^{\otimes 2}_{L}(|0,0\ket\bra 0,0|)
	&&=\big(\E^\doub\big)^{\otimes L-j}(|0,0\ket\bra 0,0|)\quad\text{and}\\
	\label{eq:ZX-MPS-YYavg}
	&\Avg\M^{\dag\otimes 2}_{j-1}\circ\dotsb\circ\M^{\dag\otimes 2}_{i+1}(\hL\otimes\hL)
	&&=\big(\E^{\doub\dag}\big)^{\otimes j-i-1}(\Avg \hL\otimes\hL)
\end{alignat}
\end{subequations}
with $\hL\stackrel{\eqref{eq:MPS_L}}{=}\bra 0_d|\,\hU_i^\dag\big[\id_m\otimes\hh\big]\hU_i\,|0_d\ket$.
In analogy to Eq.~\eqref{eq:gradVarProd}, the gradient variance is bilinear in the operators \eqref{eq:ZX-MPS-XXavg} and \eqref{eq:ZX-MPS-YYavg}. With the diagonalized form \eqref{eq:ZX-MPS_E2} they are
\begin{subequations}\label{eq:ZX-MPS-XX-YY}
\begin{align}\label{eq:ZX-MPS-XX}
	\big(\E^\doub\big)^{\otimes L-j}(|0,0\ket\bra 0,0|)
	&=\hr_1 + \mc{O}(\eta^{L-j}) \quad\text{and}\\\label{eq:ZX-MPS-YY}
	\big(\E^{\doub\dag}\big)^{\otimes j-i-1}(\Avg \hL\otimes\hL)
	&=\Tr(\hr_1 \Avg \hL\otimes\hL)\cdot\id_4 + \mc{O}(\eta^{j-i}).
\end{align}
\end{subequations}
Recall that the leading term $\sim \id_4$ from Eq.~\eqref{eq:ZX-MPS-YY} does not contribute to the gradient variance.
For generic MPS, the leading $\sim\eta^{j-i}$ term in the gradient variance [see Eq.~\eqref{eq:MPSgradDecayVar} for unconstrained MPS] results from combining the leading term ($\hr_1$) of Eq.~\eqref{eq:ZX-MPS-XX} with the subleading $\sim\eta^{j-i}$ term of Eq.~\eqref{eq:ZX-MPS-YY}. Due to particularities of the depth-1 ZX-MPS \eqref{eq:ZX-MPS-1}, the contribution of this combination vanishes exactly, and its gradient variance $\sim \eta^{L-i}$ is due to the combination of the two subleading terms in Eqs.~\eqref{eq:ZX-MPS-XX-YY}. This explains the unusual $j$-independent behavior described in theorem~3 of Ref.~\cite{Martin2023-7} and in Fig.~\ref{fig:ZX-MPS}a.
 
We recover the more generic $\sim\eta^{j-i}$ scaling of full MPS (Theorem~\ref{thrm:MPS-dist}) when sampling ZX-MPS with depths $D\geq 2$ as exemplified in Fig.~\ref{fig:ZX-MPS}b. For large $D$, $\eta$ converges to the full-MPS value \eqref{eq:MPSgradDecayVar}. It turns out that increasing $D$ is not the only option: For example, we also recover the usual $\sim\eta^{j-i}$ scaling when swapping the control and target bits of the CNOT in depth-1 ZX-MPS.

Lastly, we notice some discrepancies concerning theorem~3 in Ref.~\cite{Martin2023-7}: When $j = i$ or $j = i - 1$, our analysis suggests the following expressions for the gradient variance of depth-1 ZX-MPS.
\begin{alignat*}{3}
	&\frac{1}{4} \Big(\frac{1}{2} + \Big(\frac{3}{8}\Big)^{L-1}\Big)
	 &&\quad\text{when}\ j = i = 1,\\
	&3\cdot\Big(\frac{1}{8}\Big)^2\Big(\frac{1}{2} + \frac{11}{8}\cdot\Big(\frac{3}{8}\Big)^{L-i-1}\Big)
	 &&\quad\text{when}\ L>j = i >1,\ \ \text{and}\\
	&3\cdot\Big(\frac{1}{8}\Big)^2\Big(\frac{1}{2} + \Big(\frac{3}{8}\Big)^{L-i}\Big)
	 &&\quad\text{when}\ j=i-1>1.
\end{alignat*}
Recall that site indices are reflected ($i\leftrightarrow L-i+1$) compared to Ref.~\cite{Martin2023-7} because of the different MPS-tensor orthonormality convention \cite{Note4}.

\subsection{QMPS considered in prior work}\label{appx:QMPS}
\begin{figure*}[t]
	\label{fig:QMPS}
	\includegraphics[width=0.485\columnwidth]{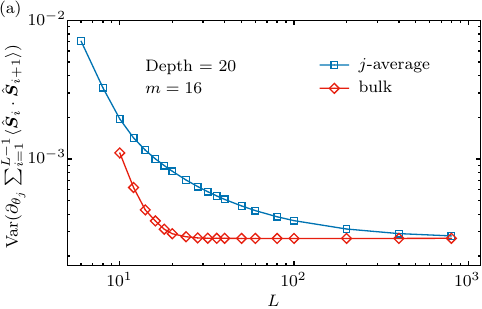}\hspace{0.02\textwidth}
	\includegraphics[width=0.485\columnwidth]{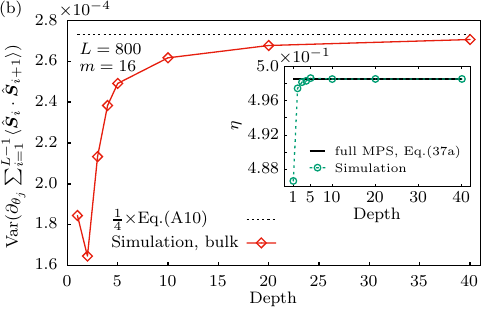}
	\caption{Gradient variance in QMPS: (a) For randomly sampled QMPS and the spin-1/2 Heisenberg-chain Hamiltonian $\hH=\sum_i\hat{\vec{S}}_i\cdot\hat{\vec{S}}_{i+1}$, we show the energy-gradient variance as studied in Ref.~\cite{Liu2019-1}. The upper (blue) curve is the average over all rotation angles of the ansatz. For the lower (red) curve, we only average over a central (bulk) region of the lattice which is not affected by the boundaries, or over a few central sites at the smaller $L$.
	The gradient variance clearly converges to a system-size independent constant in accordance with Theorem~\ref{thrm:MPSnn}. The initial decay is a finite-size effect.
	(b) With increasing depth of the circuit that defines the QMPS tensors \cite{Liu2019-1}, the expressiveness of the ansatz approaches that of full MPS. The shown (Euclidean) gradient variance closely approximates the theoretical predictions from Theorem~\ref{thrm:MPSnn}, except for an additional prefactor of $1/4$ originating from definitional differences. The parameter $\eta$ also quickly approaches the value \eqref{eq:MPSgradDecayAvg} for full MPS. For the main panels, we sampled $10^4$ QMPS with bond dimension $m=16$. For data in the inset, we sampled $10^5$ instances of the site-transition maps $\F^{\otimes 2}$ to extract the second-largest-amplitude eigenvalue $\eta$ of the doubled channel $\E^\doub$.}
\end{figure*}

Ref.~\cite{Liu2019-1} studied a subclass of MPS called QMPS and reported a power-law decay for the site-averaged gradient variance for extensive Hamiltonians as the system size $L$ increases from 6 to 20. We numerically simulate the same QMPS states for the wider range of system sizes $L=6,\dotsc,800$. The results in Fig.~\ref{fig:QMPS}a show that the initial decay is due to finite-size effects consistent with the subleading terms in Theorem~\ref{thrm:MPSnn} and that the site-averaged energy-gradient variance converges to a system-size independent value. When increasing the depth of the circuits that define the tensors of the QMPS, their expressiveness approaches that of the full (unconstrained) MPS studied in Sec.~\ref{sec:MPS}. This is corroborated by the data in Fig.~\ref{fig:QMPS}b: The average (Euclidean) gradient variance quickly approaches the theoretical predictions of Theorem~\ref{thrm:MPSnn} for the Riemannian gradient variance of full MPS, except for an additional prefactor of $1/4$ originating from definitional differences.
While our data is consistent with that in Fig.~6 of Ref.~\cite{Liu2019-1}, there are some quantitative differences and there also appears to be some inconsistency between data points in Figs.~6a and 6b of Ref.~\cite{Liu2019-1}.

\end{document}